\documentclass{aa}
\usepackage{graphics}
\usepackage{txfonts}
\usepackage{longtable}
\usepackage{graphicx}
\begin{document}

\title{The close circumstellar environment of Betelgeuse\thanks{Based on observations made with ESO telescopes at Paranal Observatory, under ESO program 082.D-0172(A).}}
\subtitle{Adaptive optics spectro-imaging in the near-IR with VLT/NACO}
\titlerunning{Adaptive optics spectro-imaging of Betelgeuse}
\authorrunning{P. Kervella et al.}
\author{
P. Kervella\inst{1}
\and
T. Verhoelst\inst{2}
\and
S. T. Ridgway\inst{3}
\and
G. Perrin\inst{1}
\and
S. Lacour\inst{1}
\and
J. Cami\inst{4}
\and
X. Haubois\inst{1}
}
\offprints{P. Kervella}
\mail{Pierre.Kervella@obspm.fr}
\institute{
LESIA, Observatoire de Paris, CNRS\,UMR\,8109, UPMC, Universit\'e Paris Diderot, 5 place
Jules Janssen, 92195 Meudon, France
\and
Instituut voor Sterrenkunde, K. U. Leuven, Celestijnenlaan 200D, B-3001 Leuven, Belgium
\and
National Optical Astronomy Observatories, 950 North Cherry Avenue, Tucson, AZ 85719, USA
\and
Physics and Astronomy Dept, University of Western Ontario, London ON N6A 3K7, Canada
}
\date{Received ; Accepted}
\abstract
% Context
{Betelgeuse is one the largest stars in the sky in terms of angular diameter.
Structures on the stellar photosphere have been detected in the visible and near-infrared as well
as a compact molecular environment called the MOLsphere. Mid-infrared observations have revealed
the nature of some of the molecules in the MOLsphere, some being the precursor of dust.}
% Aims
{Betelgeuse is an excellent candidate to understand the process of mass loss in red supergiants.
Using diffraction-limited adaptive optics (AO) in the near-infrared, we probe the photosphere
and close environment of Betelgeuse to study the wavelength dependence of its
extension, and to search for asymmetries.}
% Methods
{We obtained AO images with the VLT/NACO instrument, taking advantage of
the ``cube" mode of the CONICA camera to record separately a large number of short-exposure
frames. This allowed us to adopt a ``lucky imaging" approach for the data reduction, and
obtain diffraction-limited images over the spectral range $1.04-2.17\,\mu$m in 10 narrow-band filters.}
% Results
{In all filters, the photosphere of Betelgeuse appears partly resolved.
We identify an asymmetric envelope around the star, with in particular a relatively bright
``plume" extending in the southwestern quadrant up to a radius of approximately six times the
photosphere. The CN molecule provides an excellent match to the 1.09\,$\mu$m bandhead in absorption
in front of the stellar photosphere, but the emission spectrum of the plume is more difficult to interpret.}
% Conclusions
{Our AO images show that the envelope surrounding Betelgeuse has a complex and irregular
structure. We propose that the southwestern plume is linked either to the presence of a
convective hot spot on the photosphere, or to the rotation of the star.}

\maketitle

%__________________________________Introduction
\section{Introduction}

\object{Betelgeuse} ($\alpha$ Ori, \object{HD 39801}, \object{HR 2061}) is one of the brightest stars at infrared wavelengths and has one of the largest apparent photospheric diameters ($\sim$44~mas in the $K$ band, from Perrin et al.~\cite{perrin04}). It is a red supergiant of spectral type M2Iab with irregular flux variations. Because of its brightness and of its angular size, it has been an easy target for interferometers. It can even be resolved with large telescopes at short wavelengths, as demonstrated by Gilliland \& Dupree~(\cite{gilliland96}), who resolved its extended chromosphere using the \emph{Hubble Space Telescope} at ultraviolet wavelengths. Several mysteries remain to be solved for such an object among which the structure of its convection and the mechanism of its mass loss.
Similarities with Mira stars have been noted, including a  significant thermal infrared excess, indicative of the presence of circumstellar dust.  An important difference is that Betelgeuse is not subject to regular large amplitude pulsations which could levitate material high enough above the atmosphere where dust could form and be blown away by radiation pressure. The reason for the extension of the atmosphere is still unclear and another mechanism for dust formation and mass loss needs to be invoked.  A careful scrutiny of the object from the photospheric scale up to the distance in its circumstellar envelope (CSE) where dust can condense is the path to understanding this prototypical and enigmatic star. It is also known from interferometric observations in the thermal infrared domain by Bester et al.~(\cite{bester96}) that Betelgeuse experiences episodic mass loss.
Our new NACO observations aim at probing the close environment of Betelgeuse to etablish a link between the photosphere and mass loss. After a presentation of our observations in Sect.~\ref{observations}, we detail our analysis of the NACO images in Sect.~\ref{analysis}, and discuss the observed features in Sect.~\ref{discussion}.

%__________________________________Observations
\section{Observations and data reduction \label{observations}}

\subsection{Instrumental setup and observation log \label{log}}

We observed Betelgeuse on the nights of 2 and 3 January 2009 using the Nasmyth Adaptive Optics System (NAOS, Rousset et al.~\cite{rousset03}) of the Very Large Telescope (VLT), coupled to the CONICA infrared camera (Lenzen et al.~\cite{lenzen98}), abbreviated as NACO. NAOS is equipped with a tip-tilt mirror and a deformable mirror controlled by 185 actuators, as well as two wavefront sensors: one for visible light and one for the infrared domain. For our observations, we exclusively used the visible light wavefront sensor. The detector is a $1024 \times 1024$ pixel ALADDIN InSb array. As its name indicates, NACO is installed at the Nasmyth focus of the Unit Telescope~4 (Yepun), located on the eastern side of the VLT platform. The CONICA camera was read using the newly implemented \emph{cube} mode that allows to record very short exposures (down to 7.2\,ms) over part of the detector. As discussed in Sect.~\ref{datared}, this new capability is essential to reach truly diffraction limited imaging.

%______________ Figure
\begin{figure}[]
\centering
\includegraphics[width=\hsize]{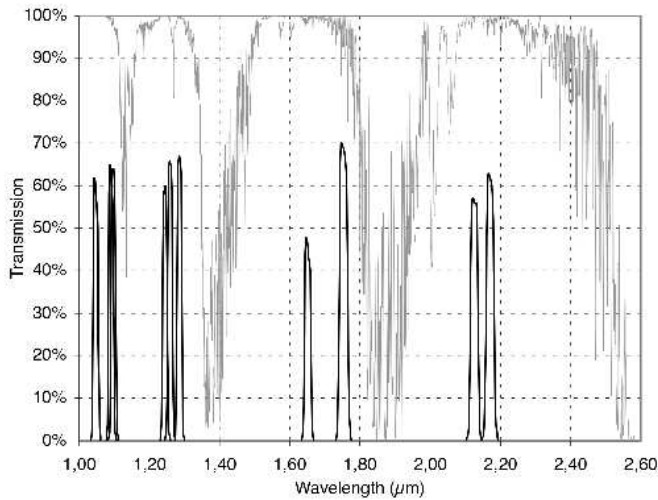} 
\caption{Transmissions of the NACO narrow-band filters (black curves) and of the atmosphere (from Lord~\cite{lord92}, thin grey curve).\label{nb_filters}}
\end{figure}

We selected the smallest available pixel scale of $13.26 \pm 0.03$\,mas/pix (Masciadri et al.~\cite{masciadri03}). Table~\ref{naco_log} gives the list of the observations of Betelgeuse and the PSF calibrator stars \object{Aldebaran} (\object{HD 29139}), 31\,Ori (\object{HD 36167}) and $\delta$\,Phe (\object{HD 9362}). 
Due to the brightness of Betelgeuse and Aldebaran, we employed narrow- and intermediate-band filters together with a neutral density filter (labeled ``{\tt ND2\_short}") with a transmission of about 1.5\%. The observations of our fainter, secondary PSF calibrators (31\,Ori and $\delta$\,Phe) were obtained without neutral density filter. The transmission curves of the 10 narrow-band filters we used for our observations are shown in Fig.~\ref{nb_filters}. They are also available through the NACO instrument web page at ESO\footnote{http://www.eso.org/sci/facilities/paranal/instruments/naco/}. 

\subsection{Data processing \label{datared}}

The individual raw images were pre-processed (bias subtraction, flat-fielding, bad pixel masking) using the Yorick\footnote{http://yorick.sourceforge.net/} and IRAF\footnote{IRAF is distributed by the NOAO, which are operated by the Association of Universities for Research in Astronomy, Inc., under cooperative agreement with the National Science Foundation.} software packages in a standard way, except that we did not subtract the negligible sky background.

\begin{figure*}[]
\centering
\includegraphics[width=4.3cm]{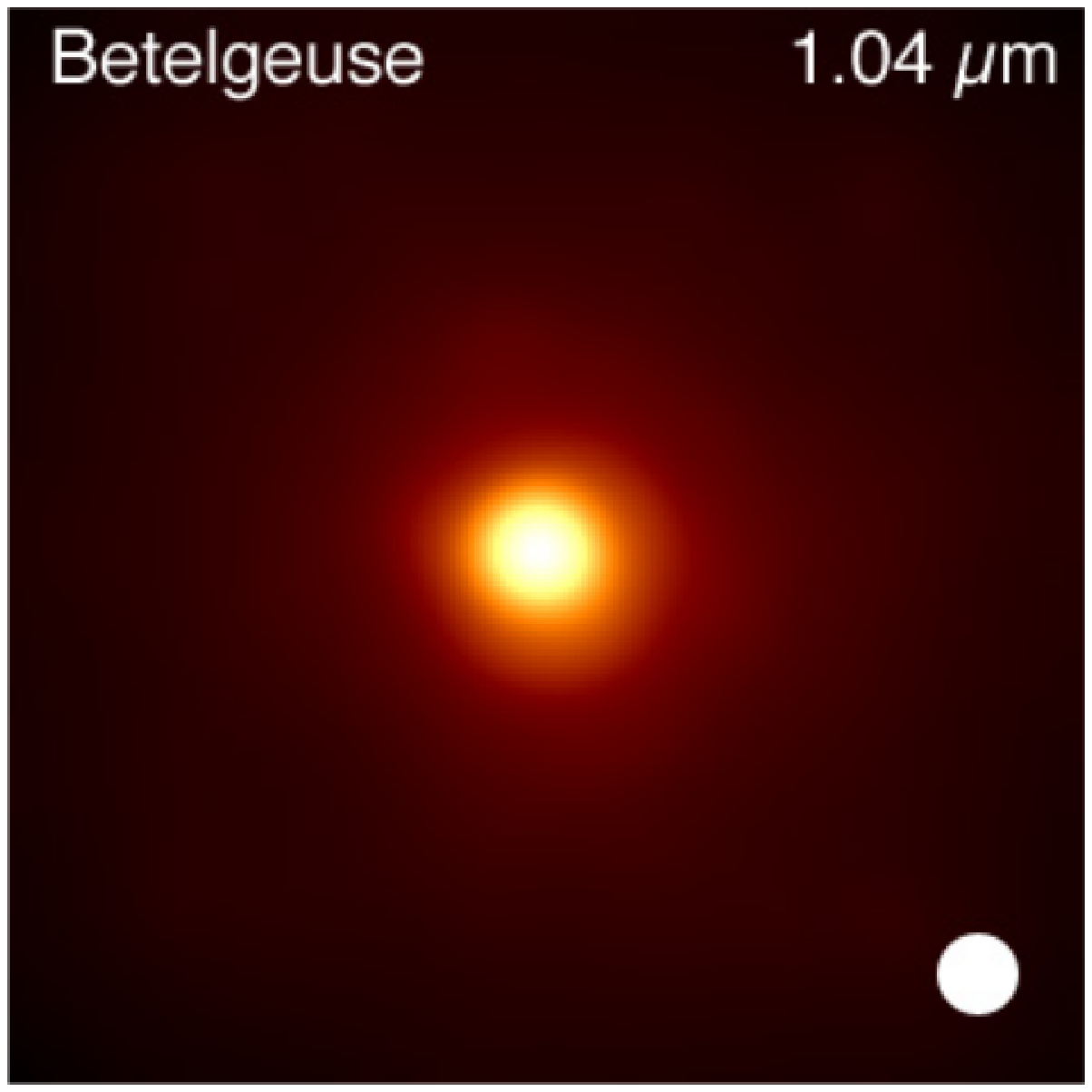} \includegraphics[width=4.3cm]{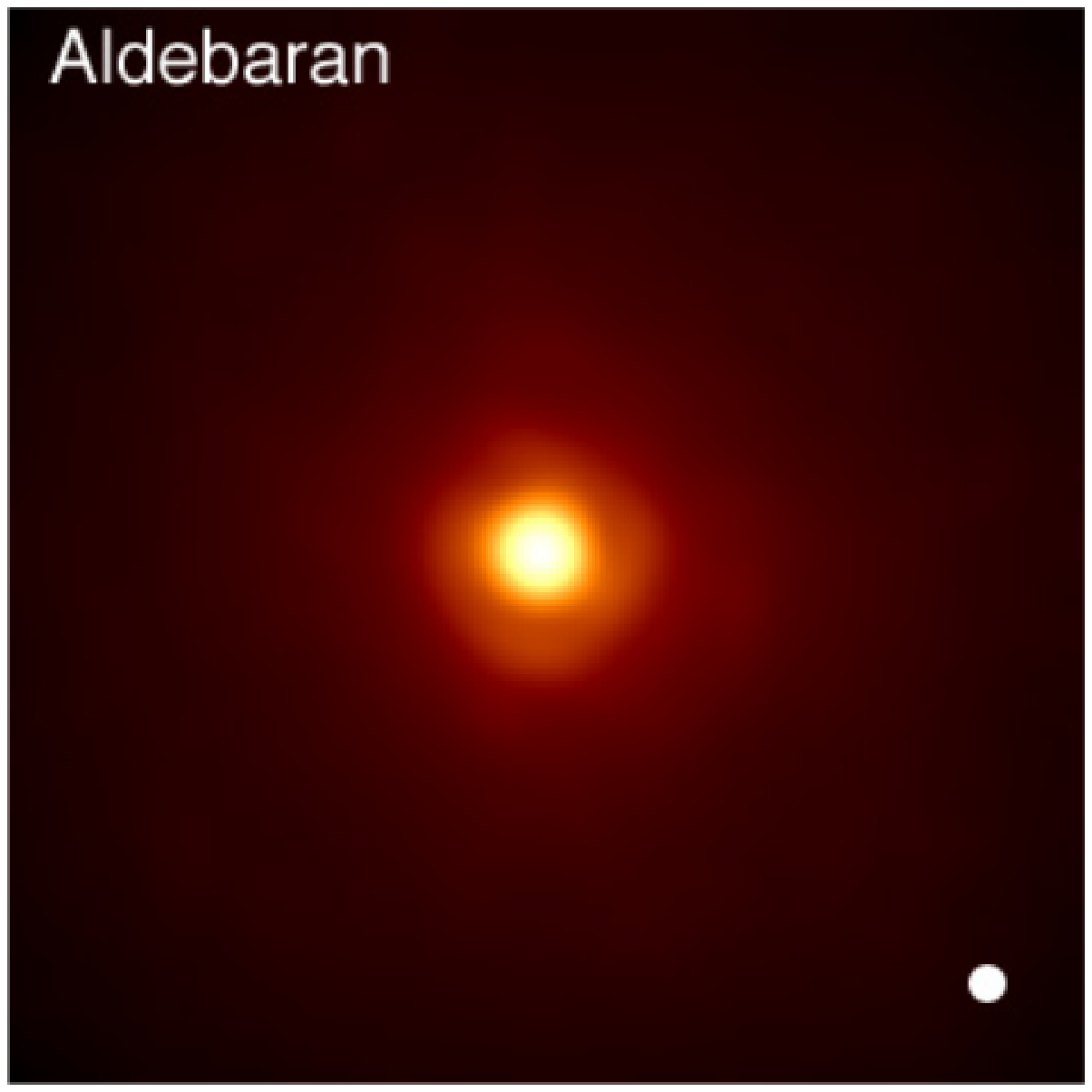} \hspace{2mm}
\includegraphics[width=4.3cm]{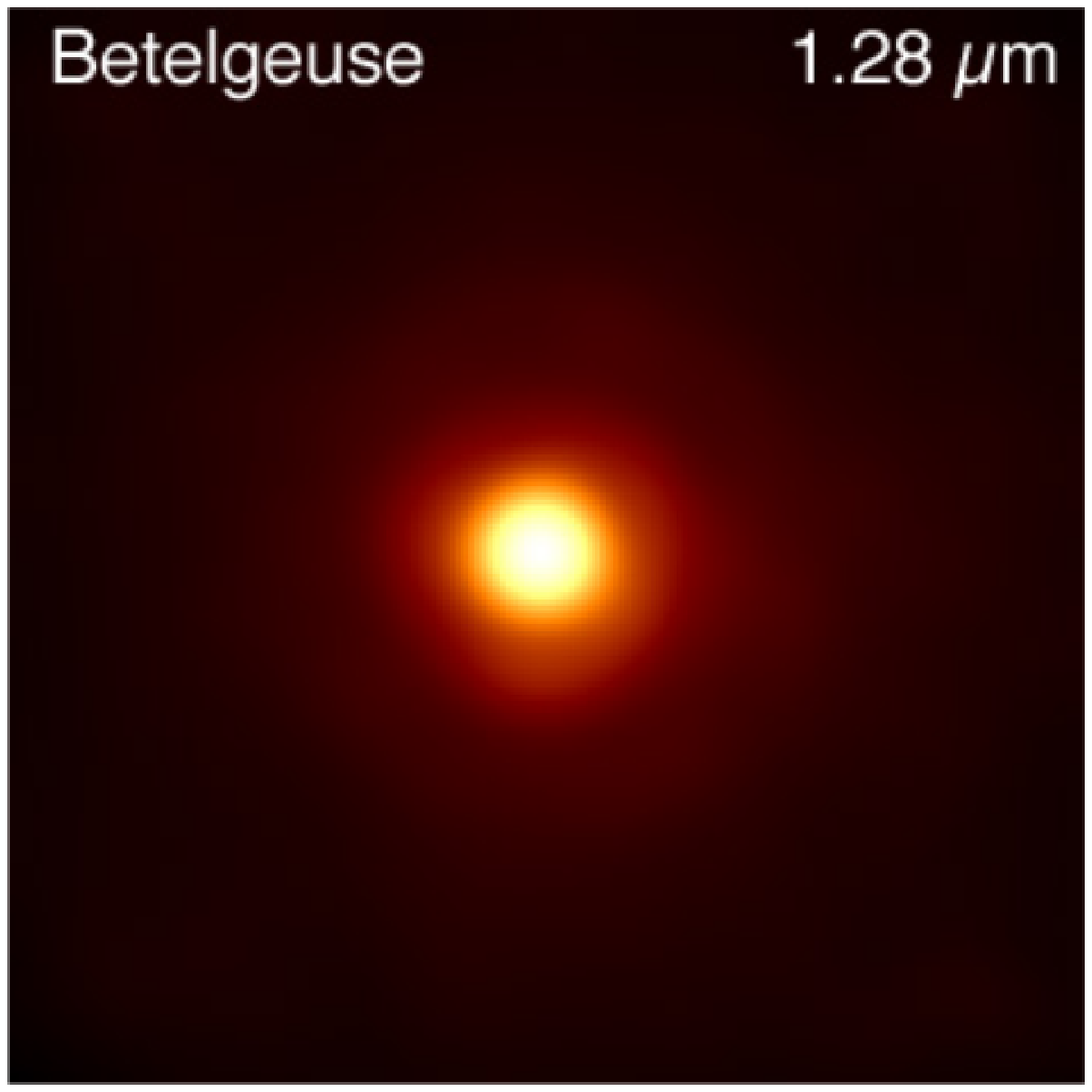} \includegraphics[width=4.3cm]{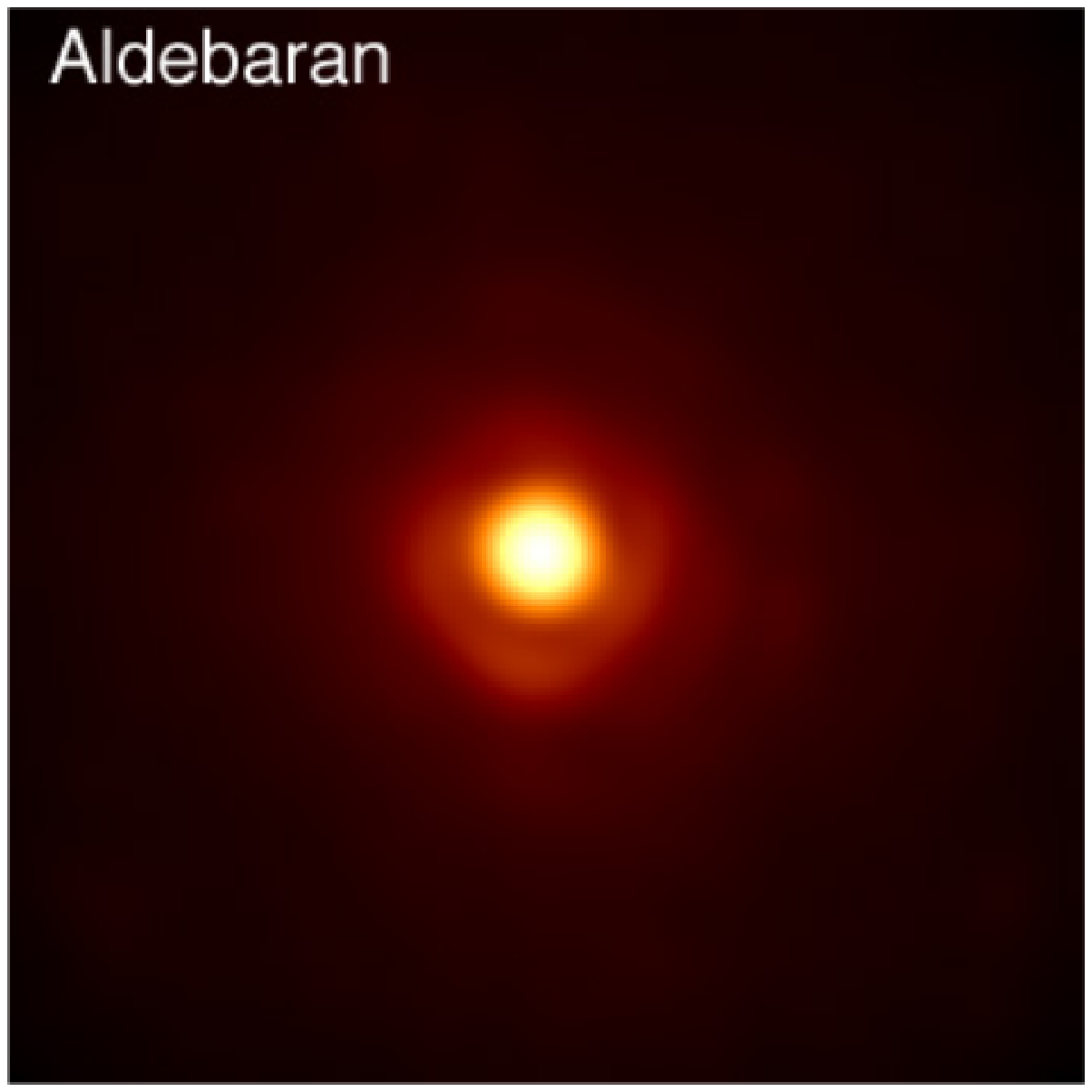}

\includegraphics[width=4.3cm]{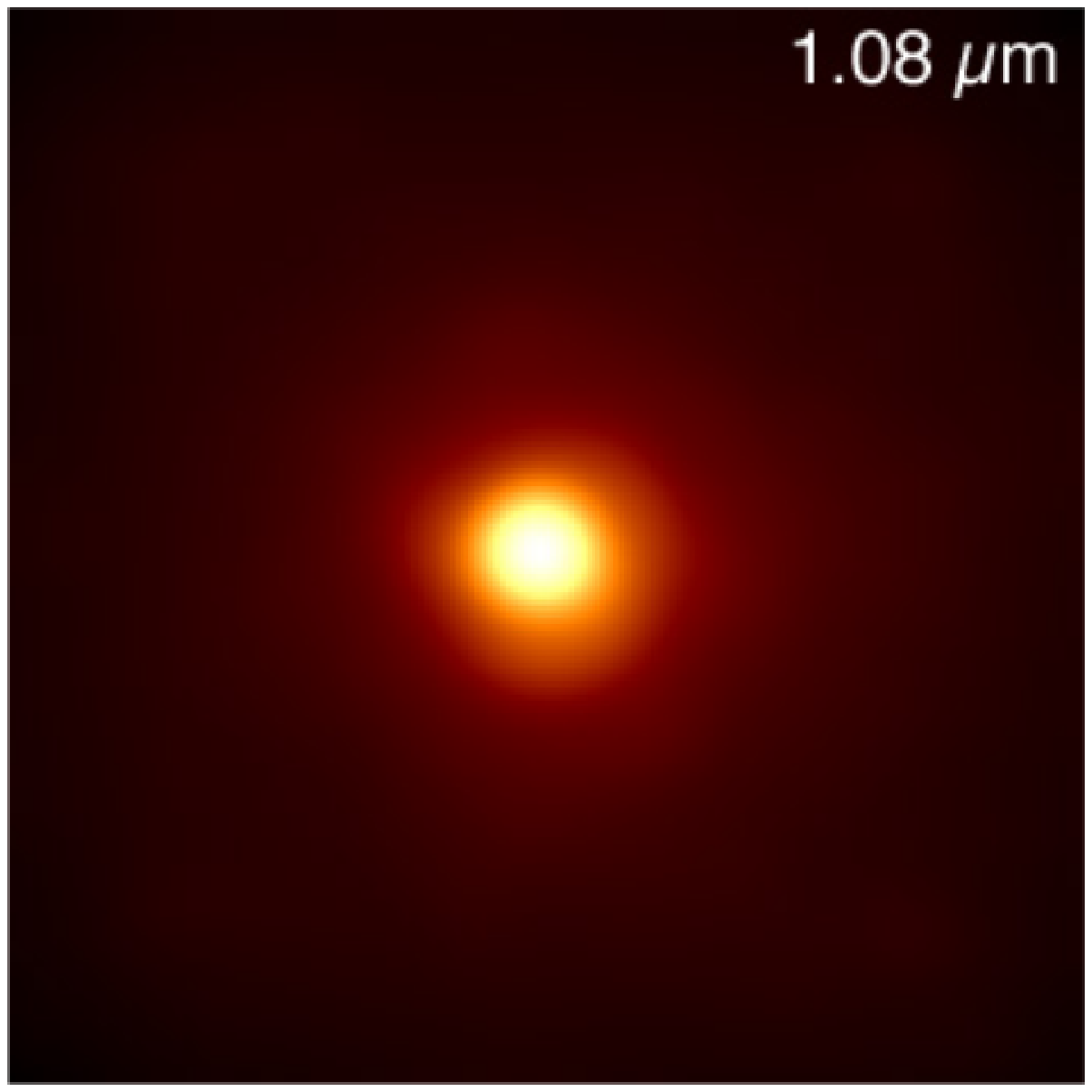} \includegraphics[width=4.3cm]{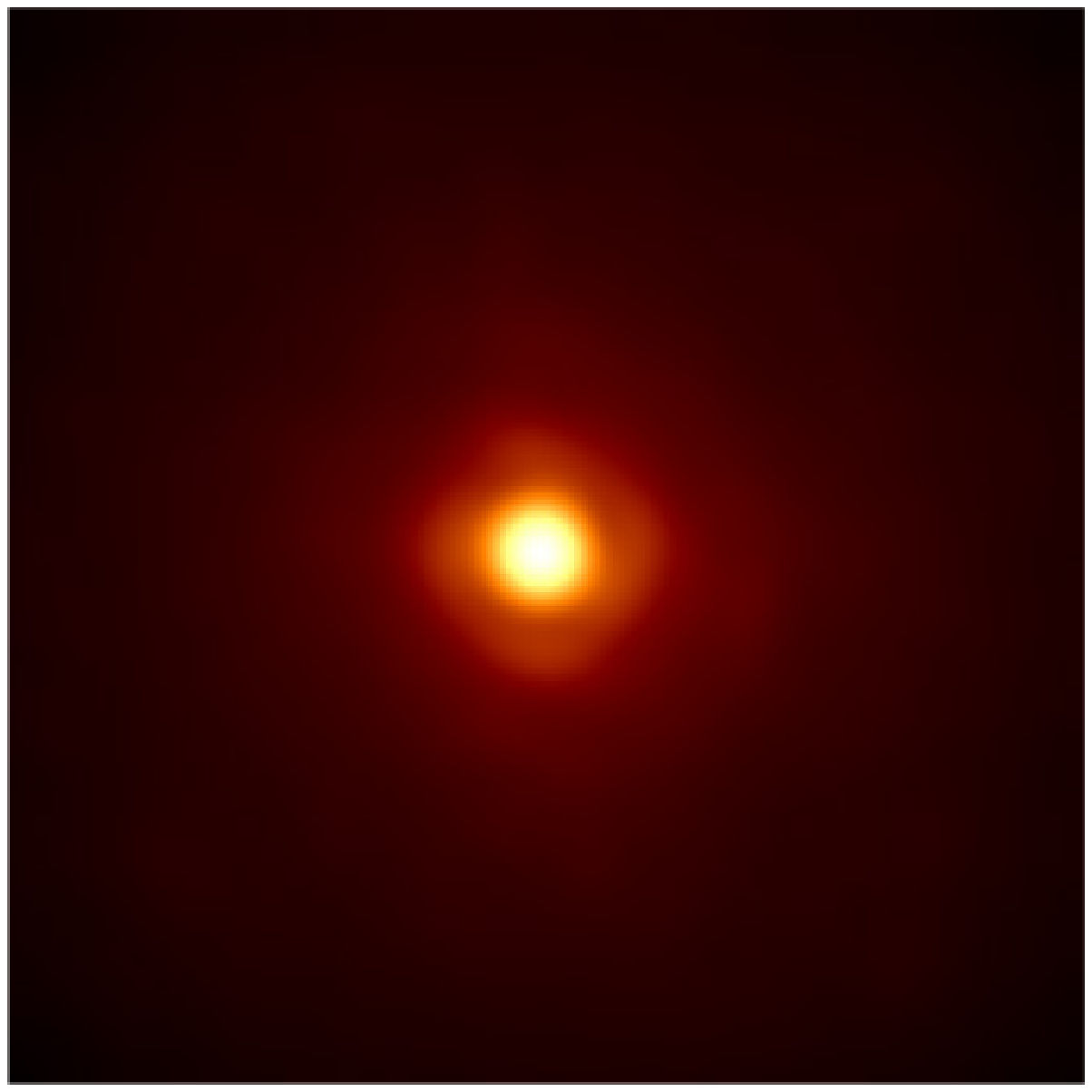} \hspace{2mm}
\includegraphics[width=4.3cm]{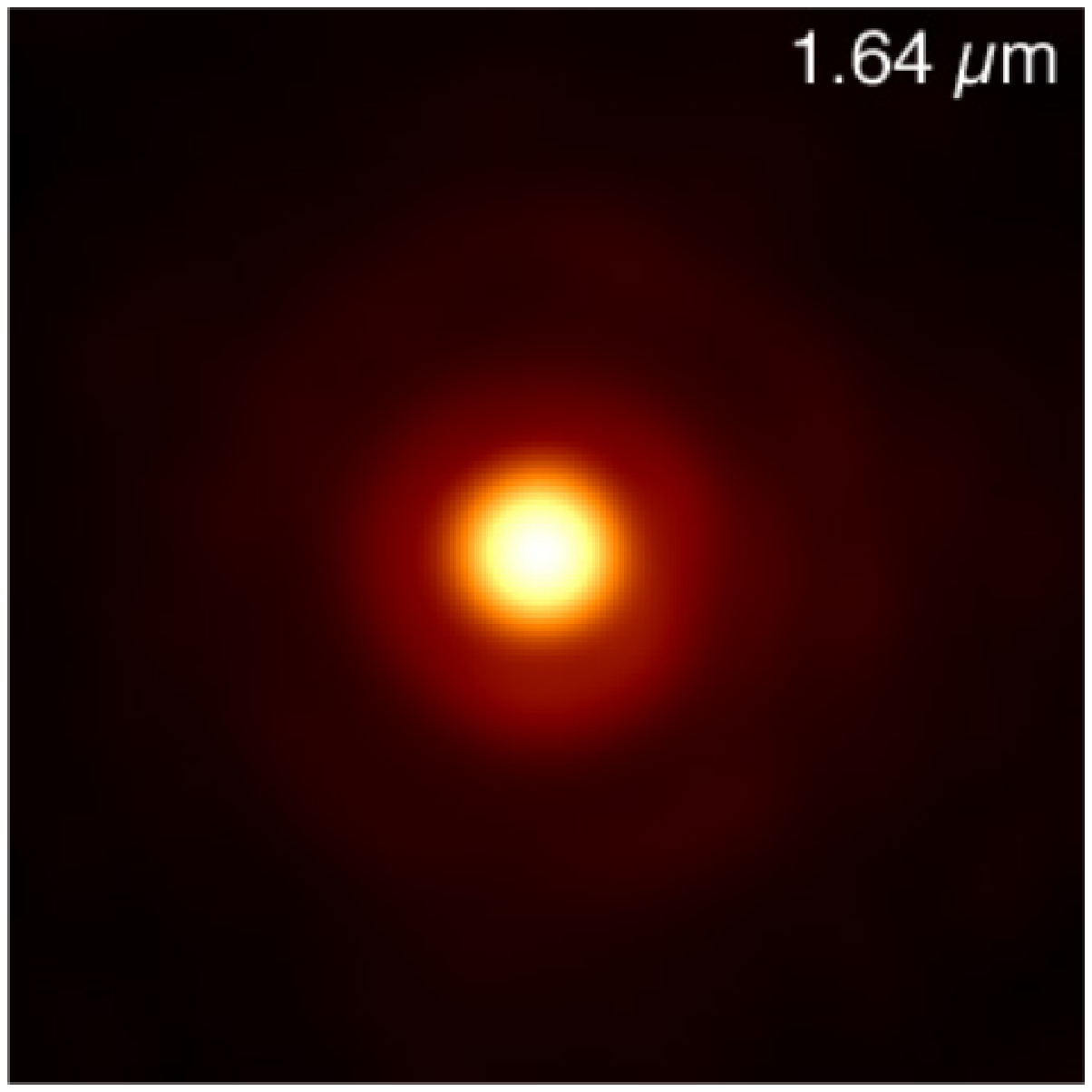} \includegraphics[width=4.3cm]{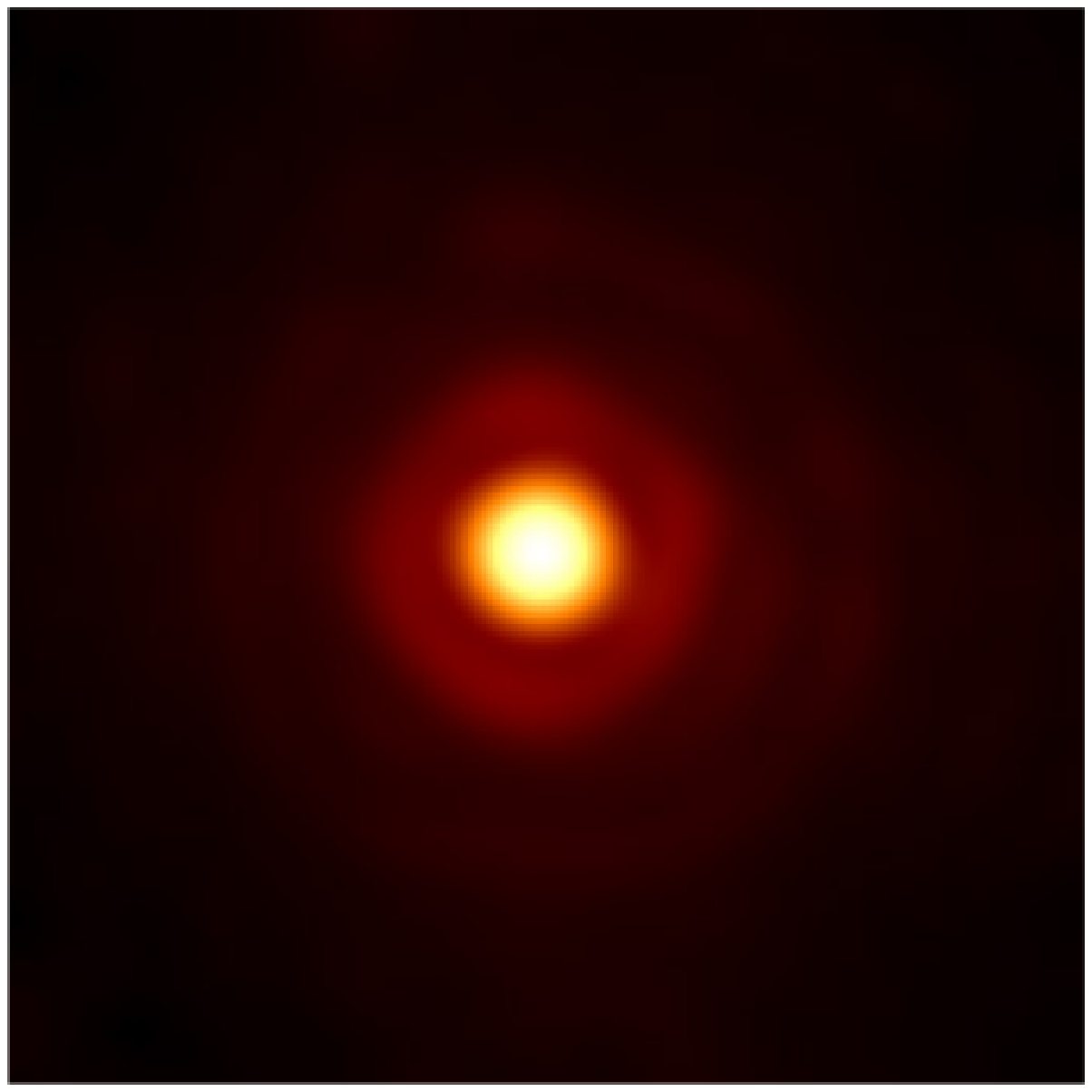}

\includegraphics[width=4.3cm]{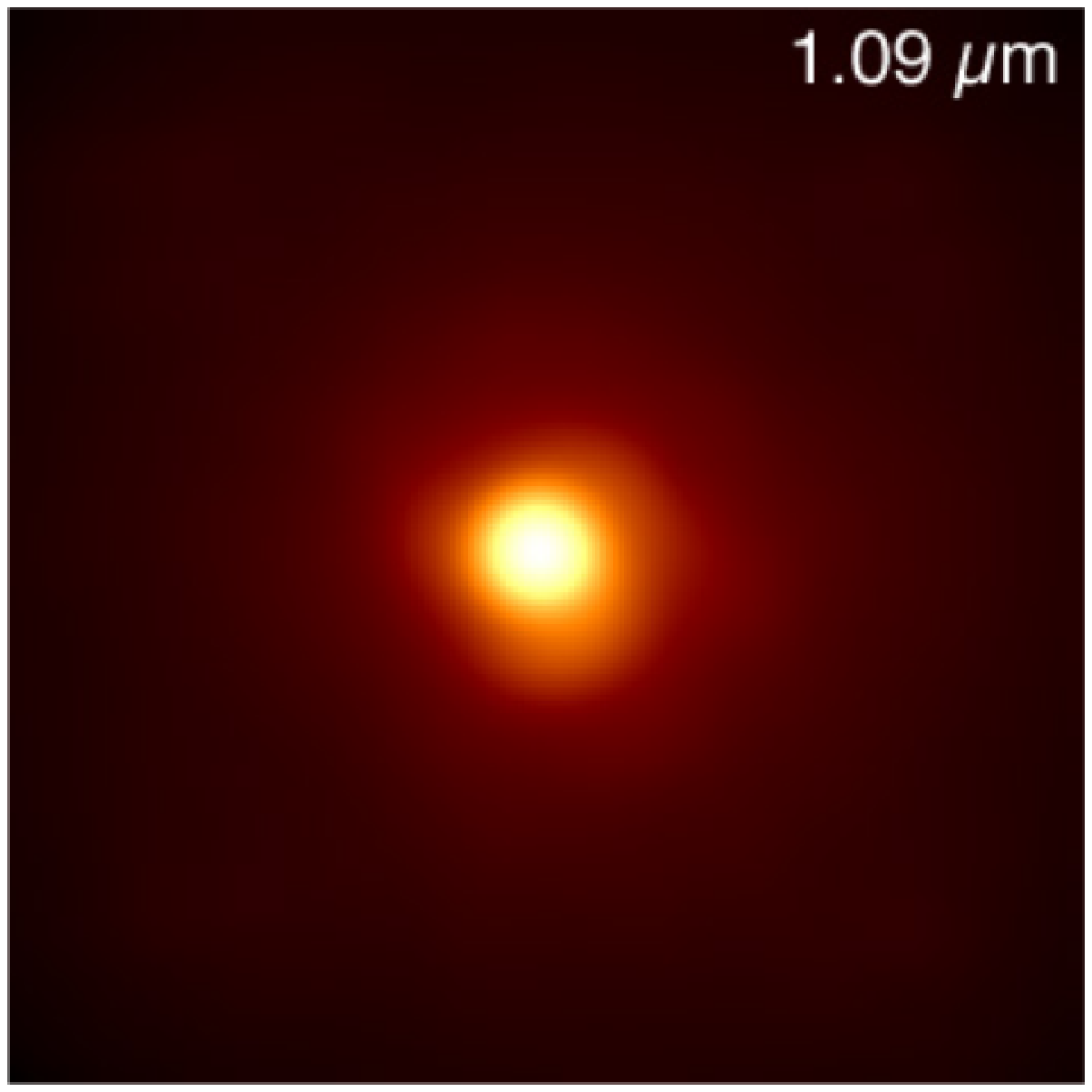} \includegraphics[width=4.3cm]{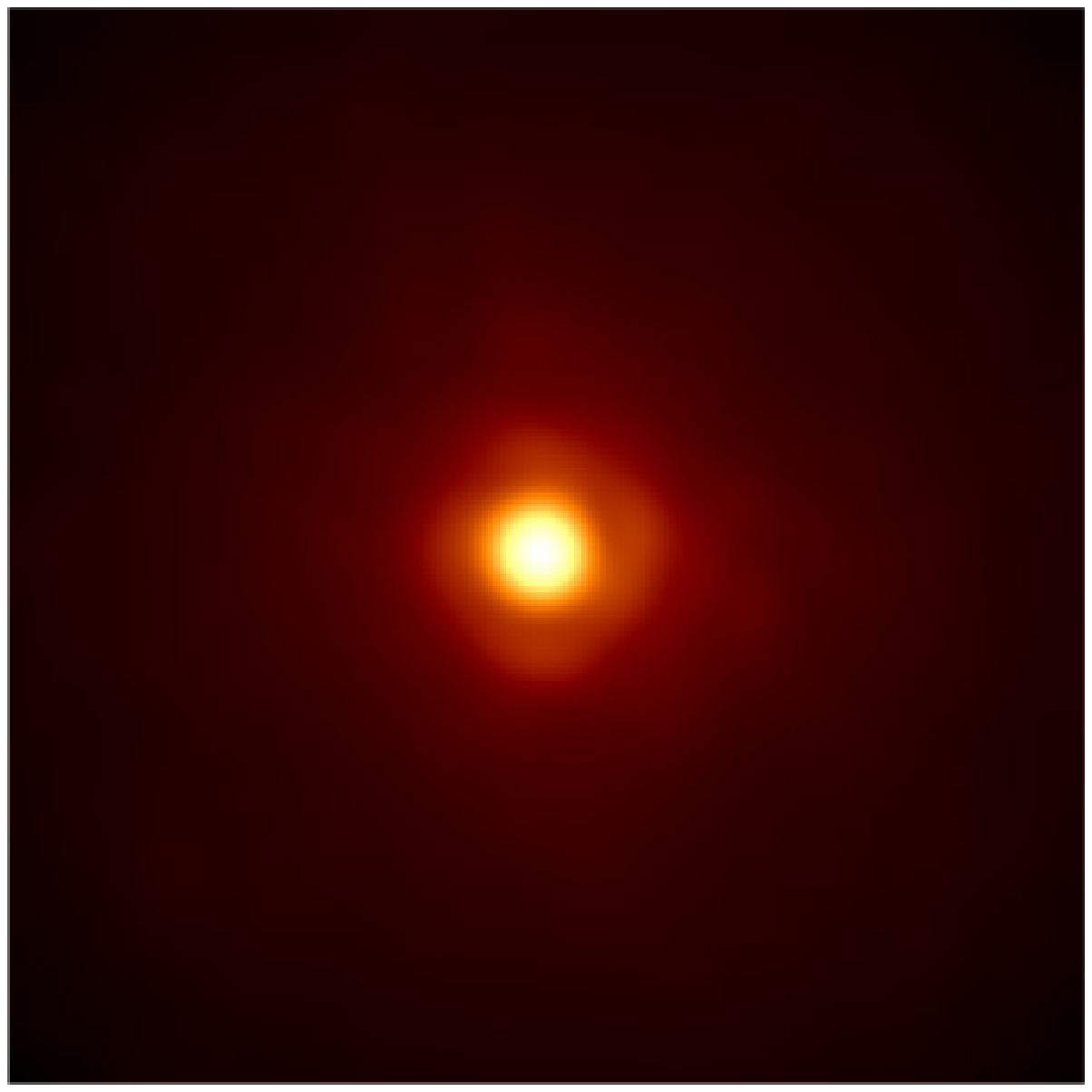} \hspace{2mm}
\includegraphics[width=4.3cm]{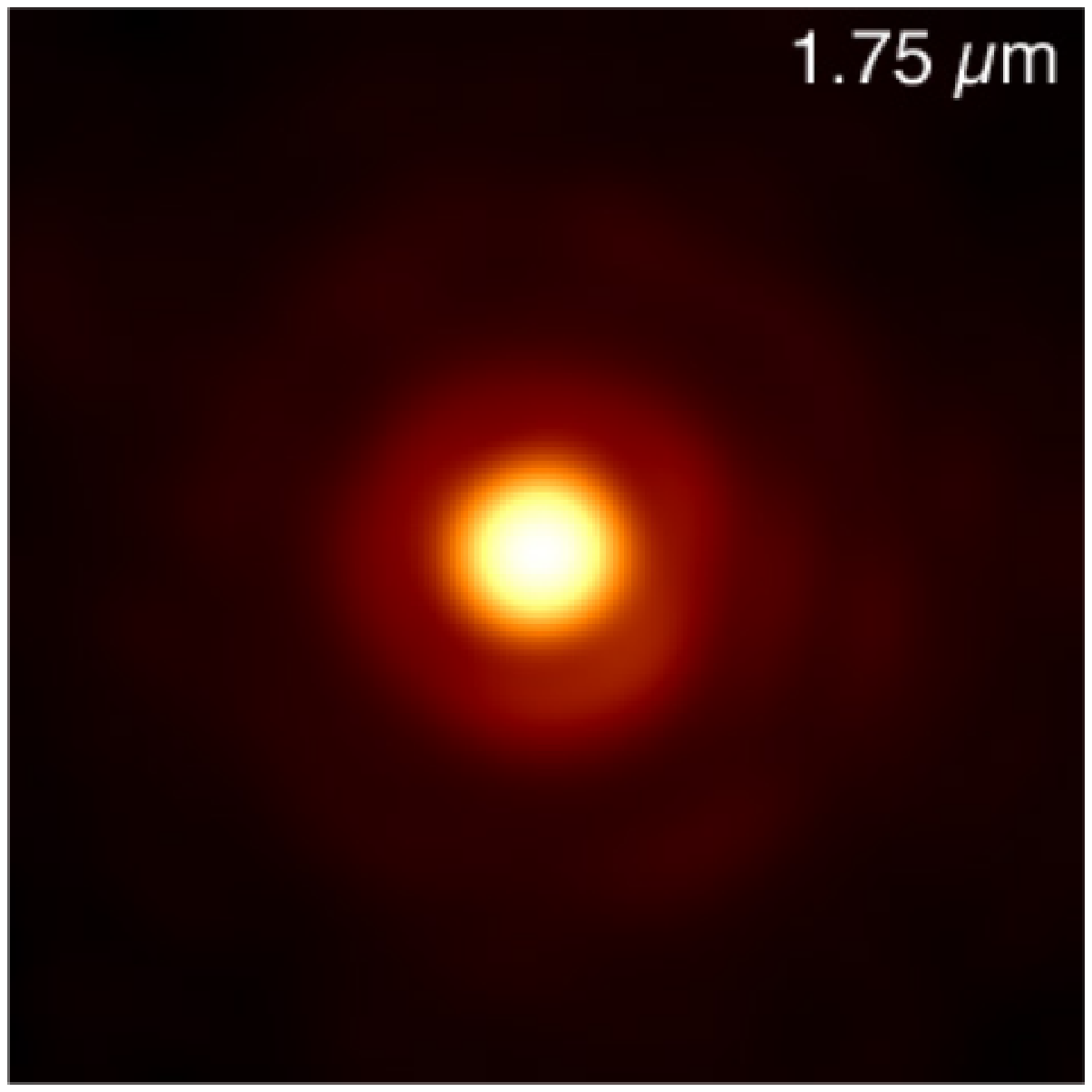} \includegraphics[width=4.3cm]{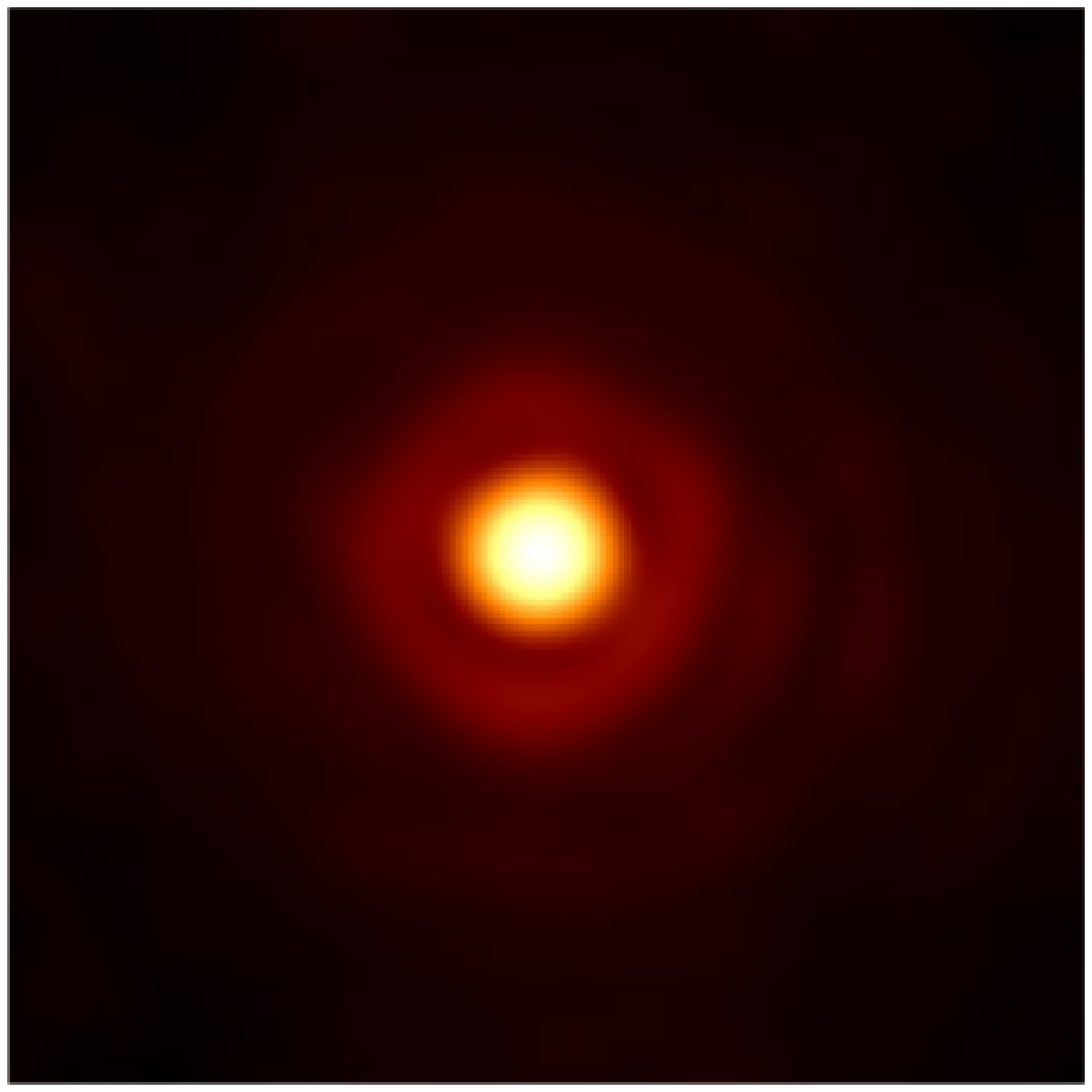}

\includegraphics[width=4.3cm]{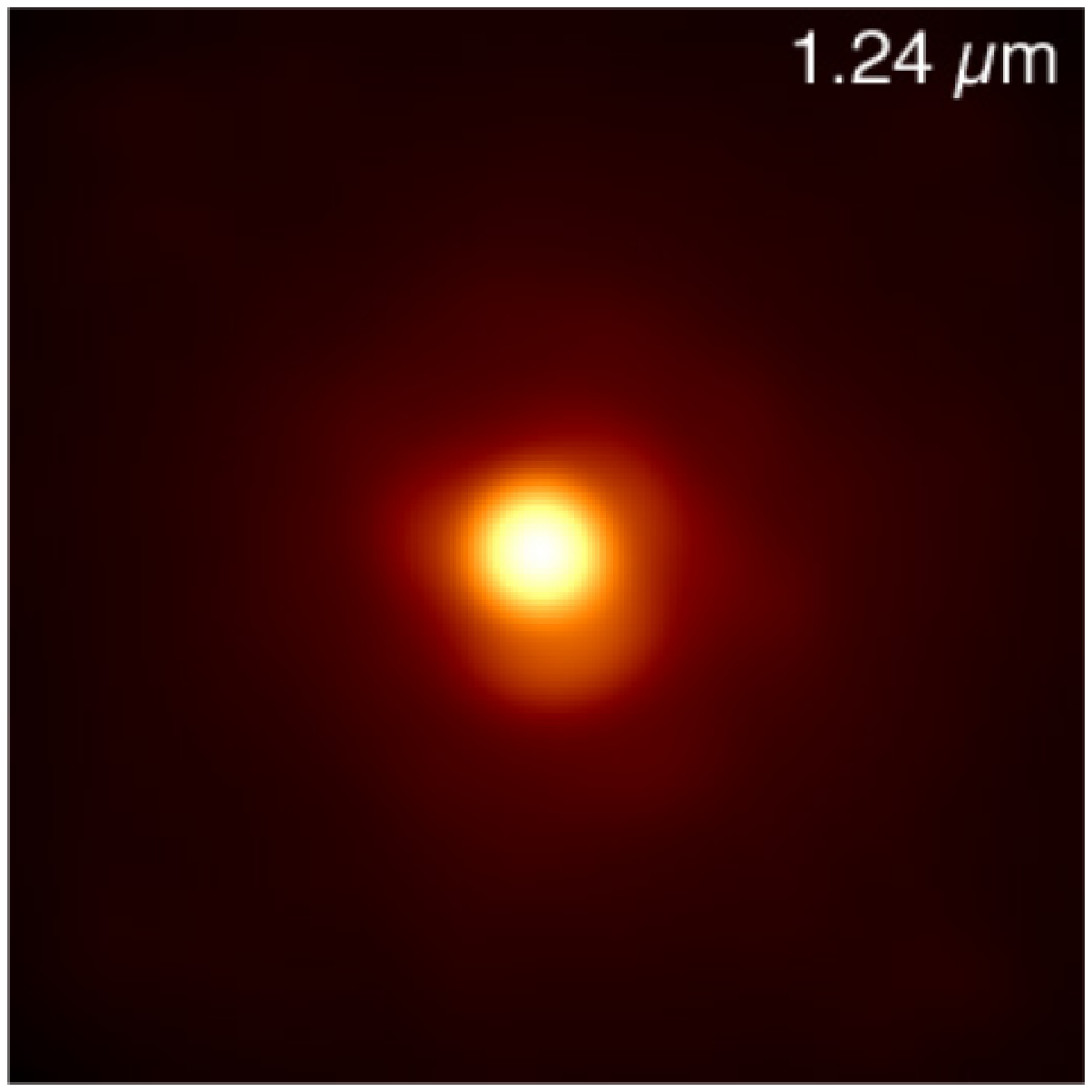} \includegraphics[width=4.3cm]{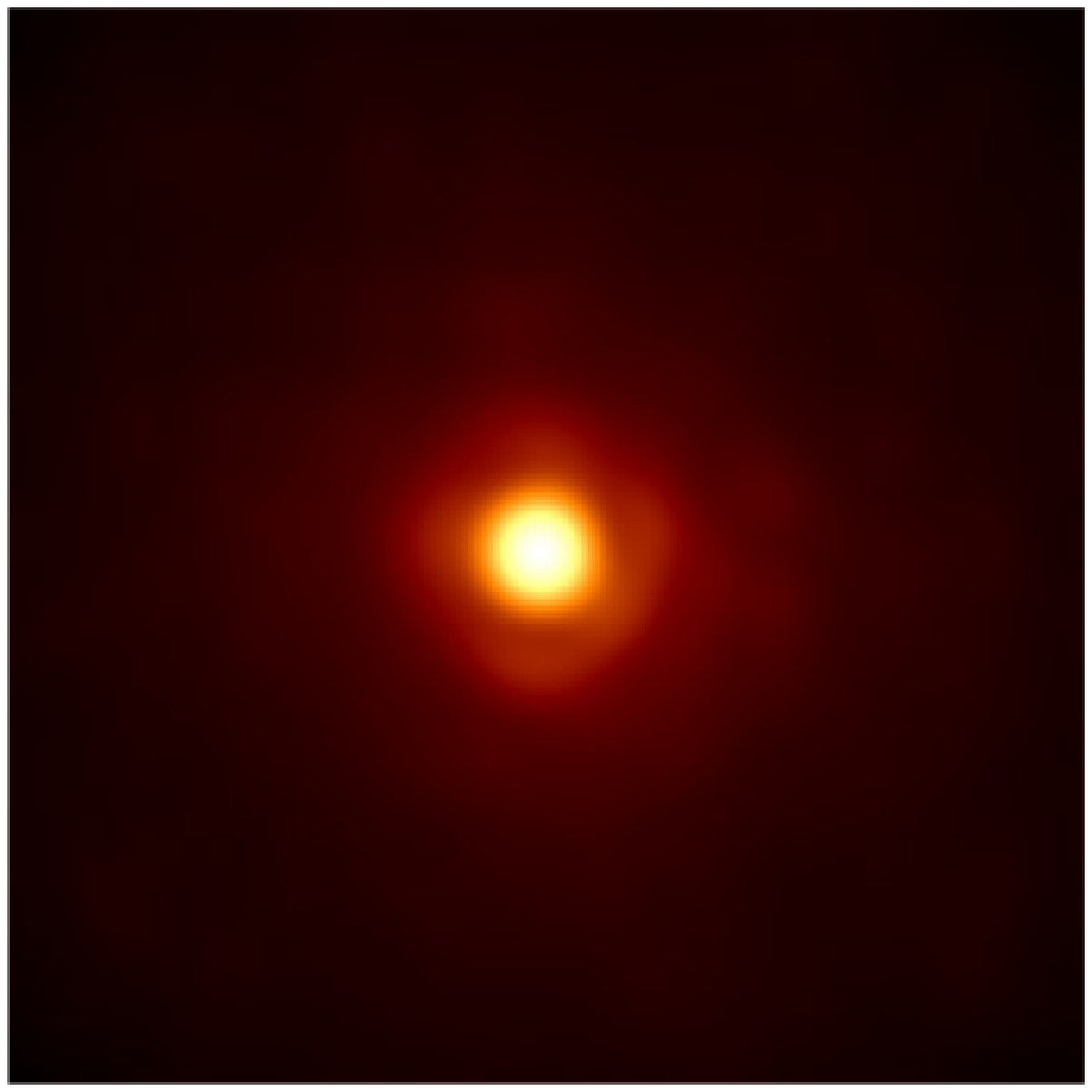} \hspace{2mm}
\includegraphics[width=4.3cm]{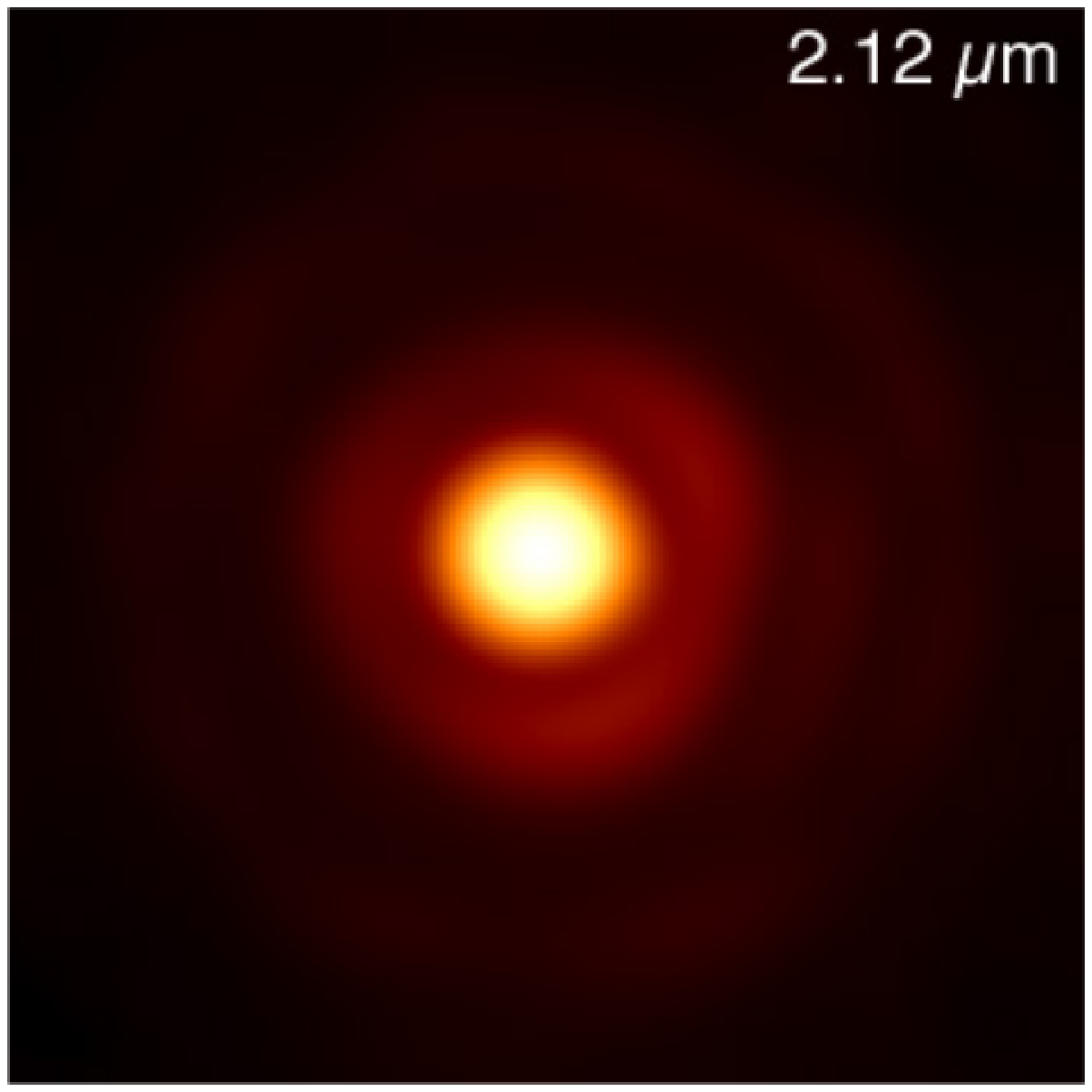} \includegraphics[width=4.3cm]{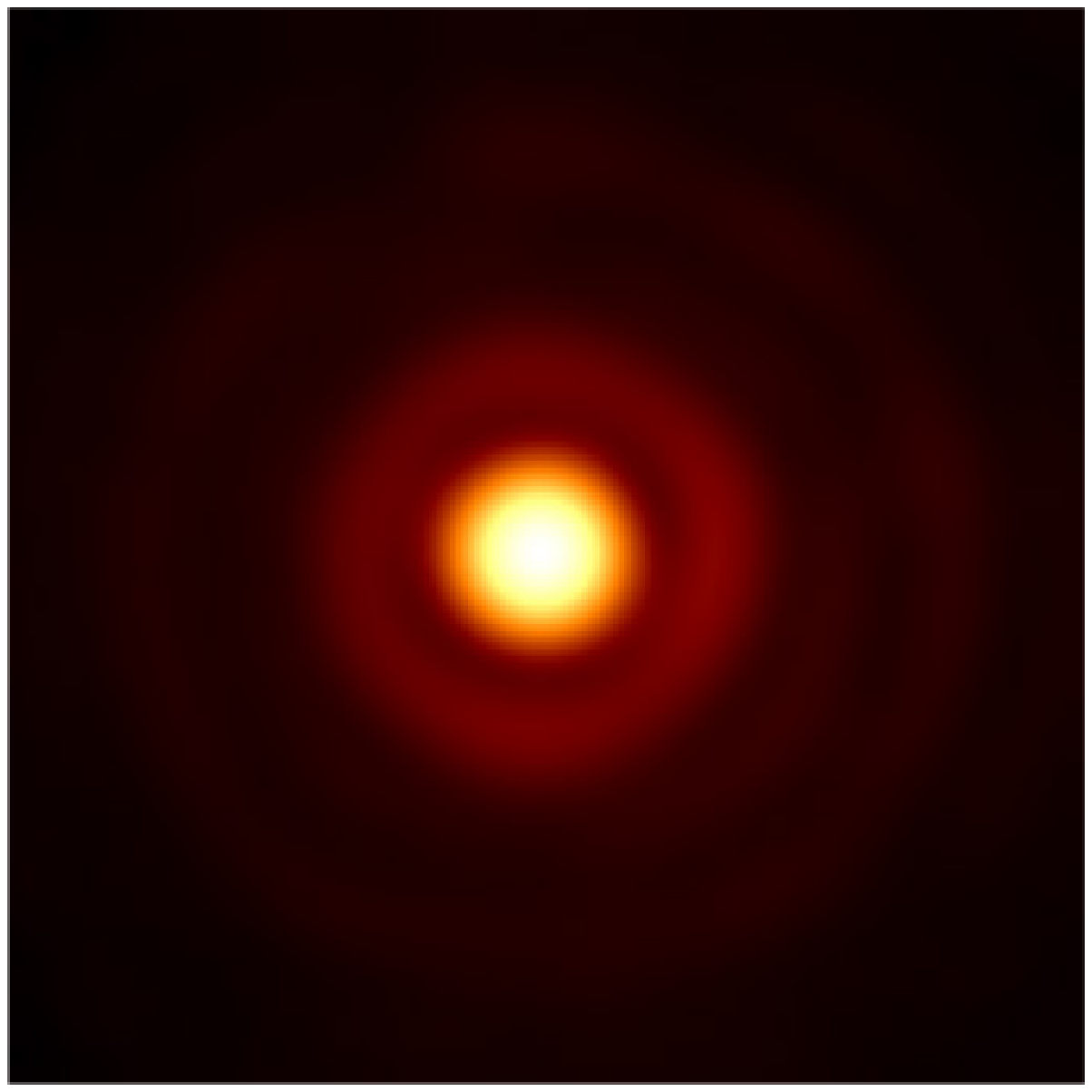}

\includegraphics[width=4.3cm]{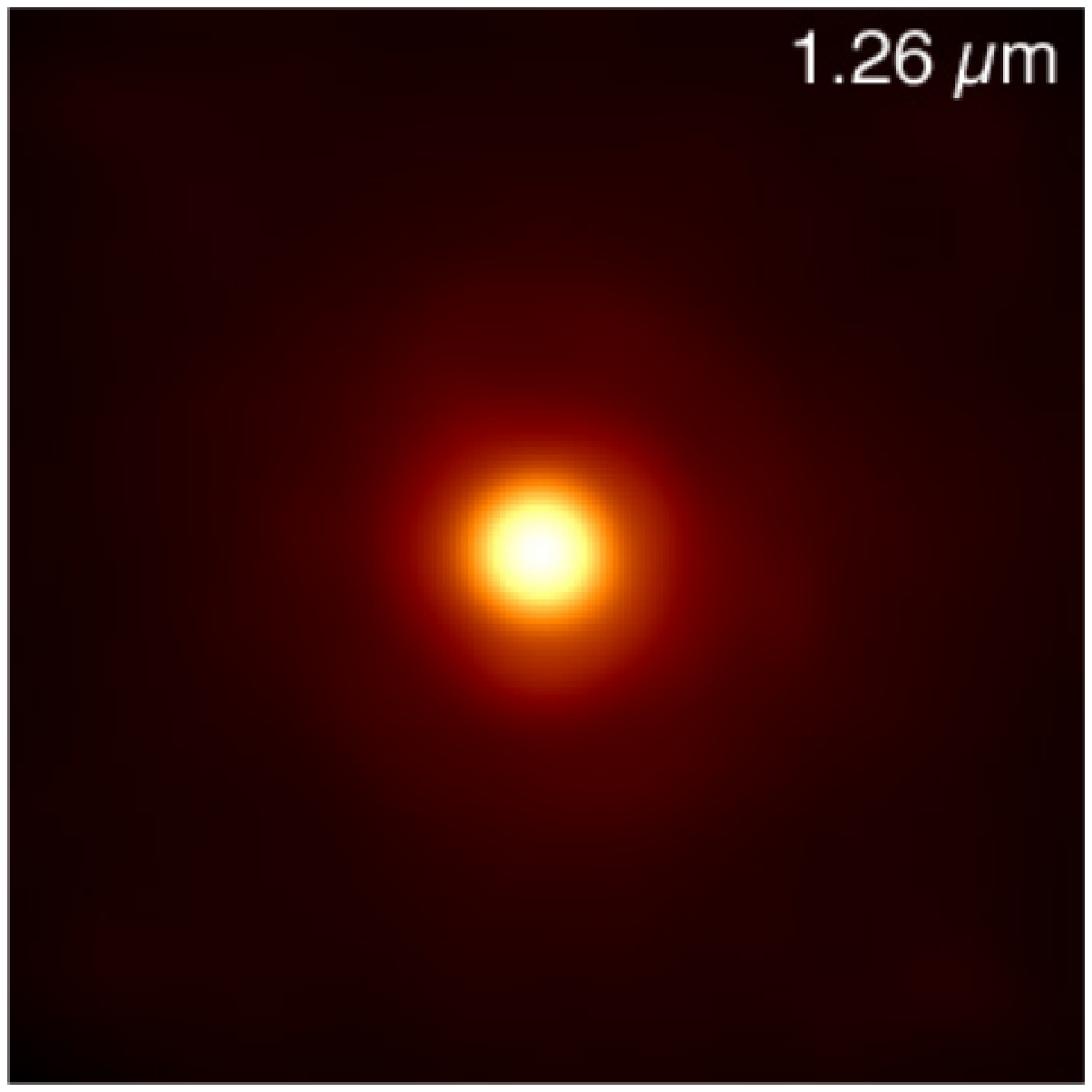} \includegraphics[width=4.3cm]{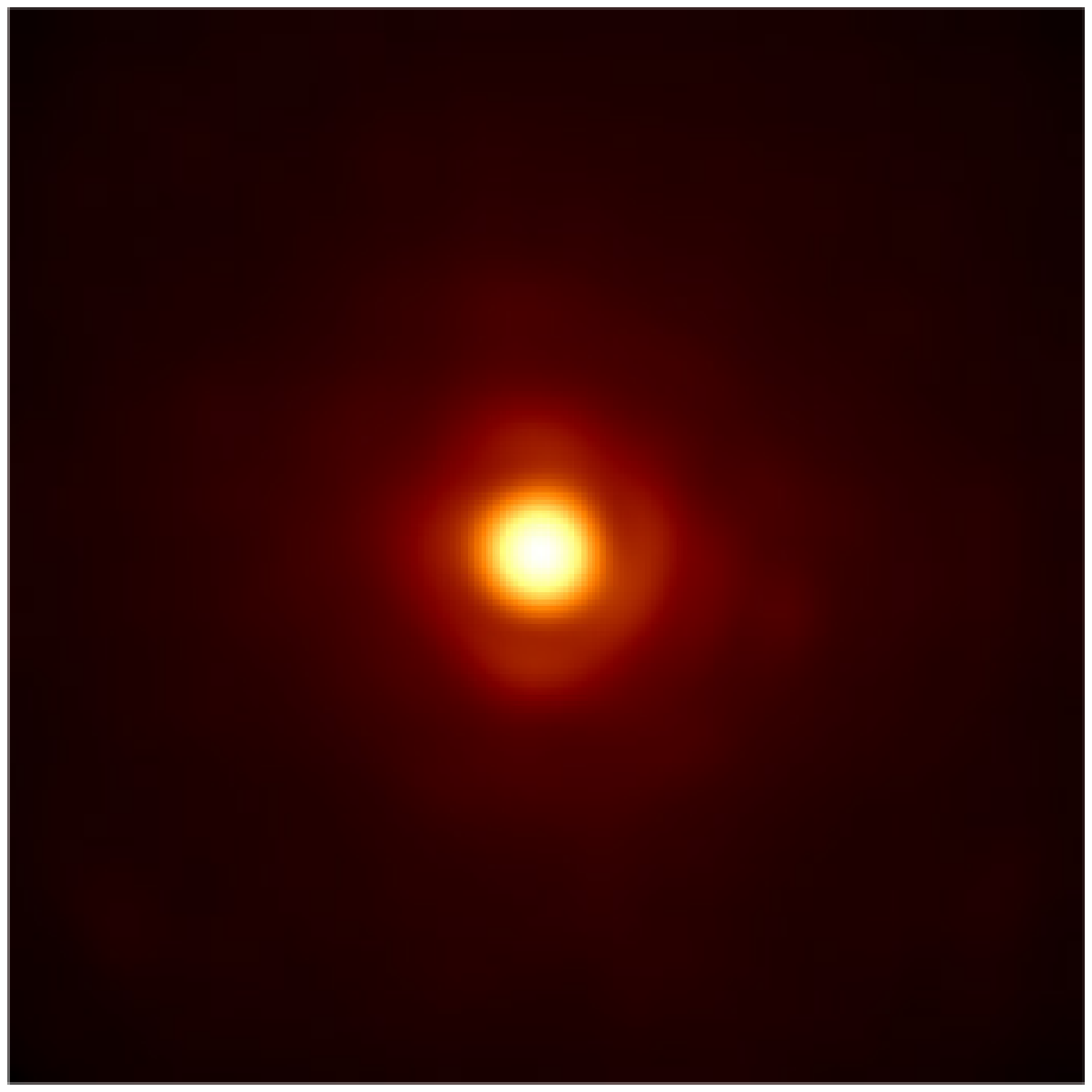} \hspace{2mm}
\includegraphics[width=4.3cm]{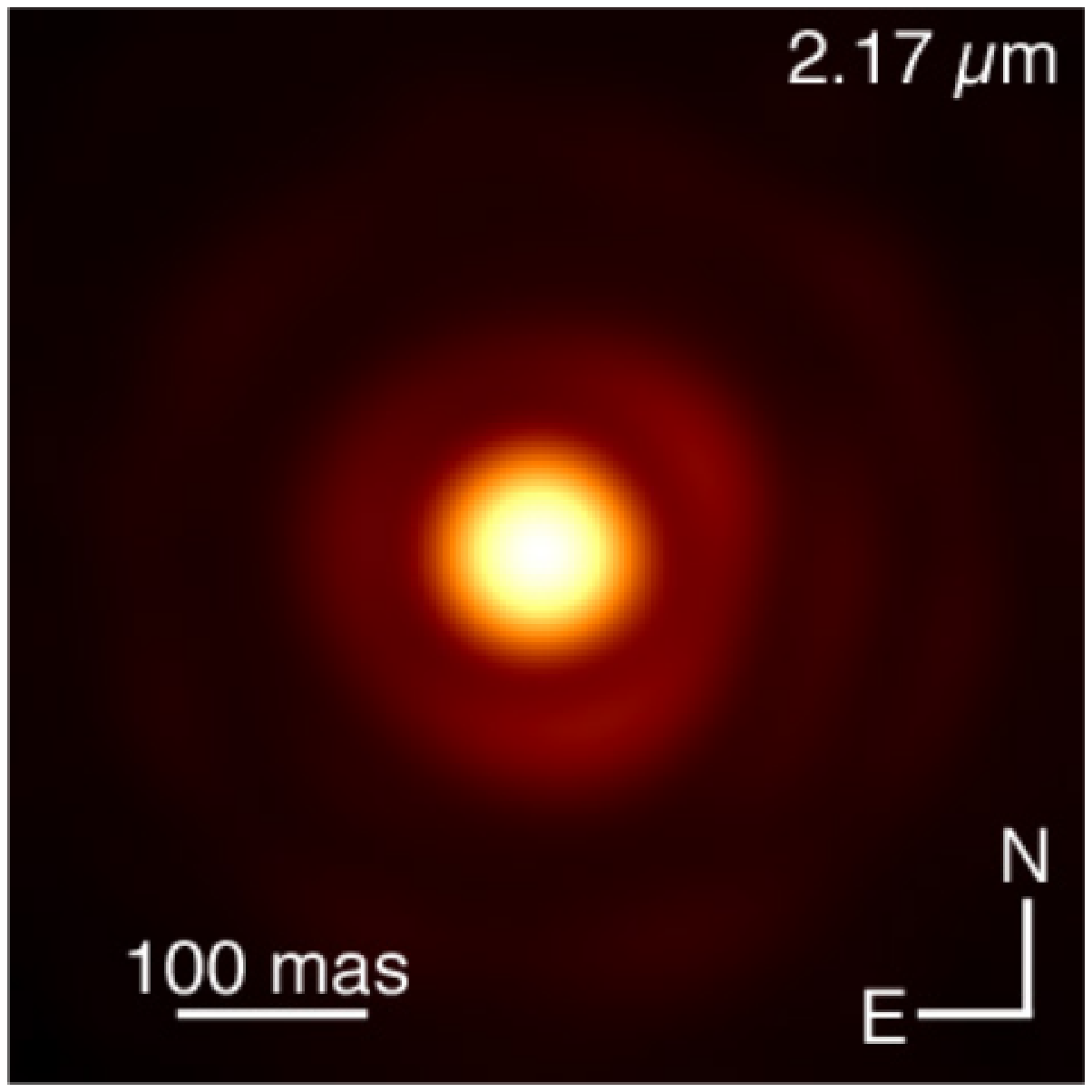} \includegraphics[width=4.3cm]{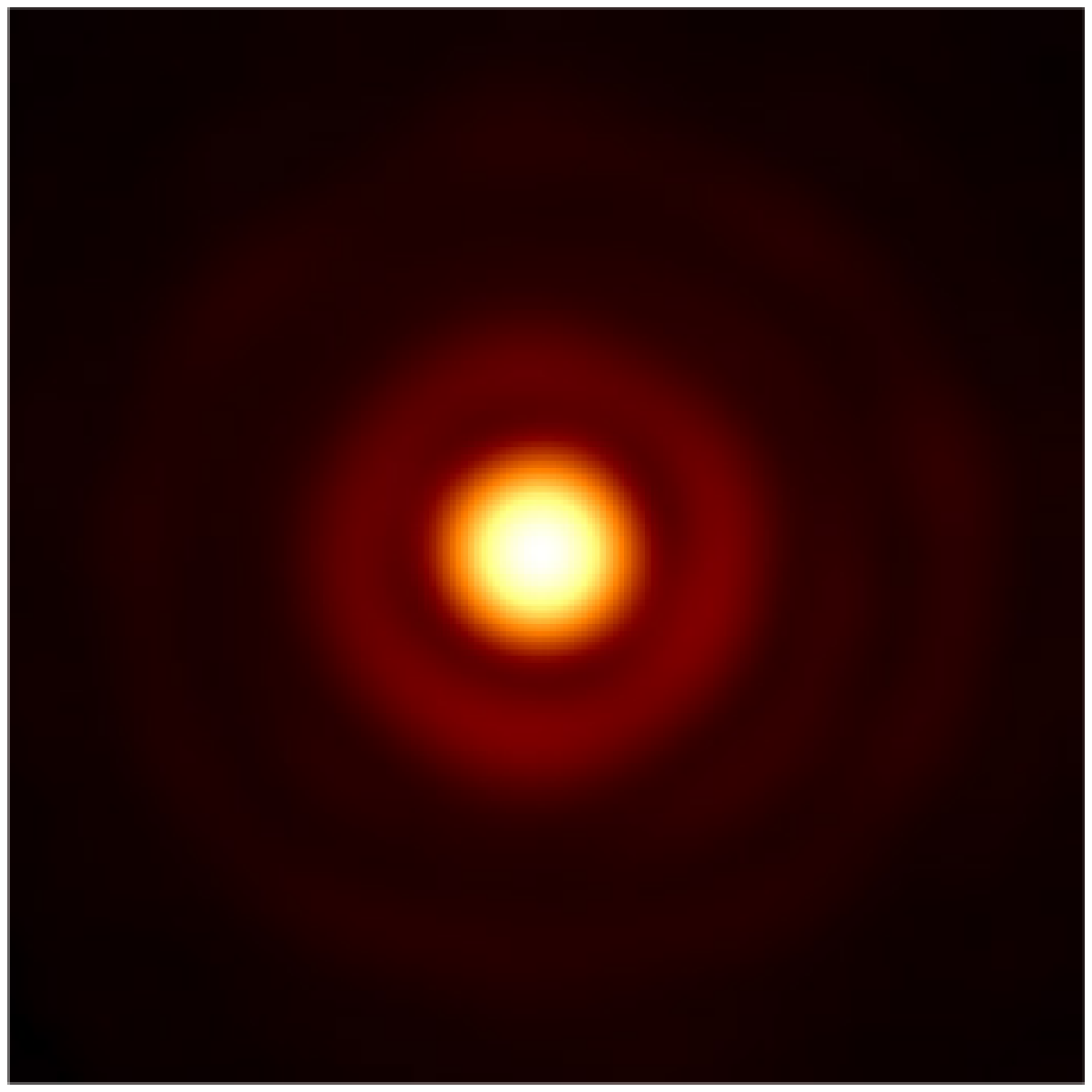}
\caption{Average NACO images of Betelgeuse (left image of each column) and the PSF calibrator Aldebaran (right image of each column) in 10 narrow-band filters from 1.04 to 2.17\,$\mu$m. The color scale is a function of the square root of the flux, normalized to the maximum and minimum value of each image, in order to emphasize the fainter parts of the images. The white disks in the lower right corners of the 1.04\,$\mu$m images represent the photospheric angular diameters of Betelgeuse and Aldebaran.\label{Avg_cubes}}
\end{figure*}

The quality of adaptive optics images is affected by a residual jitter of the star on the detector (``tip-tilt"), and by high-order residual wavefront distorsions. The star jitter is most probably caused by a combination of instrumental vibrations and atmospheric tip-tilt residuals, while the high-order distorsions are simply due to the limitations of the instrument performance. In order to obtain truly diffraction-limited image quality, especially at short wavelengths in the $J$ band, it is essential to eliminate the perturbations from these two effects.
Taking advantage of the newly implemented \emph{cube} mode of CONICA, we obtained extremely short exposures that allowed us to freeze the image jitter. After a precentering at the integer pixel level, the images were sorted based on their maximum intensity, used as a proxy of the Strehl ratio. The 10\% best images of each cube were then resampled up by a factor 4 using a cubic spline interpolation, and the star image was subsequently centered using gaussian fitting. The resulting cubes were eventually averaged to obtain the master images of each star used in the image analysis process described below. The initial oversampling of the image followed by the precise centering of the images gives an improved sampling of the image, as the residual atmospheric ``drizzle" allows to retrieve sub-pixel sampling. This is particularly important in the $J$ band, where the resolution of the telescope corresponds to only two NACO pixels. The selection of the best frames allowed us to reject the frames affected by the residual high-order distorsions.
The result of this processing is presented in Fig.~\ref{Avg_cubes} for Betelgeuse and its primary PSF calibrator Aldebaran.

%______________ Figure

\subsection{Efficiency of NACO full-pupil cube vs. standard mode}

\begin{figure}[]
\centering
\includegraphics[width=4.3cm]{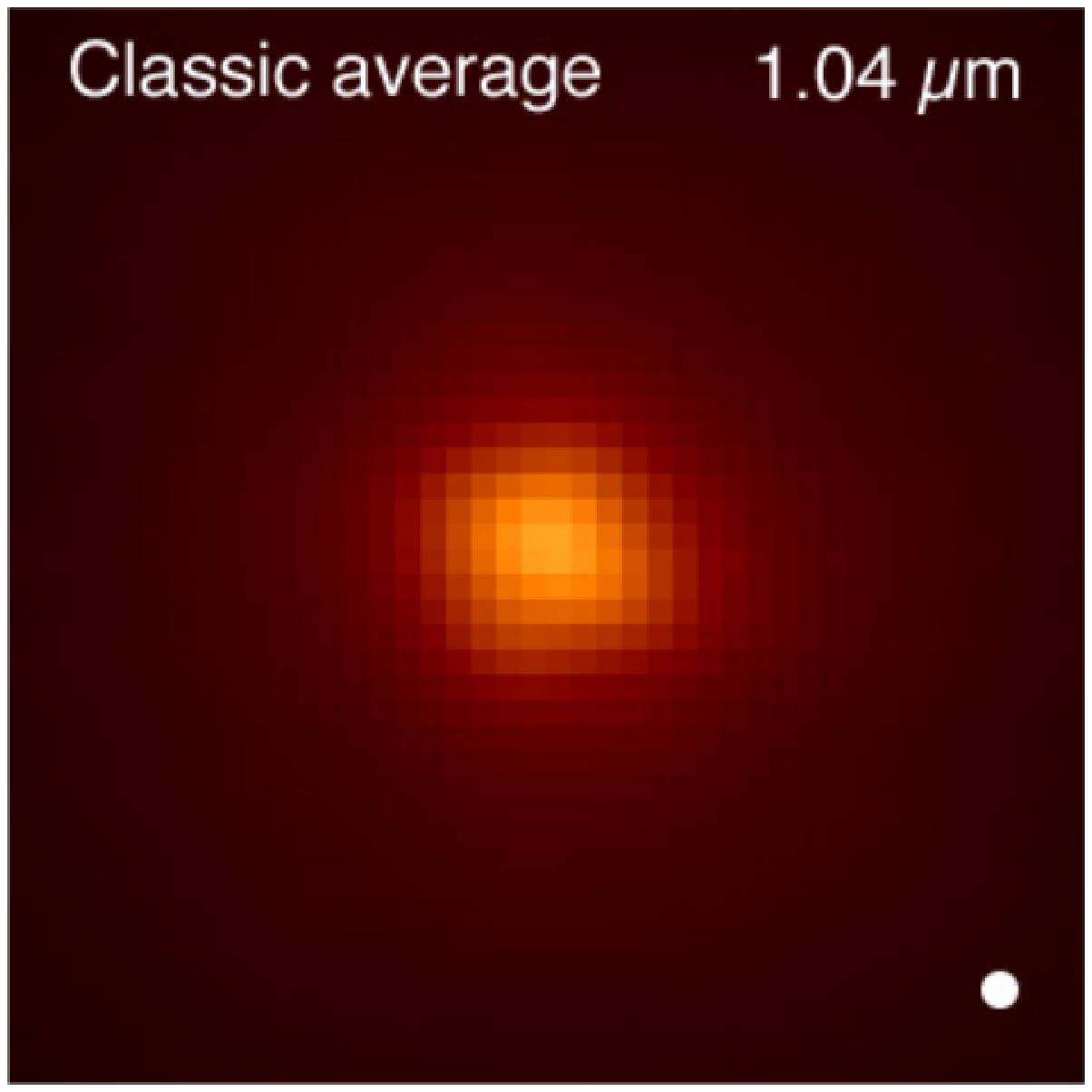} \includegraphics[width=4.3cm]{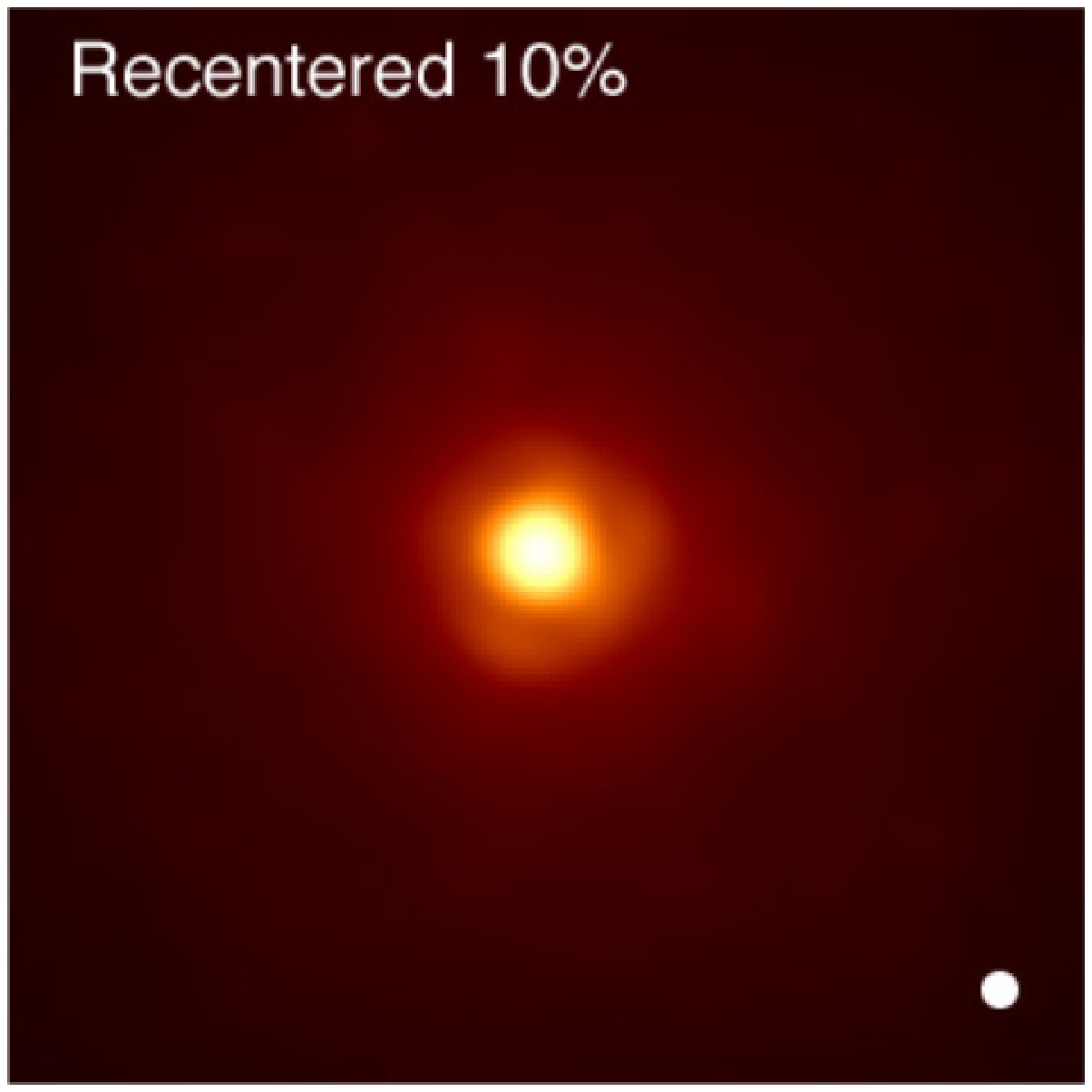}
\includegraphics[width=4.3cm]{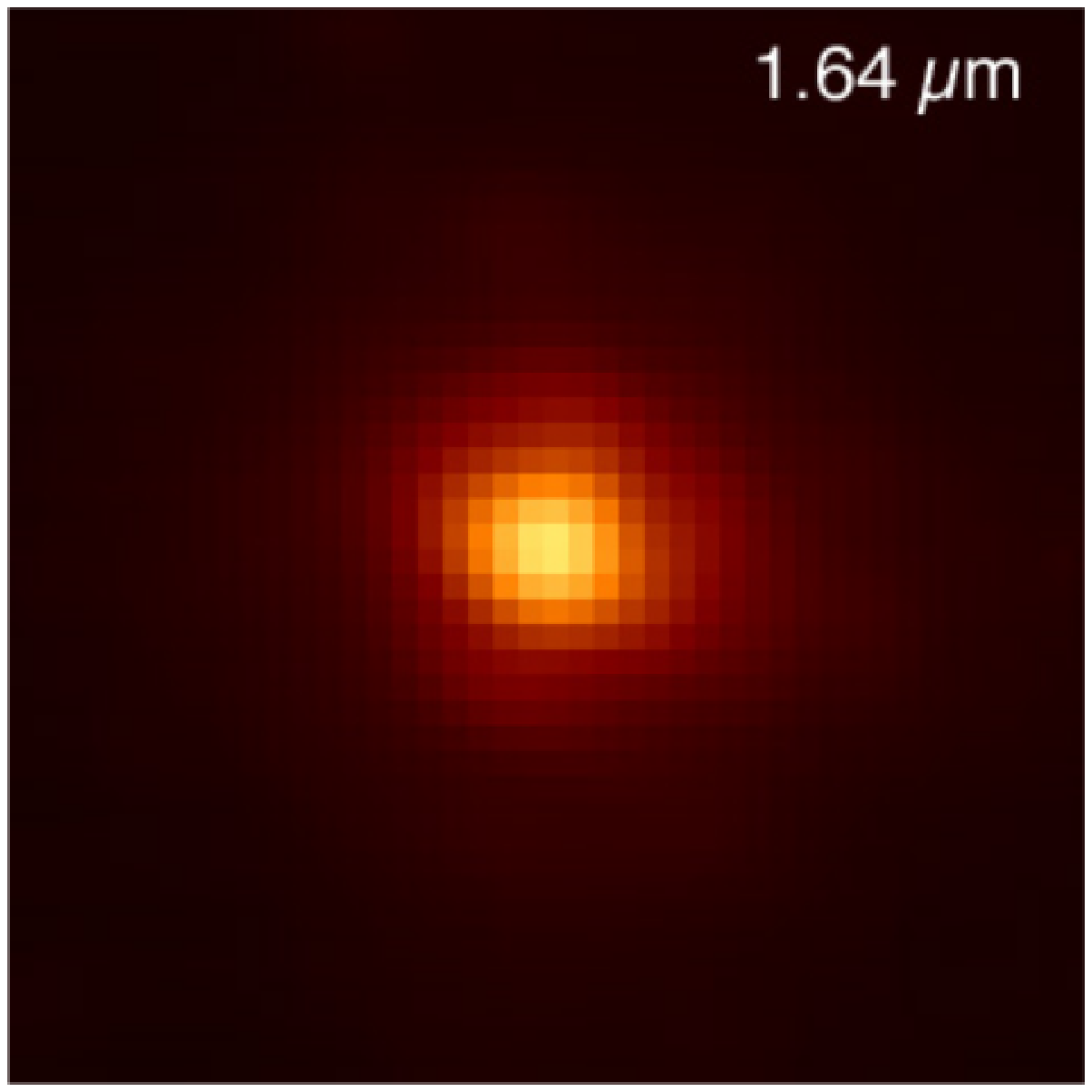} \includegraphics[width=4.3cm]{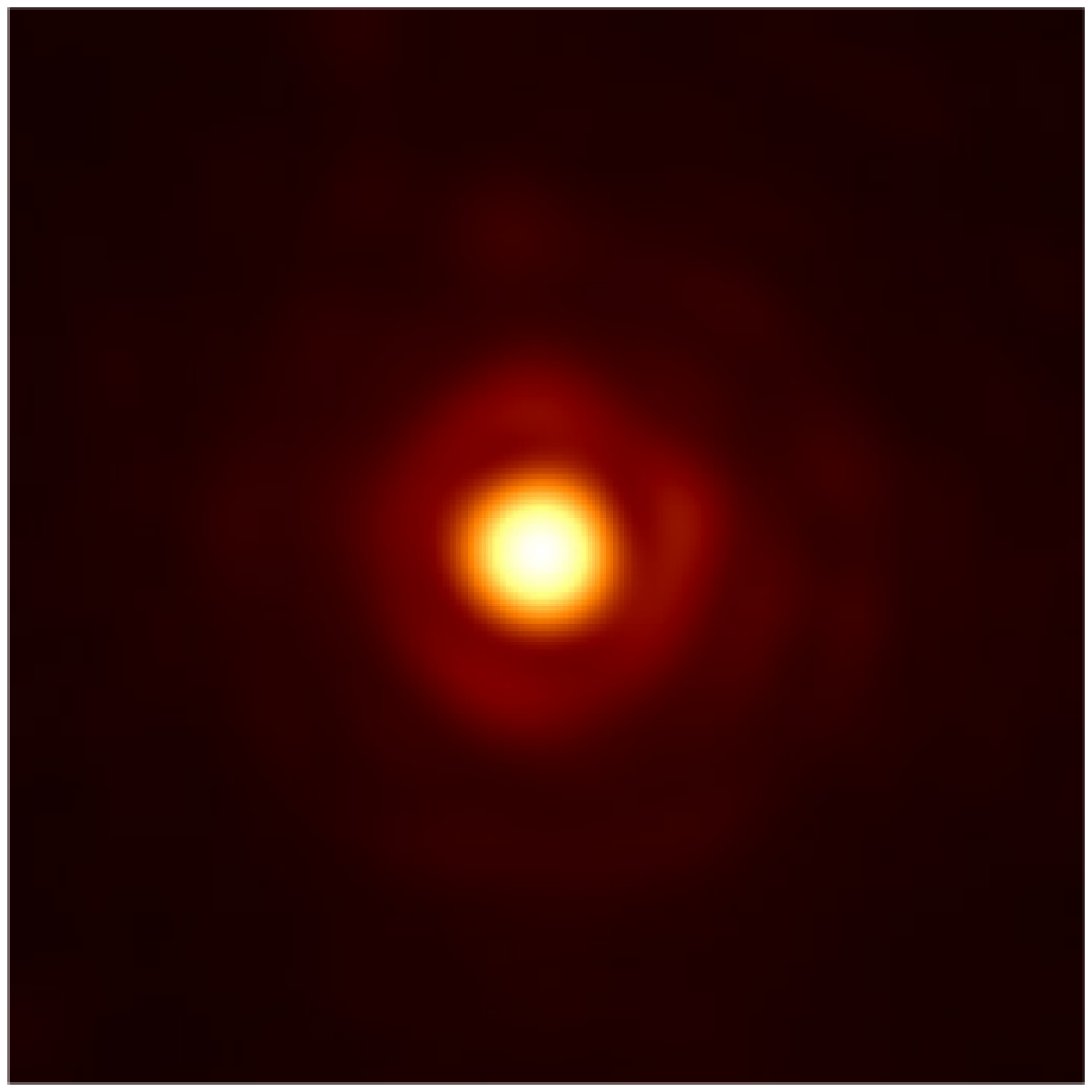} 
\includegraphics[width=4.3cm]{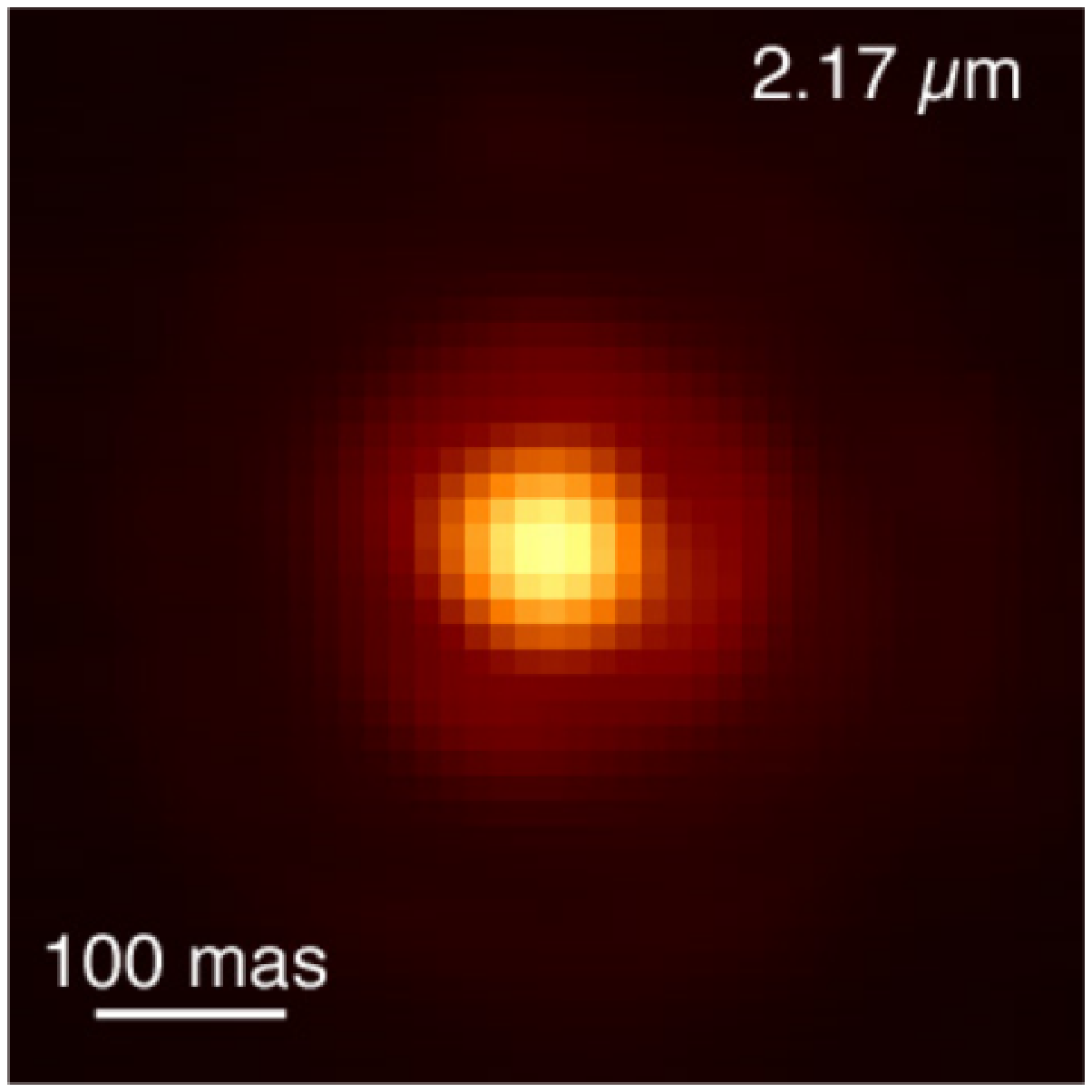} \includegraphics[width=4.3cm]{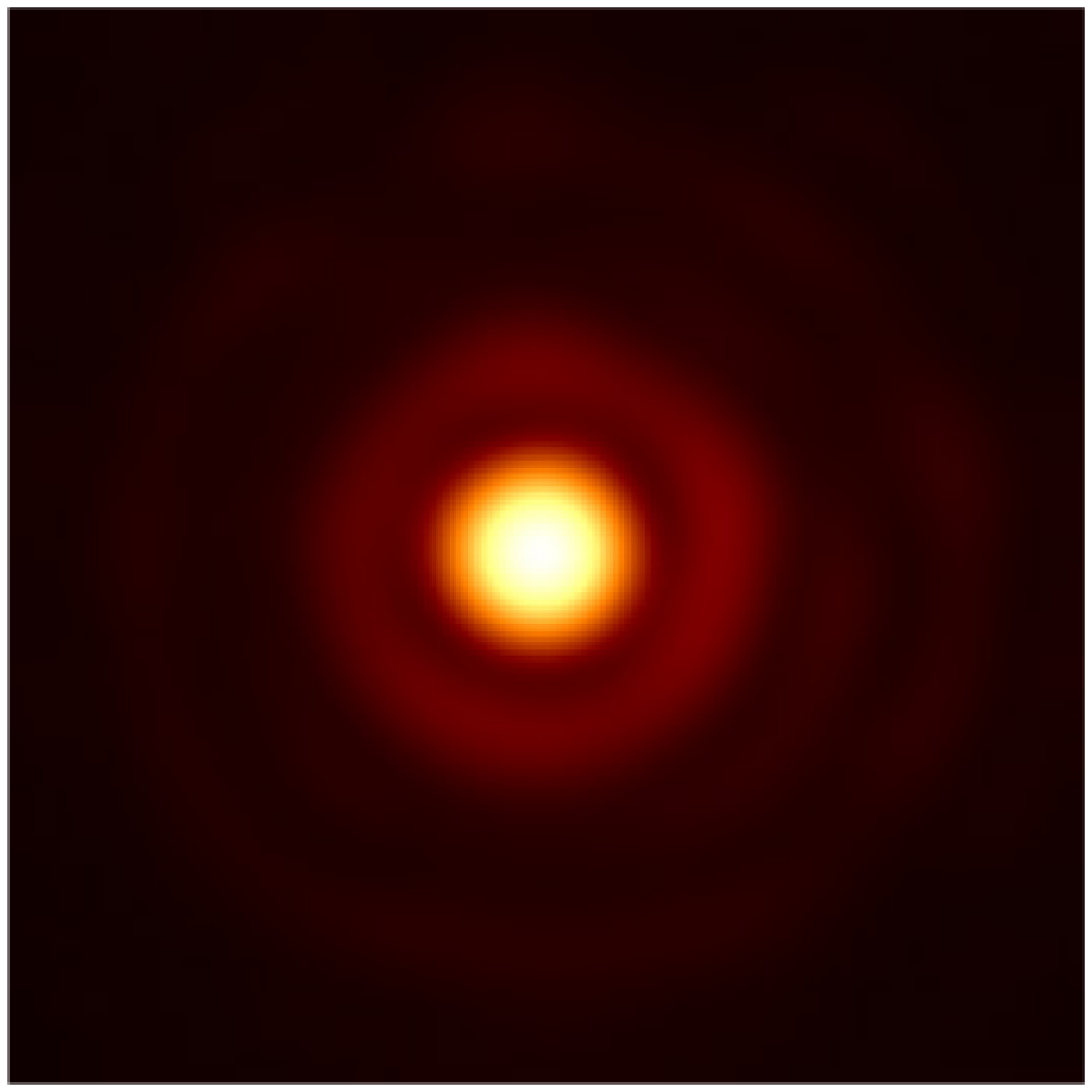}
\caption{Average frames of the NACO cubes \#001 (1.04\,$\mu$m), \#007 (1.64\,$\mu$m) and \#010 (2.17\,$\mu$m) obtained on the PSF calibrator Aldebaran without centering or frame selection (left) and with centering and 10\% best frame selection (right). The white disk in the 1.04\,$\mu$m images represents the angular size of Aldebaran's photosphere (20\,mas). The color scales are proportional to the square root of the flux, and are identical for the two images at each wavelength. \label{Comparison_PSF}}
\end{figure}

In order to check the efficiency of the image recentering and selection, we processed a series of 10 data cubes obtained on the PSF calibrator Aldebaran (\#001 to \#010 in Table~\ref{naco_log}) with and without recentering the individual frames. The resulting images for three wavelengths are shown in Fig.~\ref{Comparison_PSF}. Already clearly visible on these images, we hereafter quantify the gain in image quality by processing the same data cubes using three methods:
\begin{itemize}
\item recentering, selection of the best 10\% frames based on the peak flux in the image and co-addition,
\item recentering and co-addition of all frames (no selection),
\item coaddition without recentering of all frames (equivalent to a single long exposure of 72\,s).
\end{itemize}

\begin{figure}[t]
\centering
\includegraphics[width=\hsize]{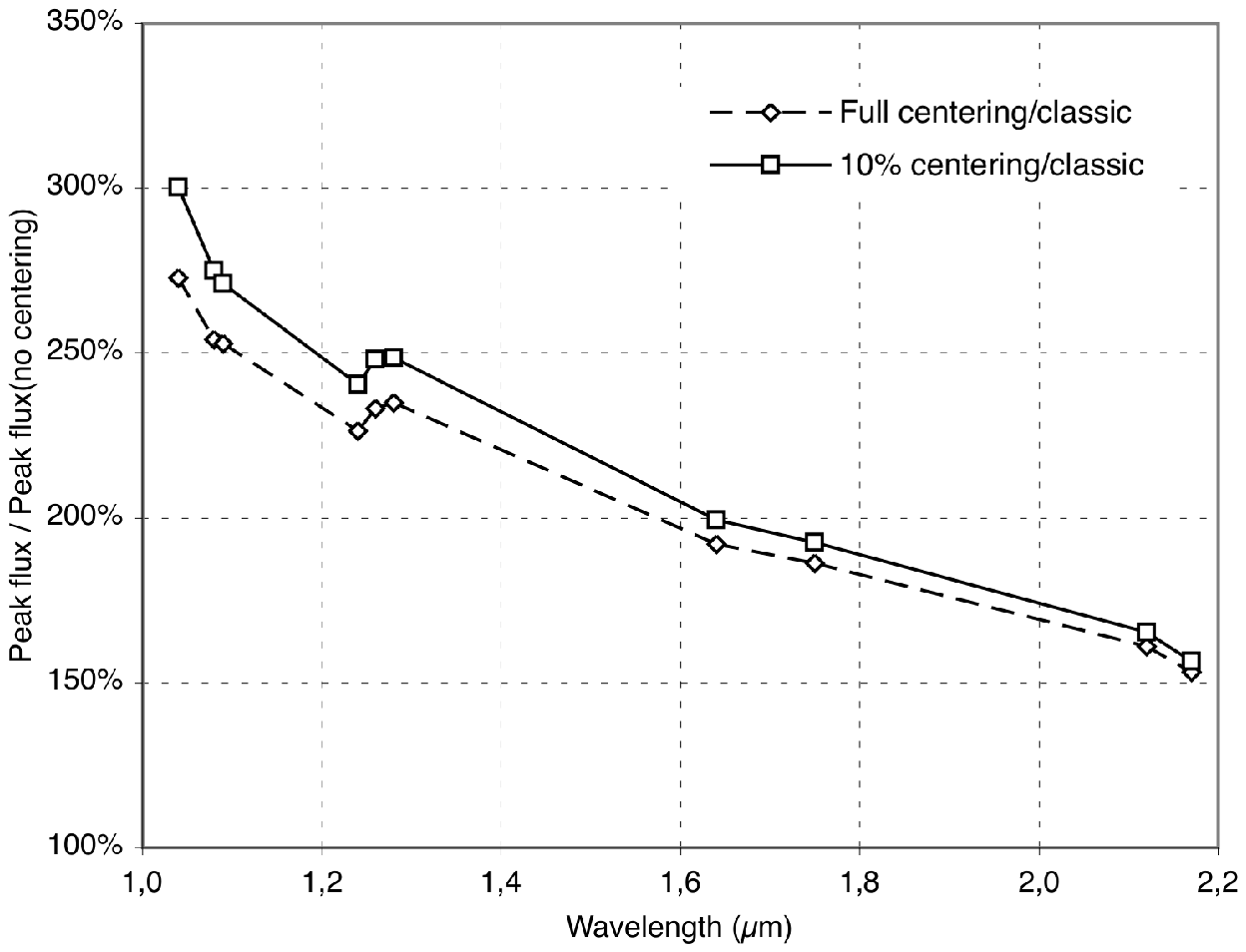} 
\includegraphics[width=\hsize]{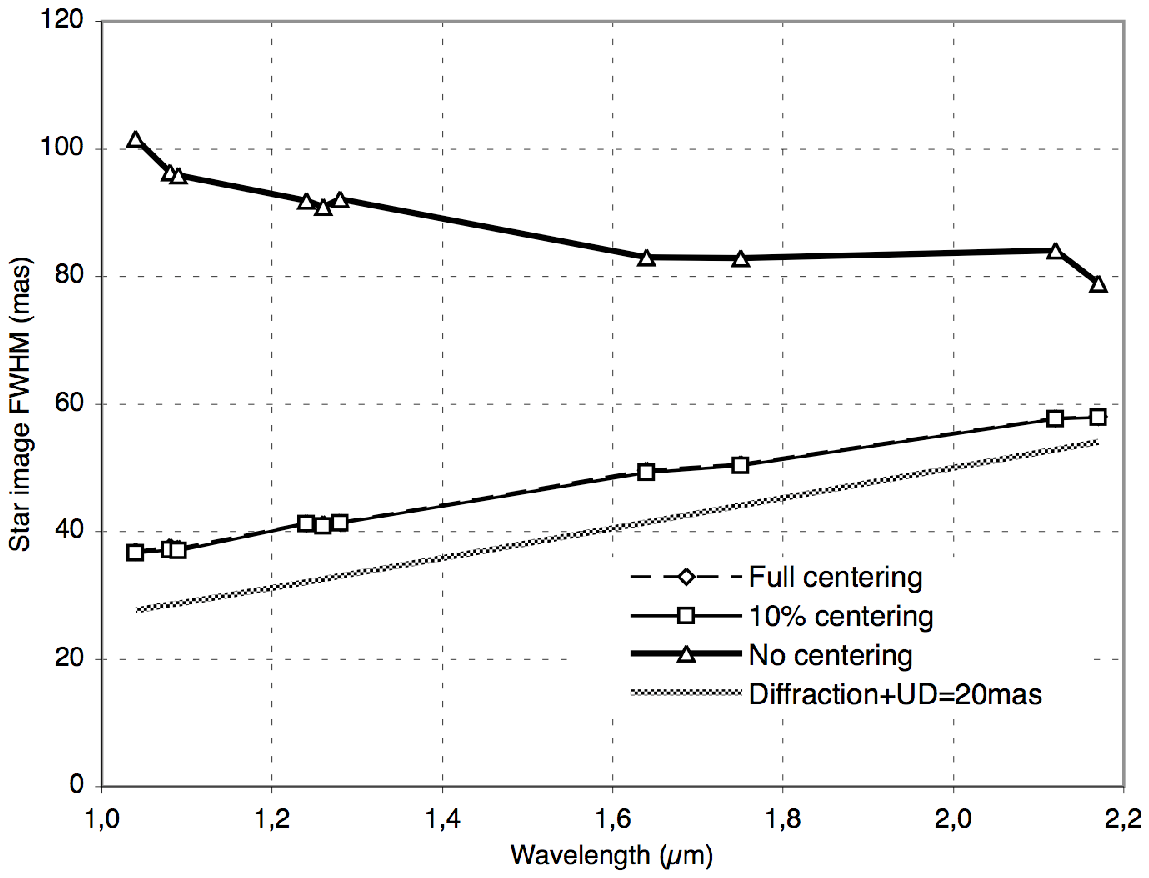}
\caption{{\it Upper panel:} Relative Strehl ratio of the recentered and co-added images compared to the classic mode of NACO, without frame selection (dashed curve) and keeping only the 10\% best peak flux frames (thin solid curve). {\it Lower panel:} Full width at half maximum of the NACO images of Aldebaran for the classic mode (black curve), frame centering without selection (dashed curve) and 10\% best frames selection and centering (thin solid curve), compared to the theoretical diffraction pattern for a UT observing a uniform disk of 20\,mas (grey curve).\label{Strehl_comparison}}
\end{figure}

From this exercise, it appears that the recentering of the frames before co-addition results in a spectacular increase of the Strehl ratio and a very significant decrease of the full width at half maximum (FWHM) of the star image (Fig.~\ref{Strehl_comparison}). As expected, this effect is stronger at shorter wavelengths, where the adaptive optics performance is comparatively lower. In the 1.04-1.09\,$\mu$m range, the improvement reaches a factor three in Strehl ratio and a factor two in FWHM. The improvement in FWHM brought by the selection of the frames with the best Strehl appears very small compared to the effect of the recentering. Selecting the 10\% best frames improves the average Strehl by approximately 10\% (relatively to centering without selection) for the shortest wavelengths. For our very bright targets, we chose to keep this selection criterion as the signal-to-noise ratio of the average images is essentially limited by the AO residuals, and not by the number of frames.

%______________ Figure

As shown in Fig.~\ref{Strehl_comparison}, we can also compare the image FWHM with the theoretical diffraction limit of the telescope. Using a very simplified model of an 8\,m primary mirror with a central 1.12\,m obscuration, and a 20\,mas uniform disk (UD) for Aldebaran, we obtained the theoretical FWHM shown as a thick grey curve.  We chose this UD angular size as the angular diameter of Aldebaran was measured in the $K$ band by Richichi \& Roccatagliata~(\cite{richichi05}), who found $\theta_{\rm UD}(K)=19.96 \pm 0.03$\,mas and $\theta_{\rm LD}(K)=20.58 \pm 0.03$\,mas. As a remark, the FWHM of our images is measured over a large sub-window that includes part of the adaptive optics photometric incoherent residuals (also called ``pedestal"). The resulting values are thus slightly overestimated, but have the advantage of being directly comparable between real and theoretical images. 

From these results, the frame-by-frame recentering of adaptive optics observations appears as a very efficient way to improve the angular resolution compared to single long exposures. The spectacular increase in Strehl ratio observed at short wavelengths should also result in an improved sensitivity for faint object, but the necessity of combining a large number of frames results in a proportionally higher detector readout noise. For this reason, this data reduction strategy may not be optimal for $K$ band observations where the gain is ``only" 50\% in Strehl ratio. However, in the $J$ band, whenever a bright enough source is available to recenter the frames, the \emph{cube} mode of NACO with short exposures will in most cases be more sensitive than classical long exposures.

%______________ Figure

%__________________________________Analysis / deconvolution
\section{Image analysis \label{analysis}}

\subsection{Point spread function \label{psf}}

We obtained images of two unresolved and relatively fainter stars (31\,Ori and $\delta$\,Phe, \#022-025 and \#051-054 in Table~\ref{naco_log}) without neutral density filter, to check the influence of the ND filter on the PSF shape. We do not find evidence of an effect of this filter on the PSF shape. However, considering the faintness of the CSE of Betelgeuse, we chose to strictly keep the same instrumental setup between Betelgeuse and the PSF to avoid any contamination in the deconvolved images. We therefore used Aldebaran as the PSF reference.
To retrieve the true PSF of the telescope from the images of Aldebaran, we took into account the angular extension of the star. Although this star is significantly smaller ($\theta_{\rm LD}=20.6$\,mas) than Betelgeuse ($\theta_{\rm LD}=43.7$\,mas), it is not negligible compared to the resolution of the telescope in the $J$ band. We therefore computed synthetic images of Aldebaran using its known LD size and model limb darkening coefficients in the J, H, and K bands from Claret's~(\cite{claret00}) tables using Teff = 4 000K and log g = 1.5. We then computed a classical Lucy-Richardson (L-R) deconvolution of the NACO images of Aldebaran (single iteration) using these synthetic images as the ``dirty beams". This procedure gives us the true PSF of the telescope for our ten narrow-band filters, i.e. the image of a point source.

\subsection{Photometry of Betelgeuse\label{photometry}}

For the photometric calibration of our images, we used a classical aperture photometry approach using Aldebaran as the calibrator. As we used non-standard narrow-band filters, we have to refer our measurements to a synthetic spectral energy distribution (SED) of Aldebaran, normalized by its interferometrically measured angular diameter.

The shape of the SED of Aldebaran was read from the theoretical spectra table by Castelli \& Kurucz~(\cite{castelli03}), for an effective temperature of 4\,000\,K, solar metallicity, $V_T=2$\,km.s$^{-1}$ and $\log g = 1.5$ (parameters approximated from Cayrel de Strobel et al.~\cite{cayrel01}). We then used the limb darkened disk angular diameter of 20.6\,mas measured by Richichi \& Roccatagliata~(\cite{richichi05}) to normalize this synthetic spectrum. We checked that the existing broadband photometric measurements are consistent with the resulting SED, but we did not used them to fit the SED. Considering the true profiles of the NACO filters we used for our observations (Fig.~\ref{nb_filters}), we integrated the flux from Aldebaran in W.m$^{-2}$.$\mu$m$^{-1}$ in each band. We then normalized the flux integrated on the Betelgeuse images by the flux measured on Aldebaran using the same relatively narrow aperture (50 to 70\,mas in radius, slightly increasing with wavelength). This approach allows us to integrate the full flux from the resolved Betelgeuse disk, while avoiding the adaptive optics halo. We did not consider the airmass difference between the two stars and the negligible infrared background.

%______________ Figure
\begin{figure}
\centering
\includegraphics[width=\hsize]{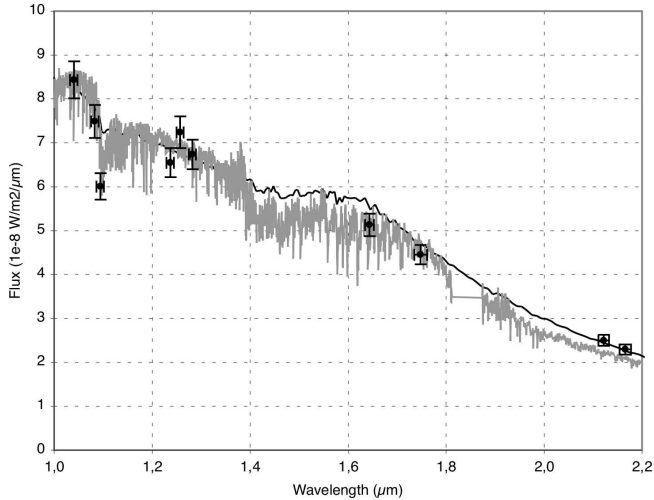} 
\caption{Calibrated photometry of Betelgeuse. The dots represent our photometry in the 10 narrow-band filters of NACO. The thin black curve represents the model spectrum of a $T_{\rm eff}=3\,500$\,K normal supergiant (Castelli \& Kurucz~\cite{castelli03}) normalized to a limb darkened disk angular diameter of 43.7\,mas. The grey curve is a spectrum of Betelgeuse from the IRTF database (Cushing et al.~\cite{cushing05}), scaled by a factor 1.2. \label{photometryNB}}
\end{figure}

The resulting photometric measurements are shown in Fig.~\ref{photometryNB} for the ten narrow-band flters. In this figure, we plotted the synthetic spectrum of a normal supergiant with $T_{\rm eff}=3\,500$\,K, solar metallicity, $V_T=2$\,km.s$^{-1}$ and $\log g =0.0$  from the Castelli \& Kurucz~(\cite{castelli03}) library. We normalized the expected flux using the 43.7\,mas angular diameter measured by Perrin et al.~(\cite{perrin04}). The overall agreement with our photometry is satisfactory, but a strong absorption feature around 1.09-1.24\,$\mu$m is present in our data and absent from the model spectrum. We also plotted in Fig.~\ref{photometryNB} a calibrated spectrum of Betelgeuse (Cushing et al.~\cite{cushing05}) taken from the IRTF database\footnote{http://irtfweb.ifa.hawaii.edu/$\sim$spex/IRTF\_Spectral\_Library/}, scaled by a factor 1.2 to match the brightness of Betelgeuse at the epoch of our observations. This spectrum also shows the absorption feature that is particularly strong at $1.09\,\mu$m.

\subsection{Image deconvolution}

\begin{figure*}[ht]
\centering
\includegraphics[width=4.3cm]{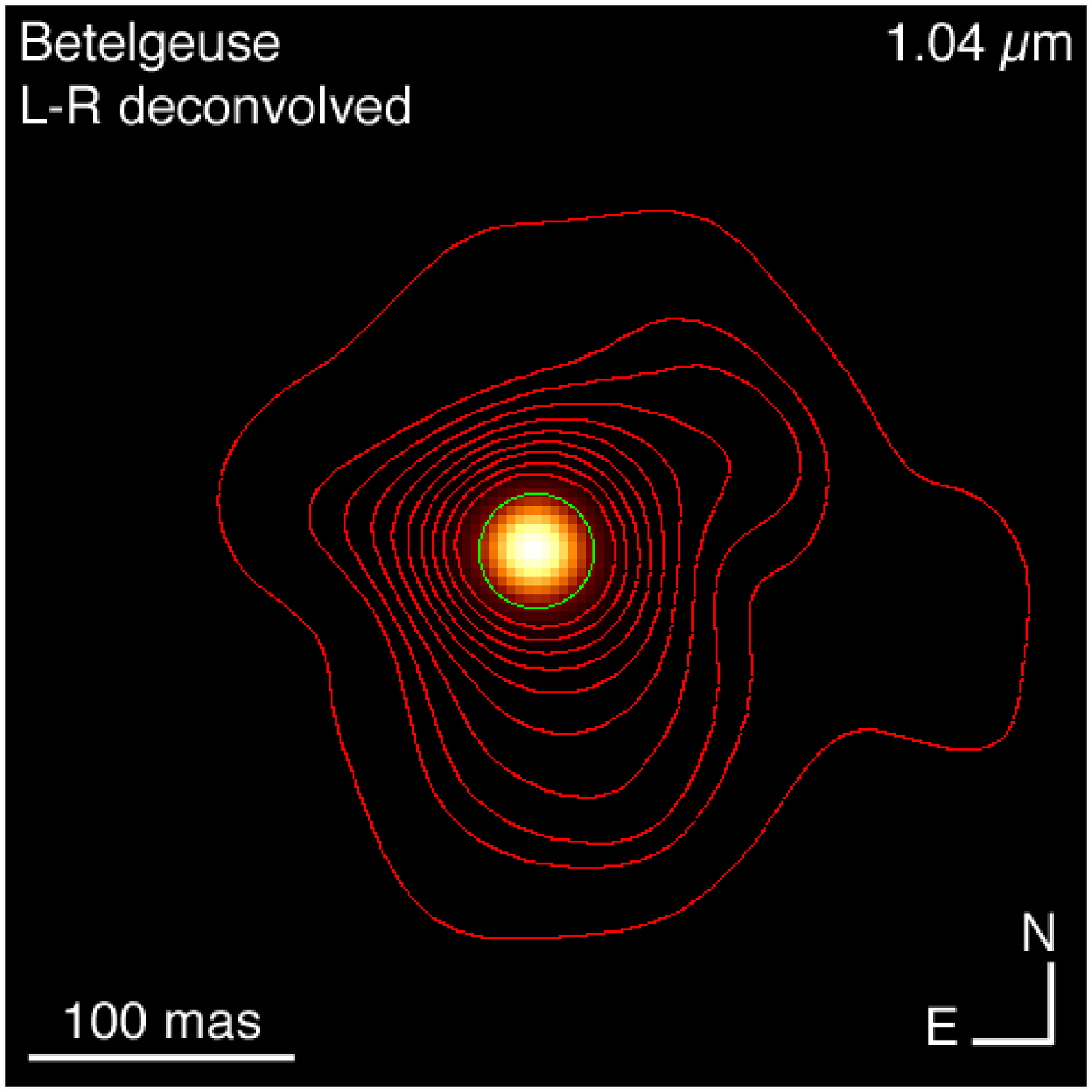} \includegraphics[width=4.3cm]{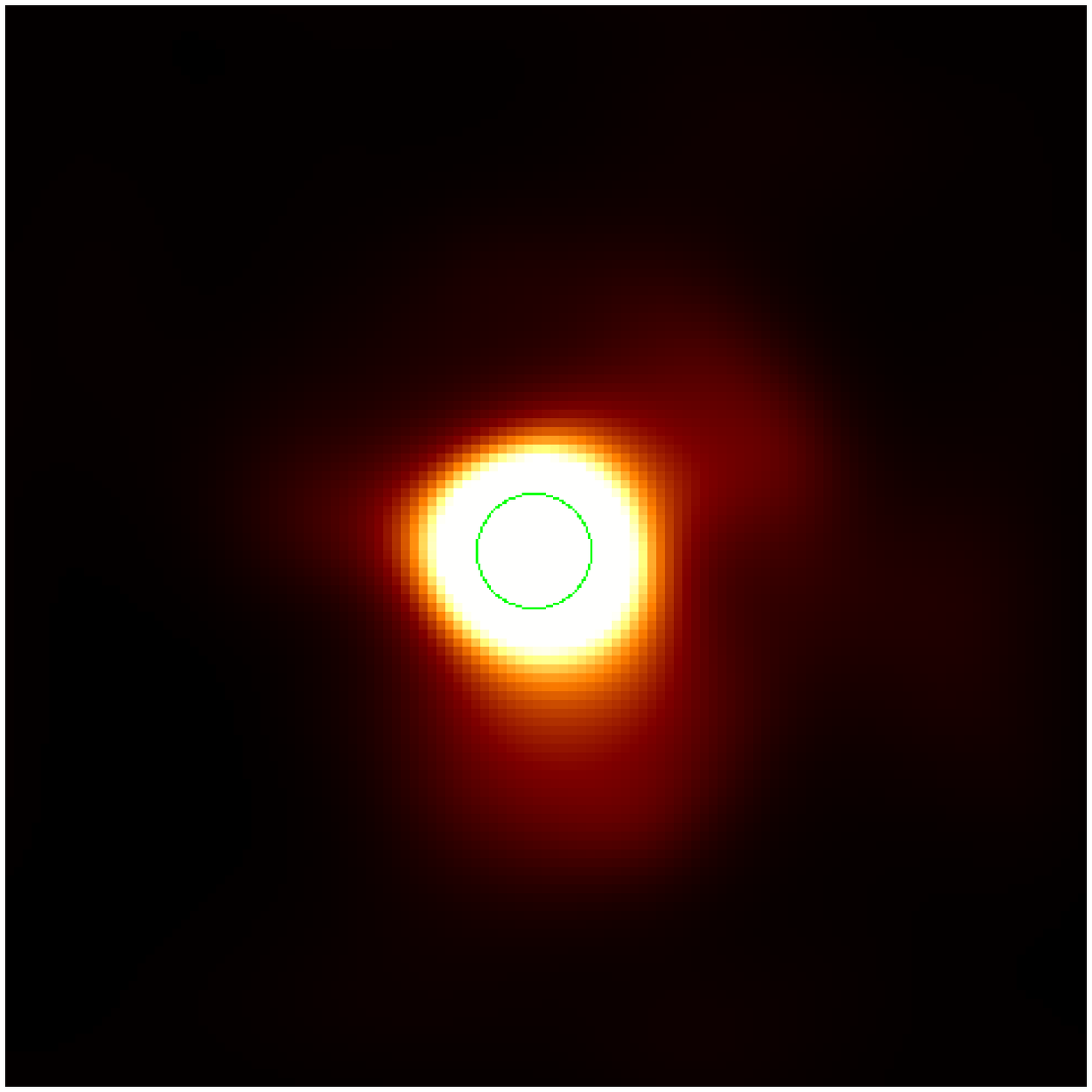} \hspace{2mm}
\includegraphics[width=4.3cm]{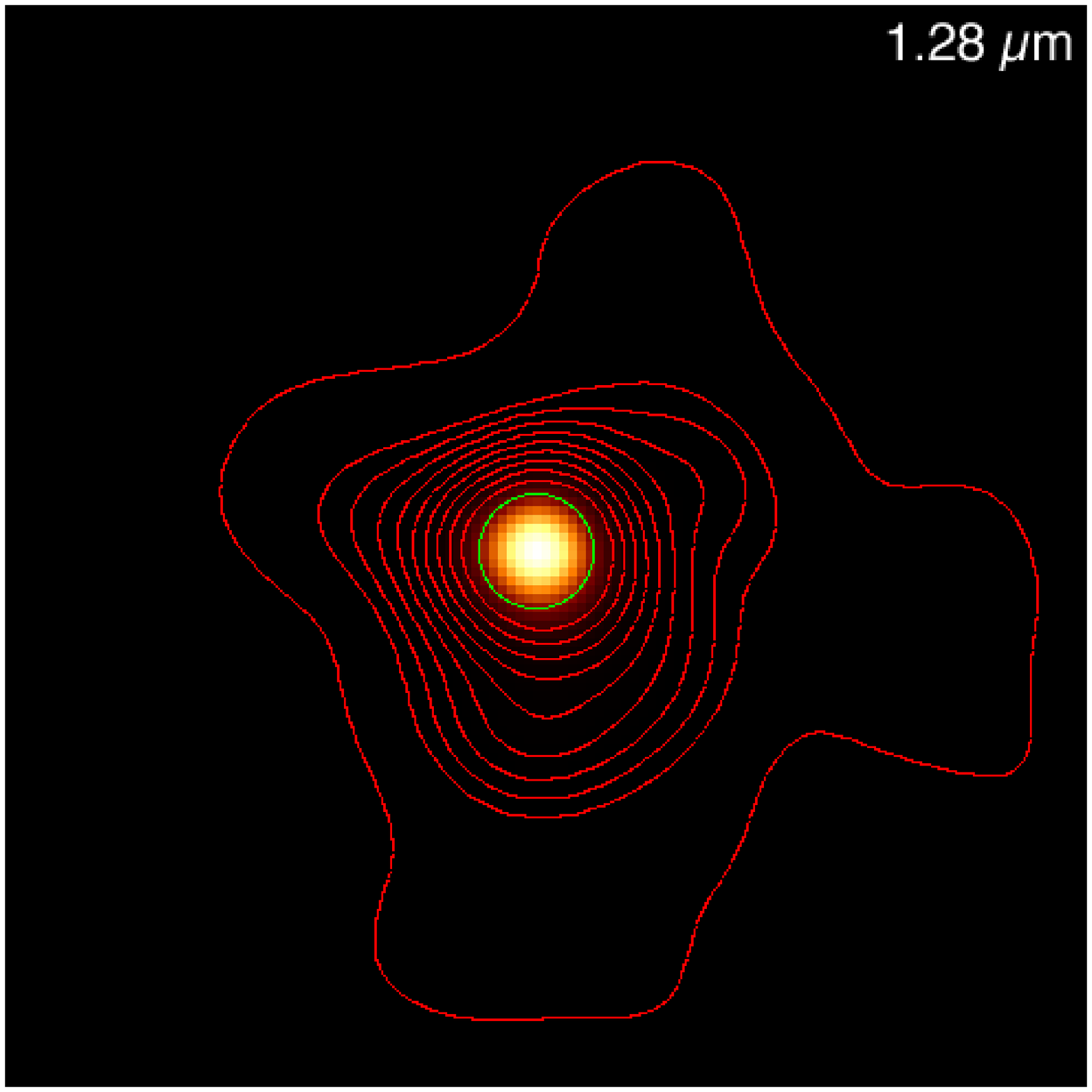} \includegraphics[width=4.3cm]{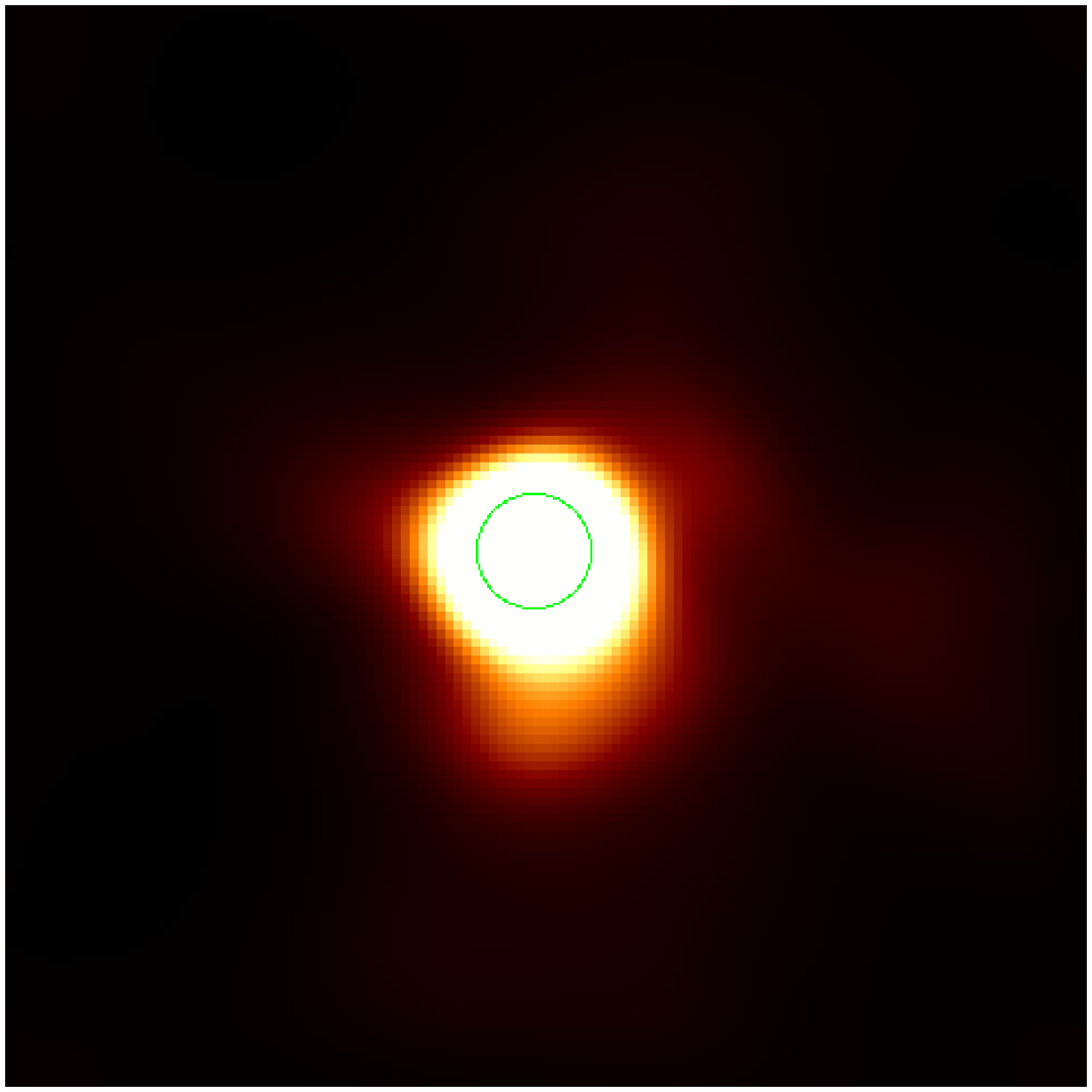}

\includegraphics[width=4.3cm]{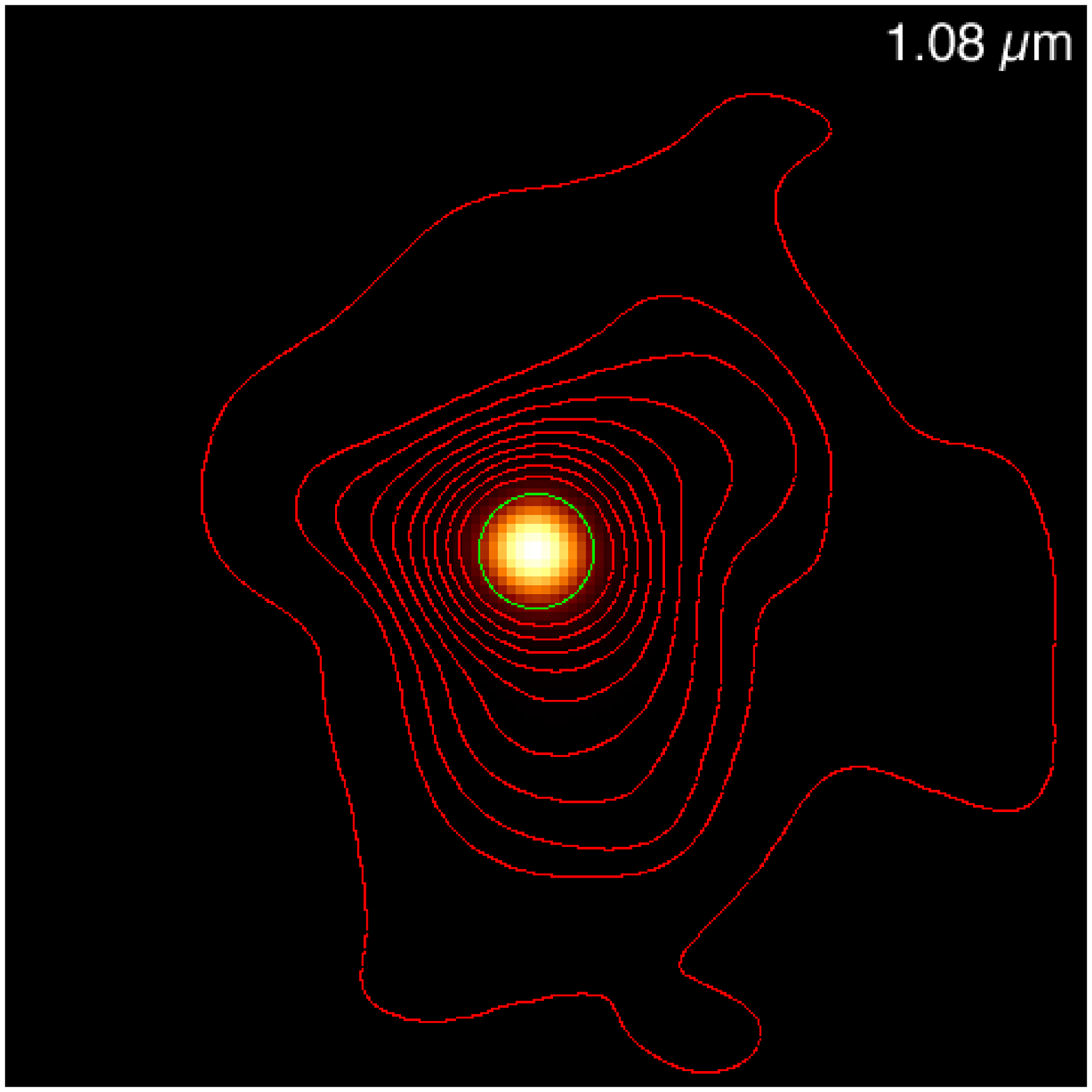} \includegraphics[width=4.3cm]{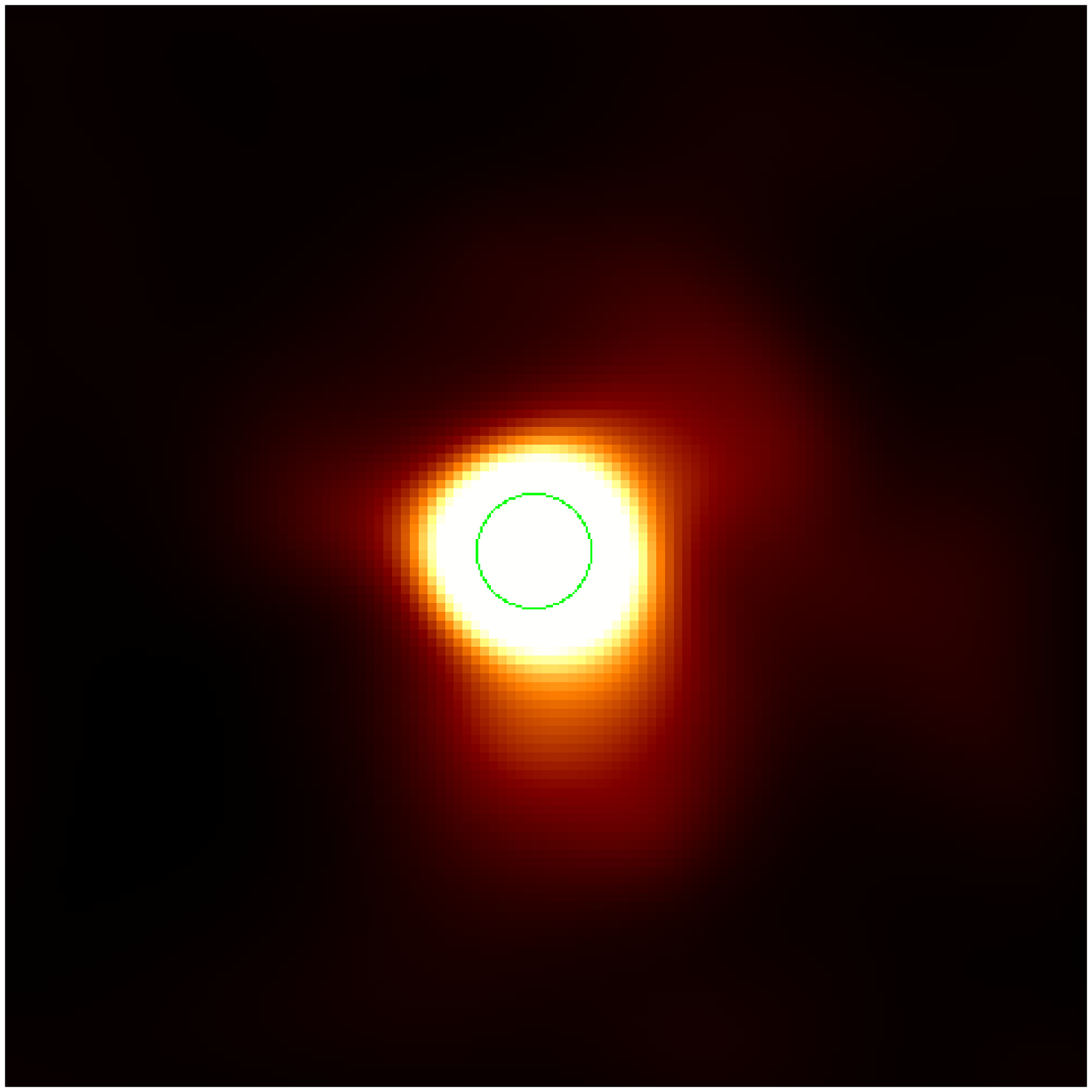} \hspace{2mm}
\includegraphics[width=4.3cm]{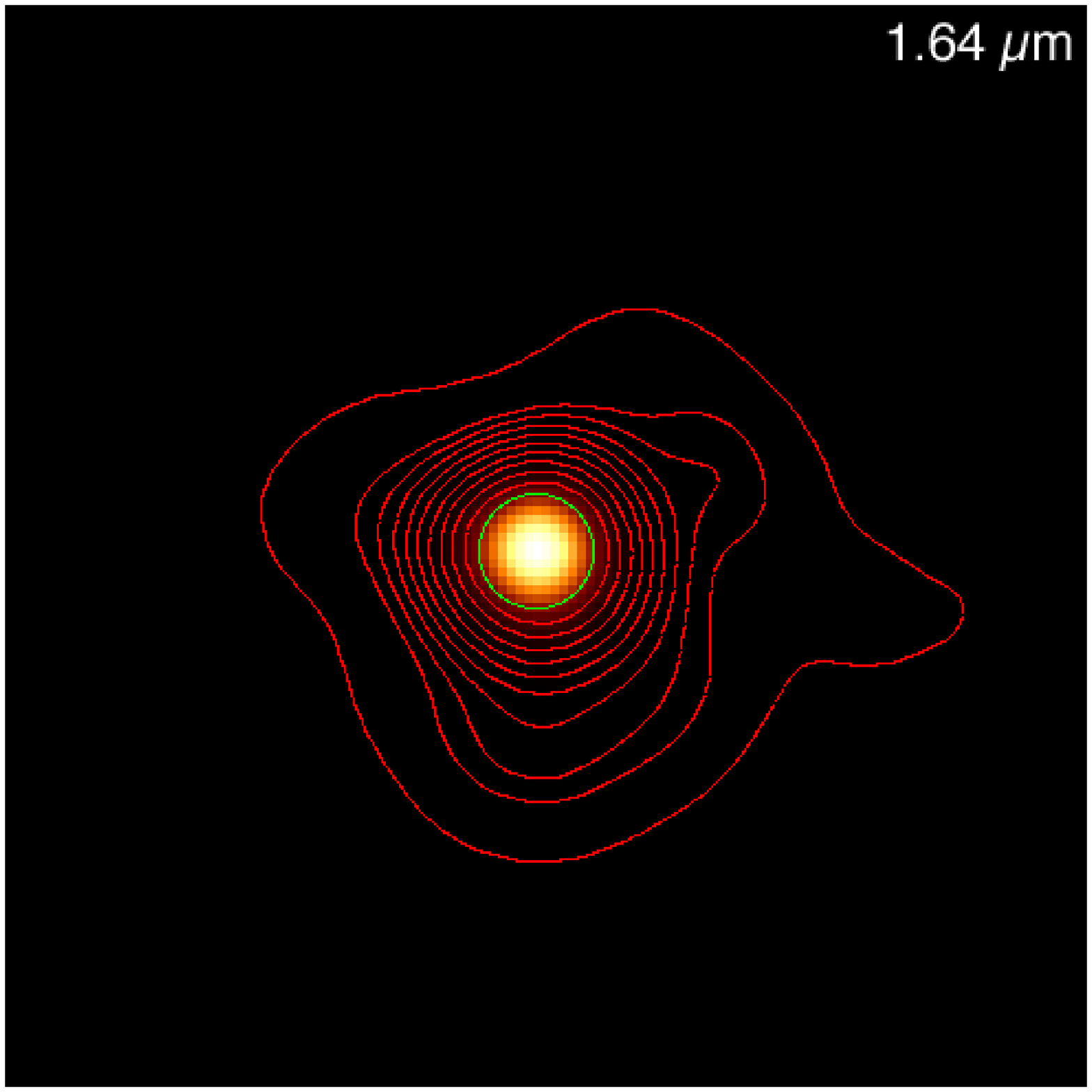} \includegraphics[width=4.3cm]{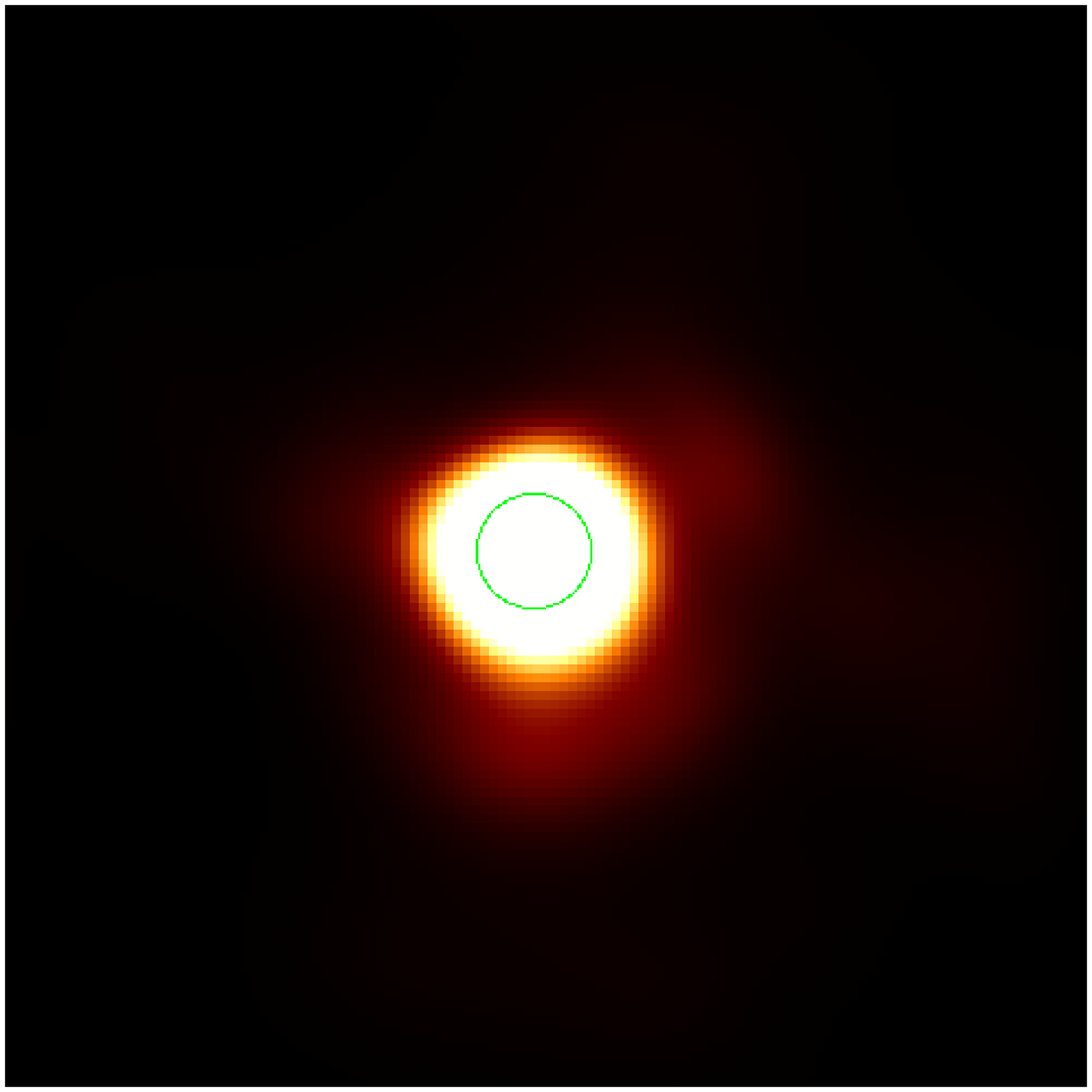}

\includegraphics[width=4.3cm]{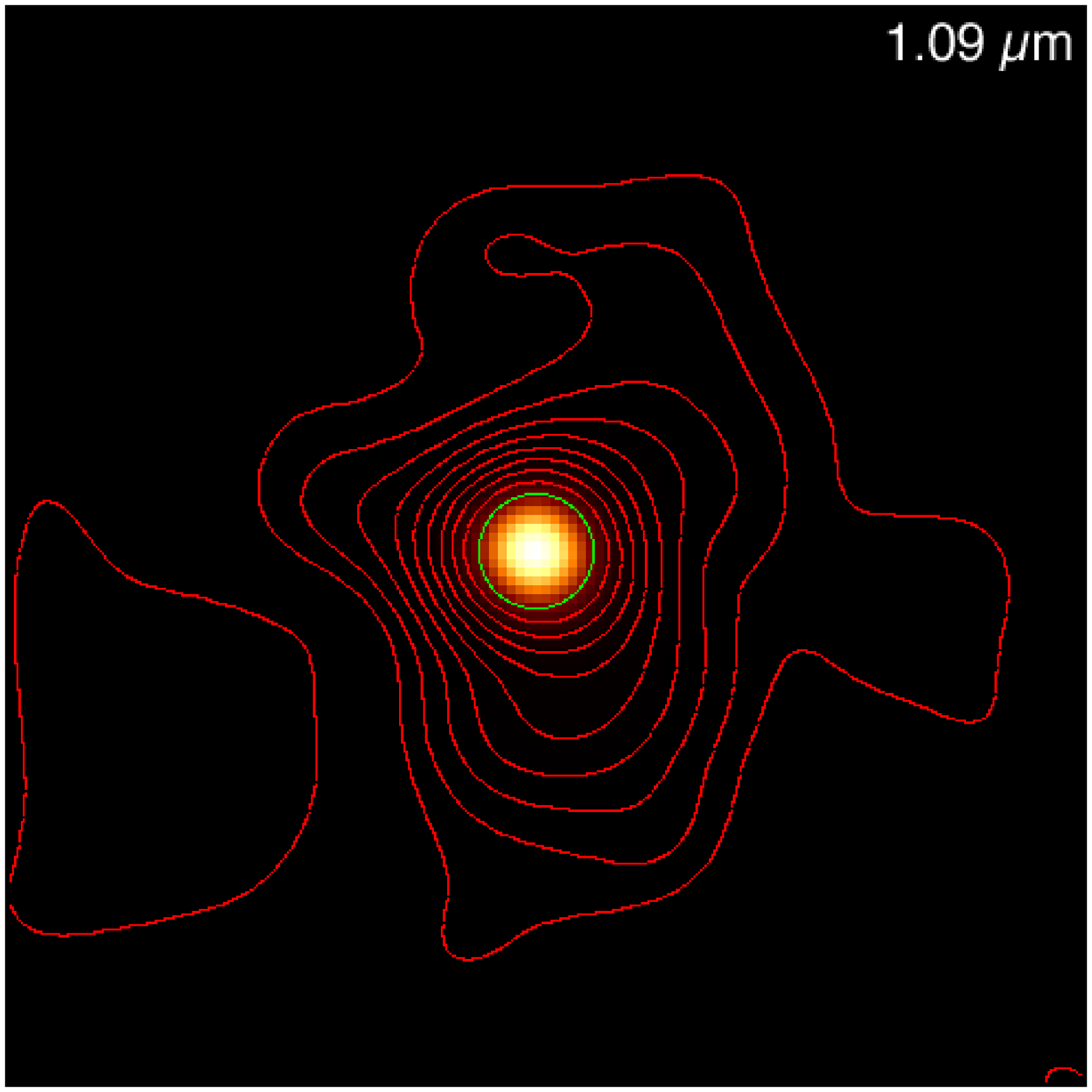} \includegraphics[width=4.3cm]{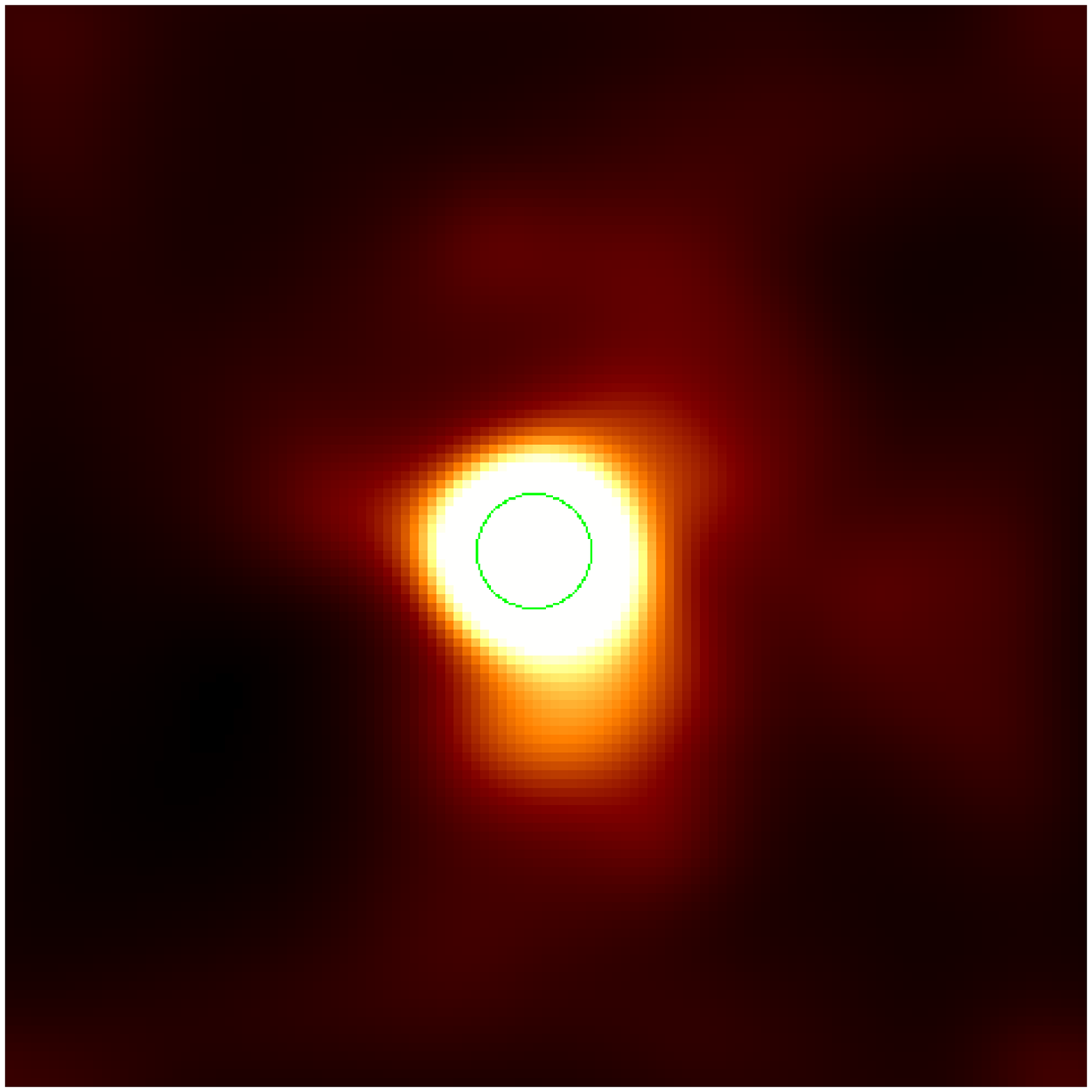} \hspace{2mm}
\includegraphics[width=4.3cm]{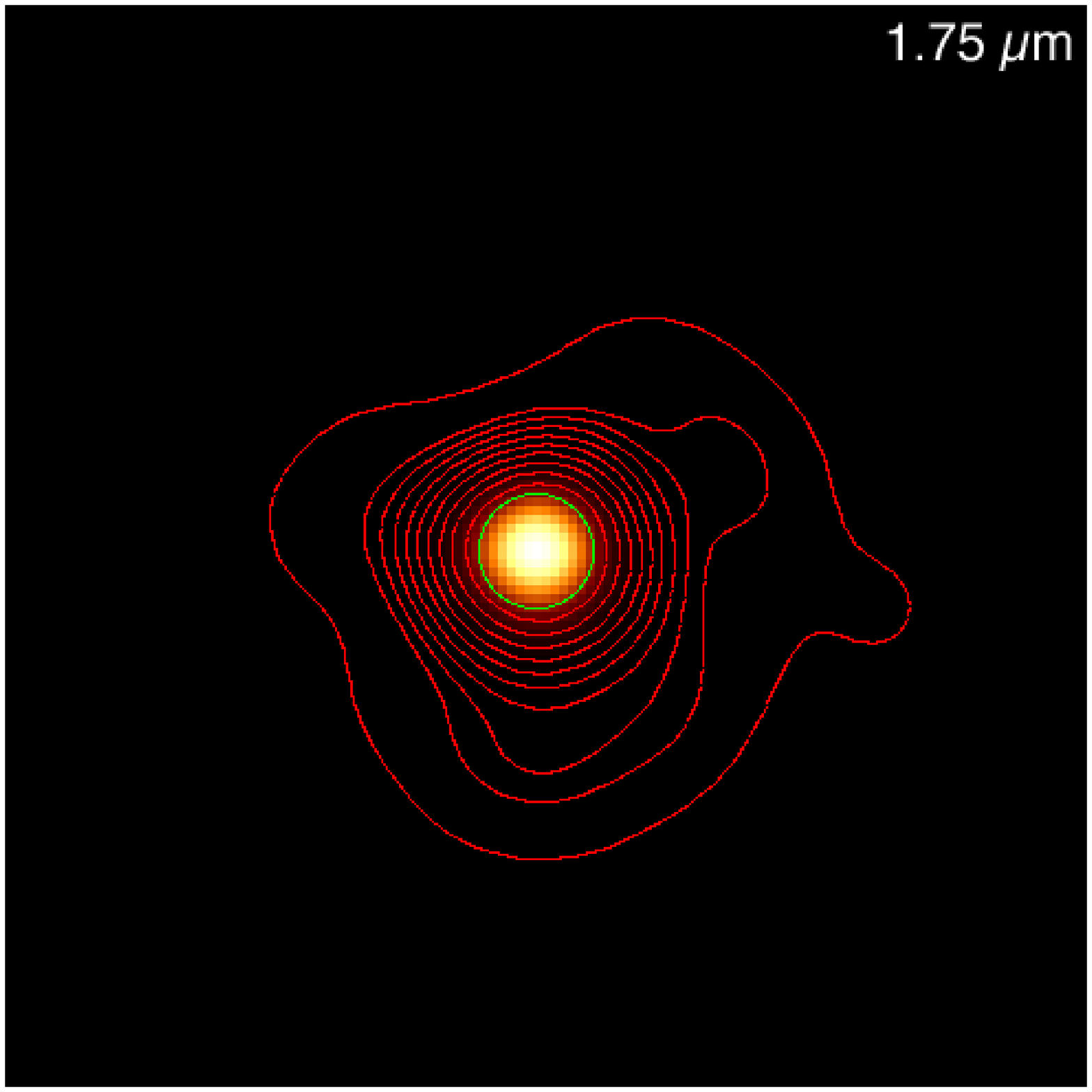} \includegraphics[width=4.3cm]{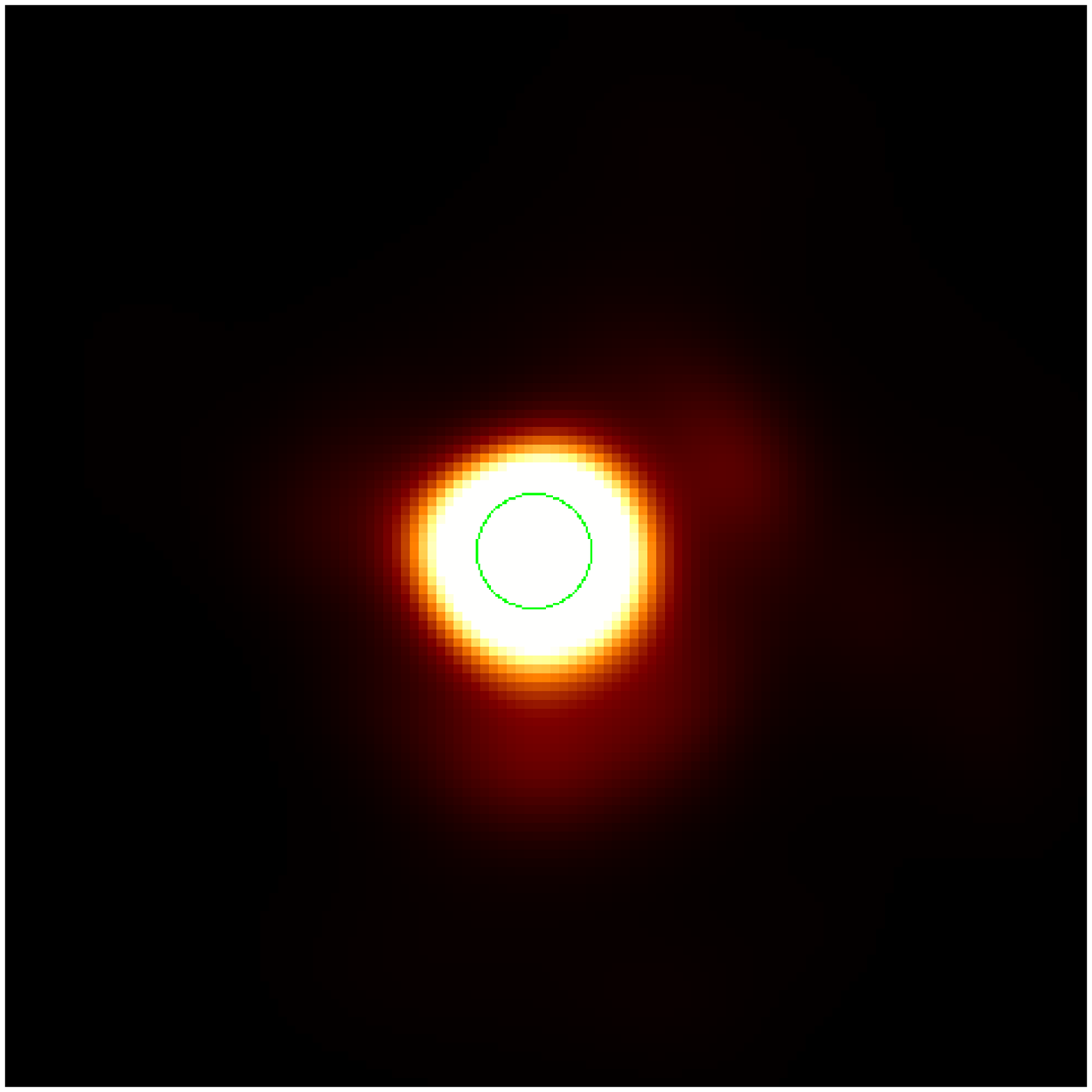}

\includegraphics[width=4.3cm]{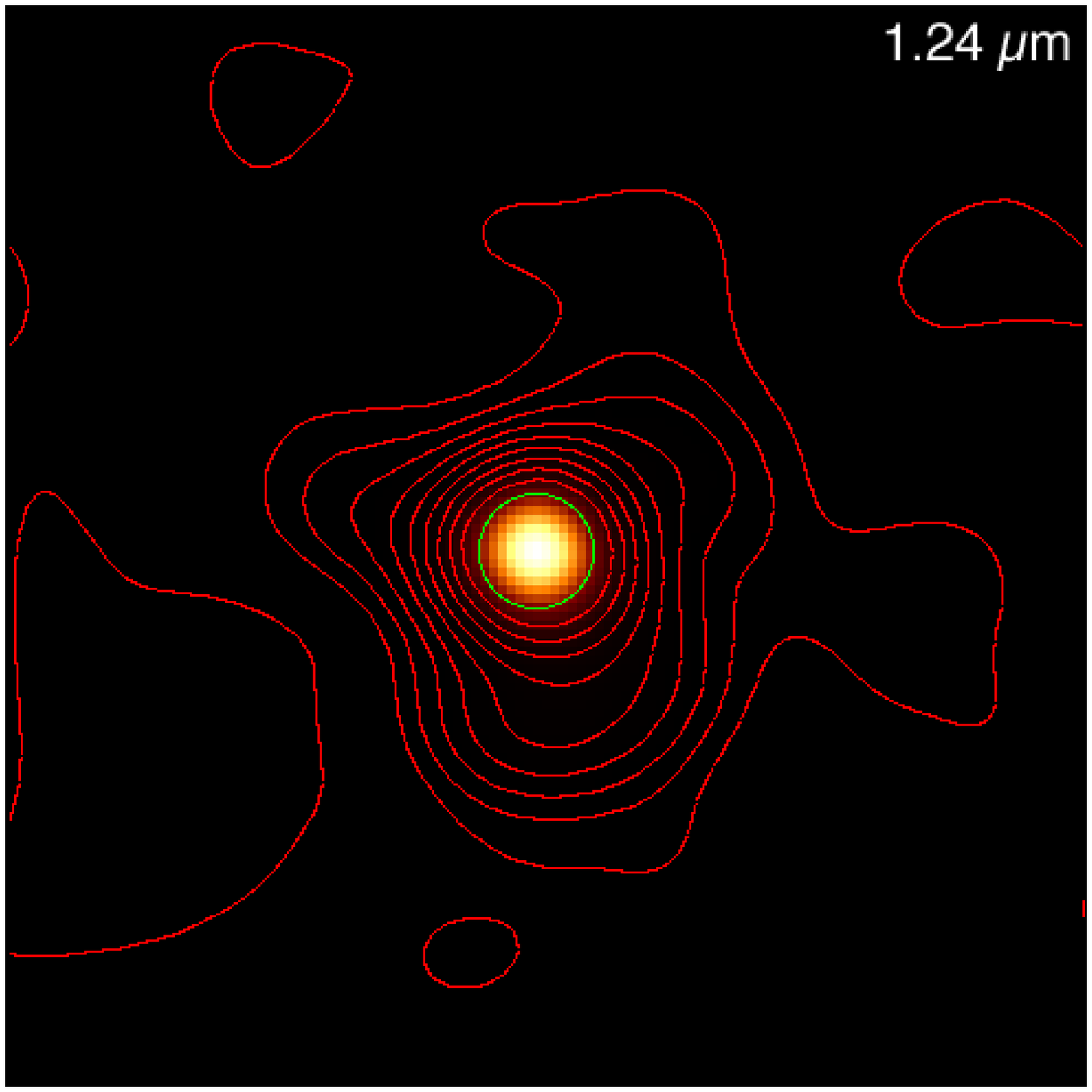} \includegraphics[width=4.3cm]{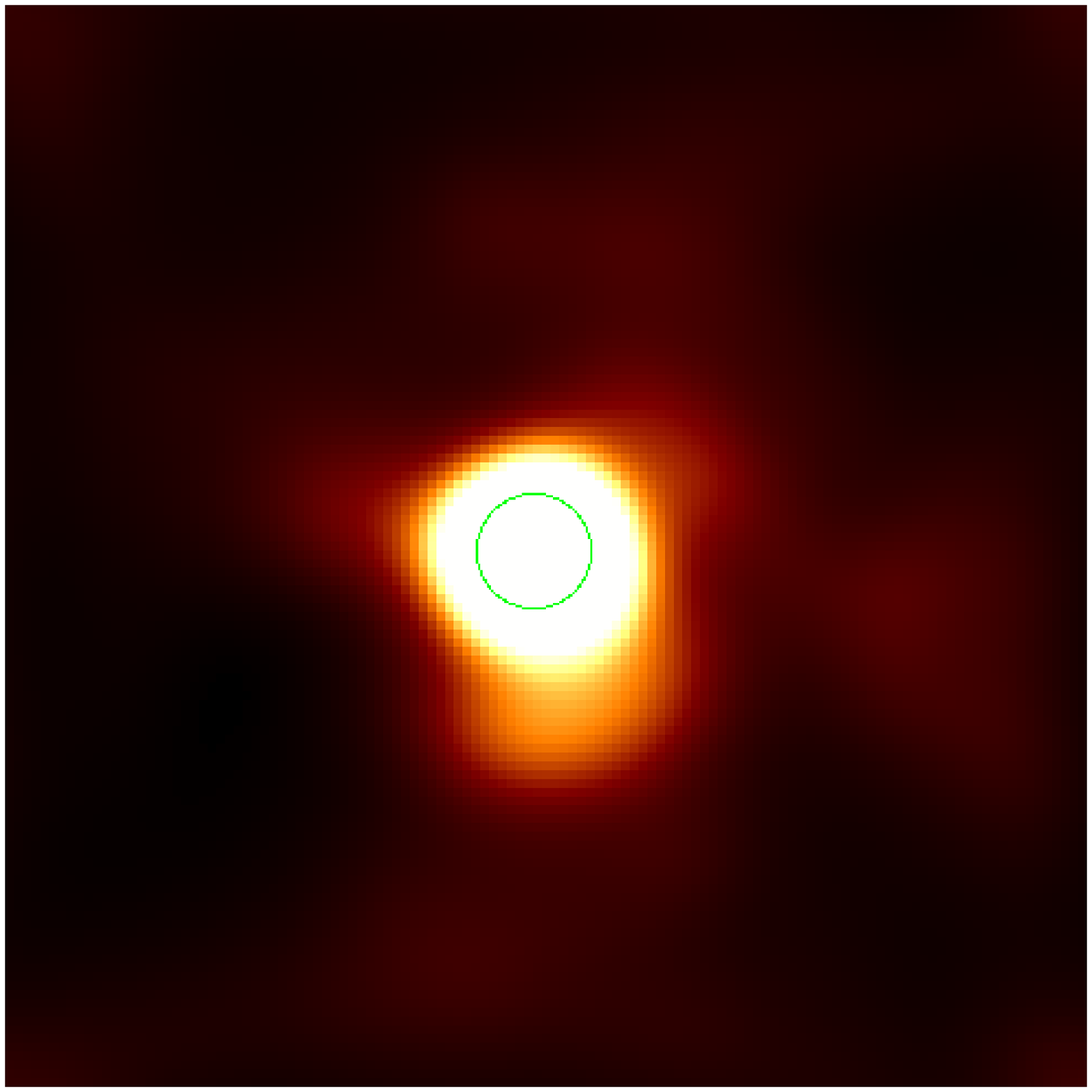} \hspace{2mm}
\includegraphics[width=4.3cm]{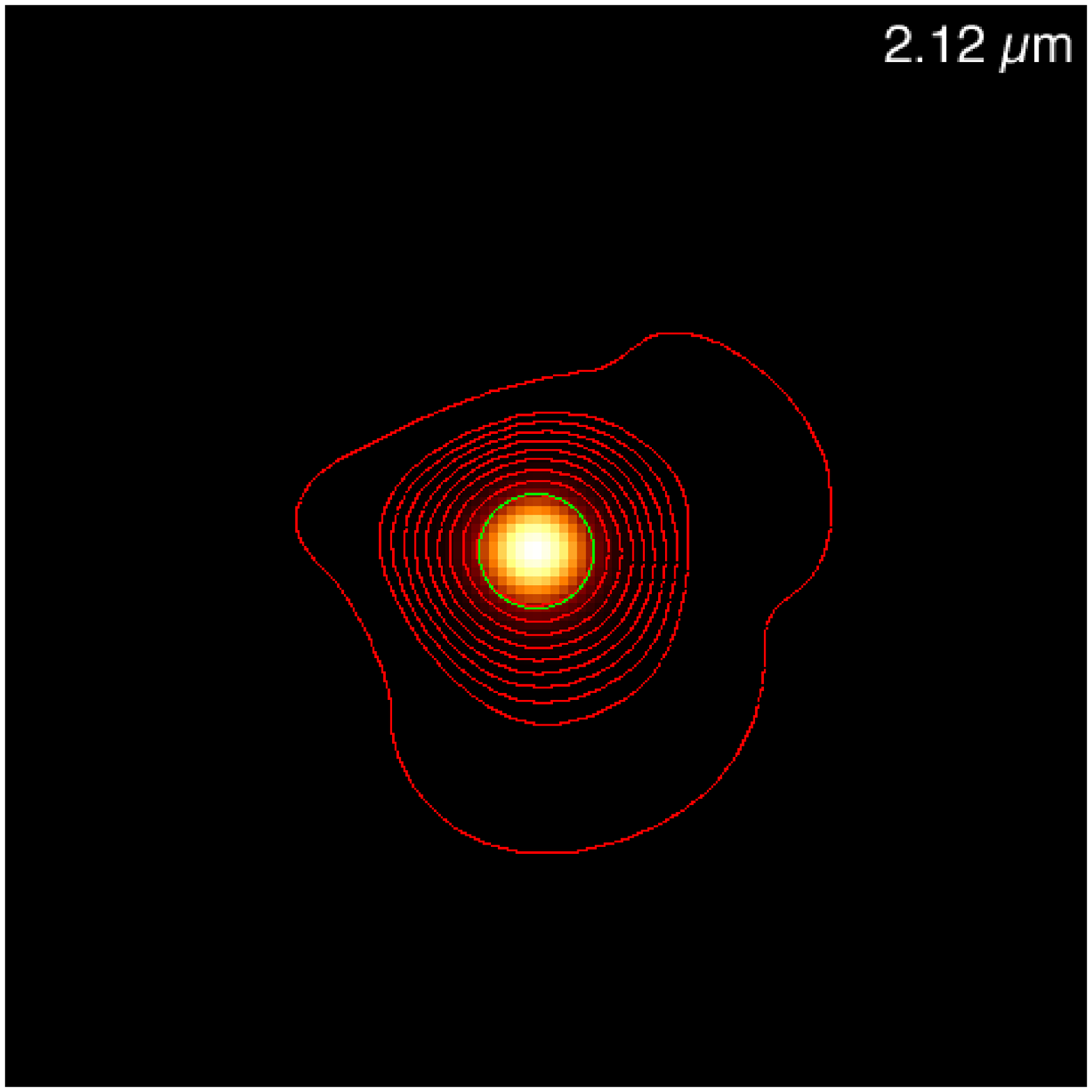} \includegraphics[width=4.3cm]{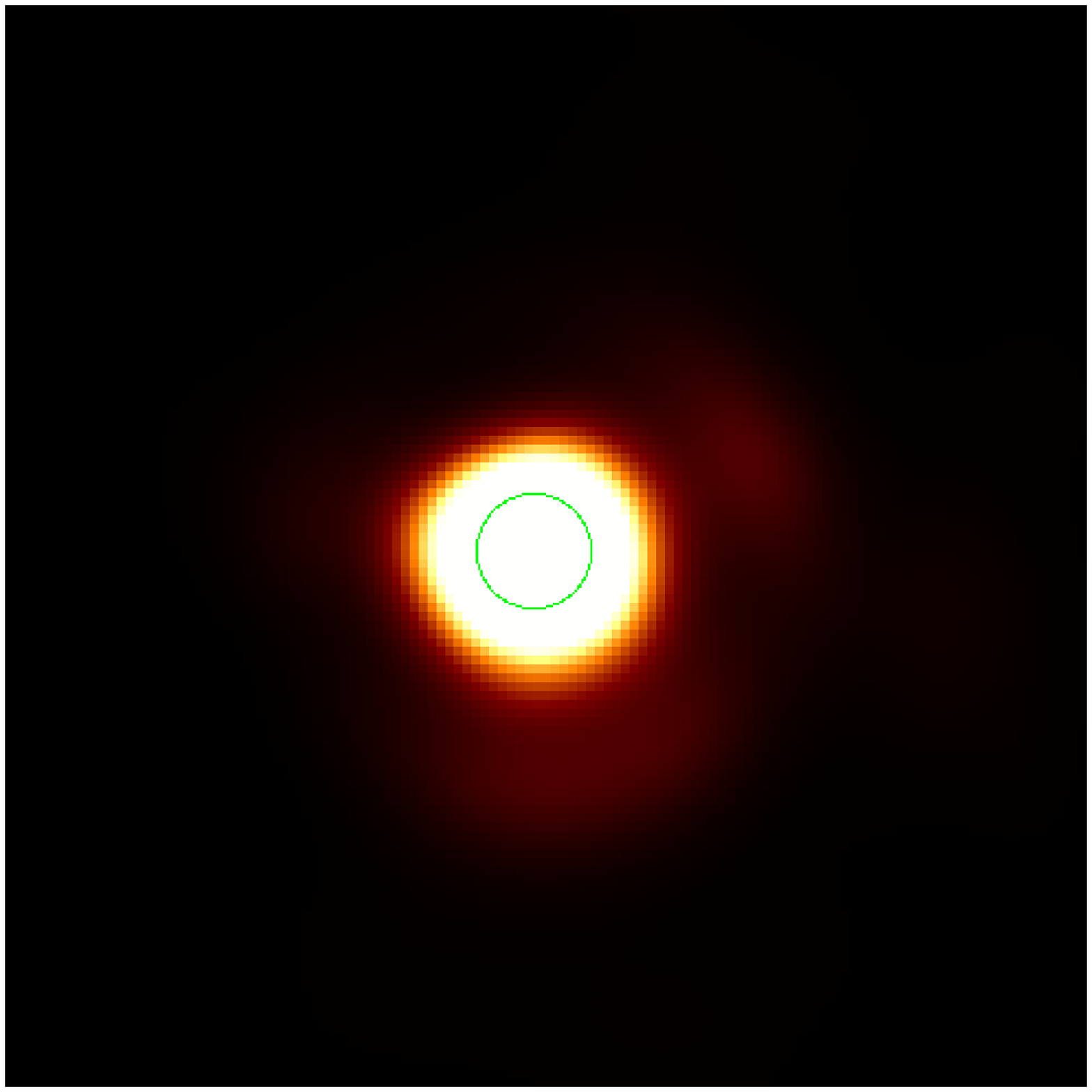}

\includegraphics[width=4.3cm]{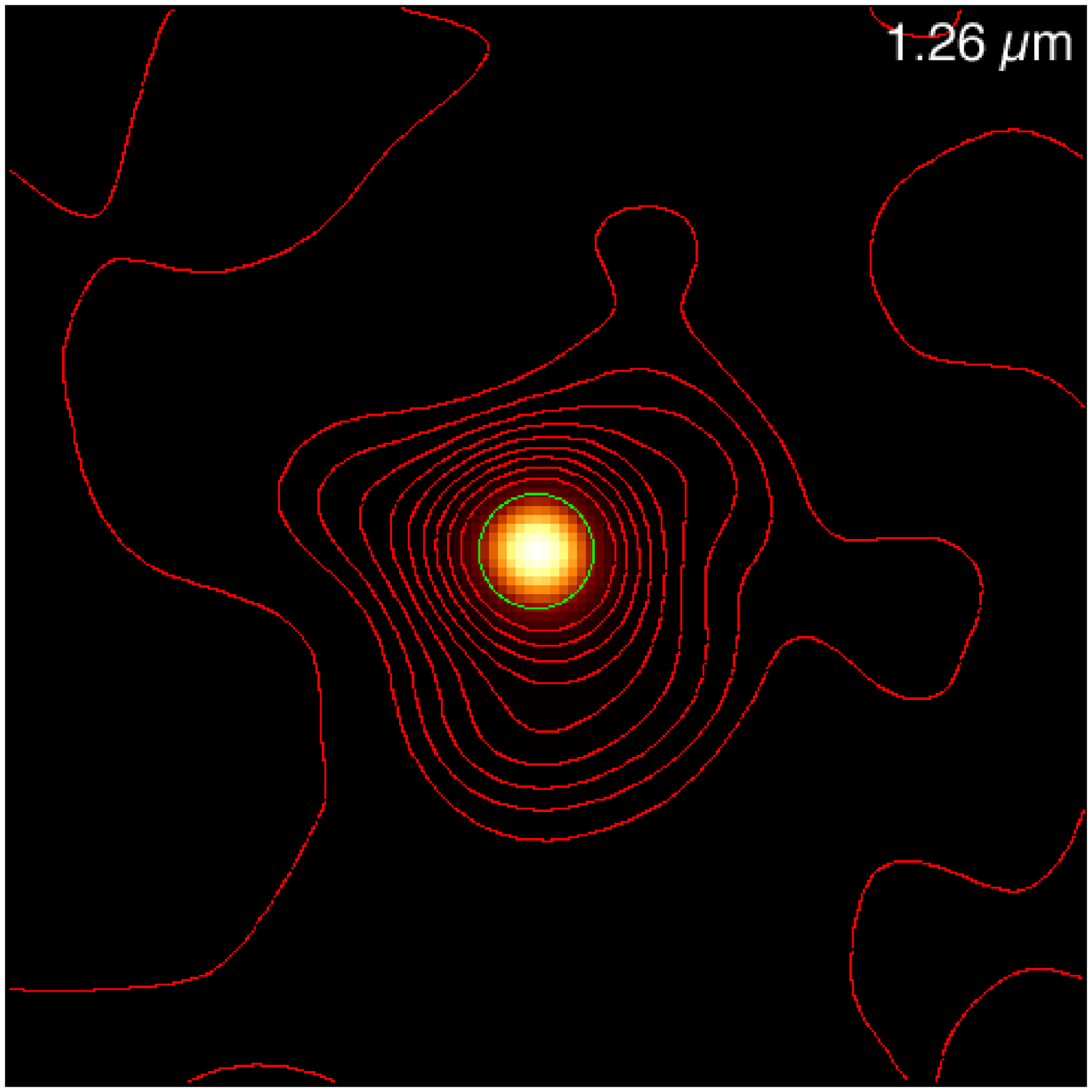} \includegraphics[width=4.3cm]{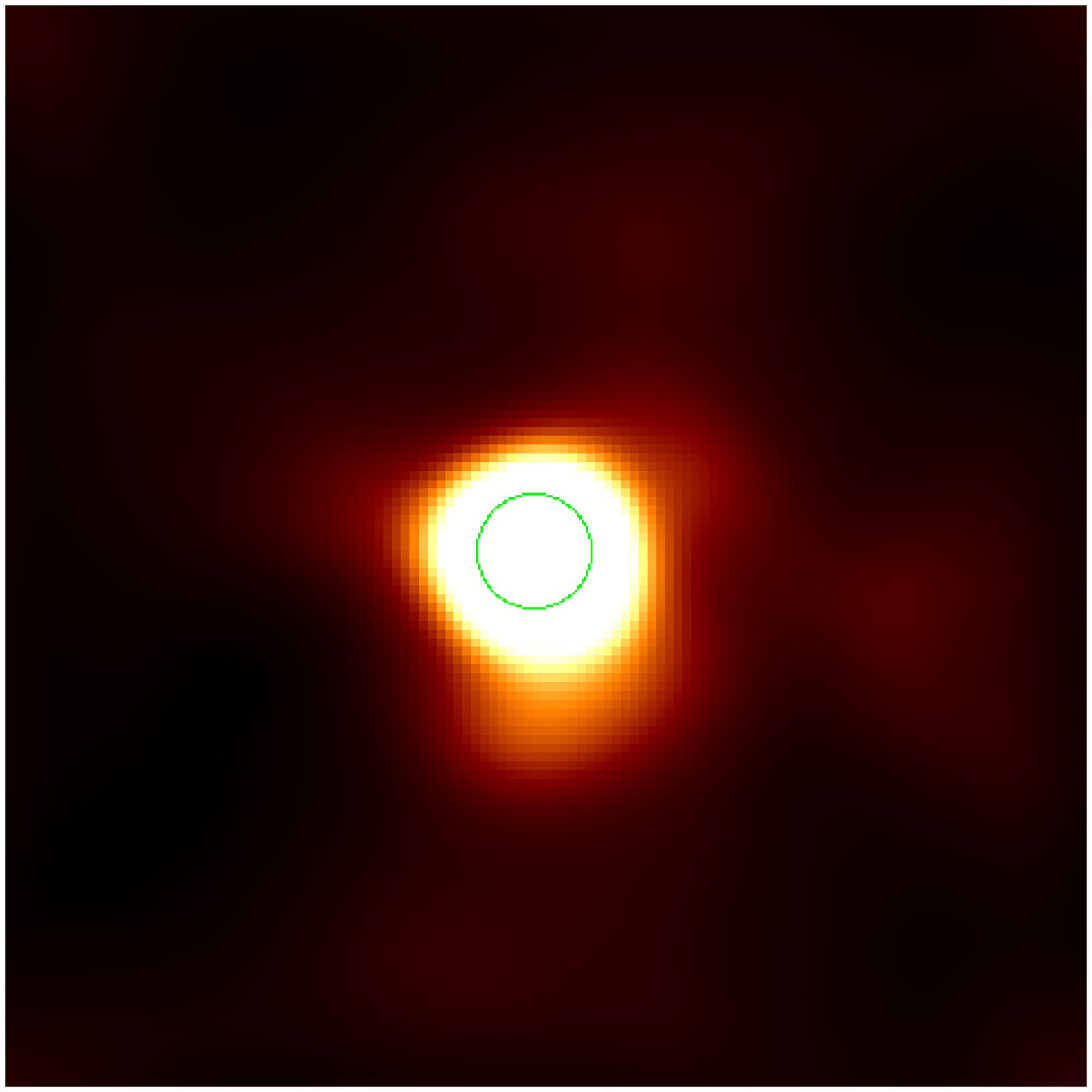} \hspace{2mm}
\includegraphics[width=4.3cm]{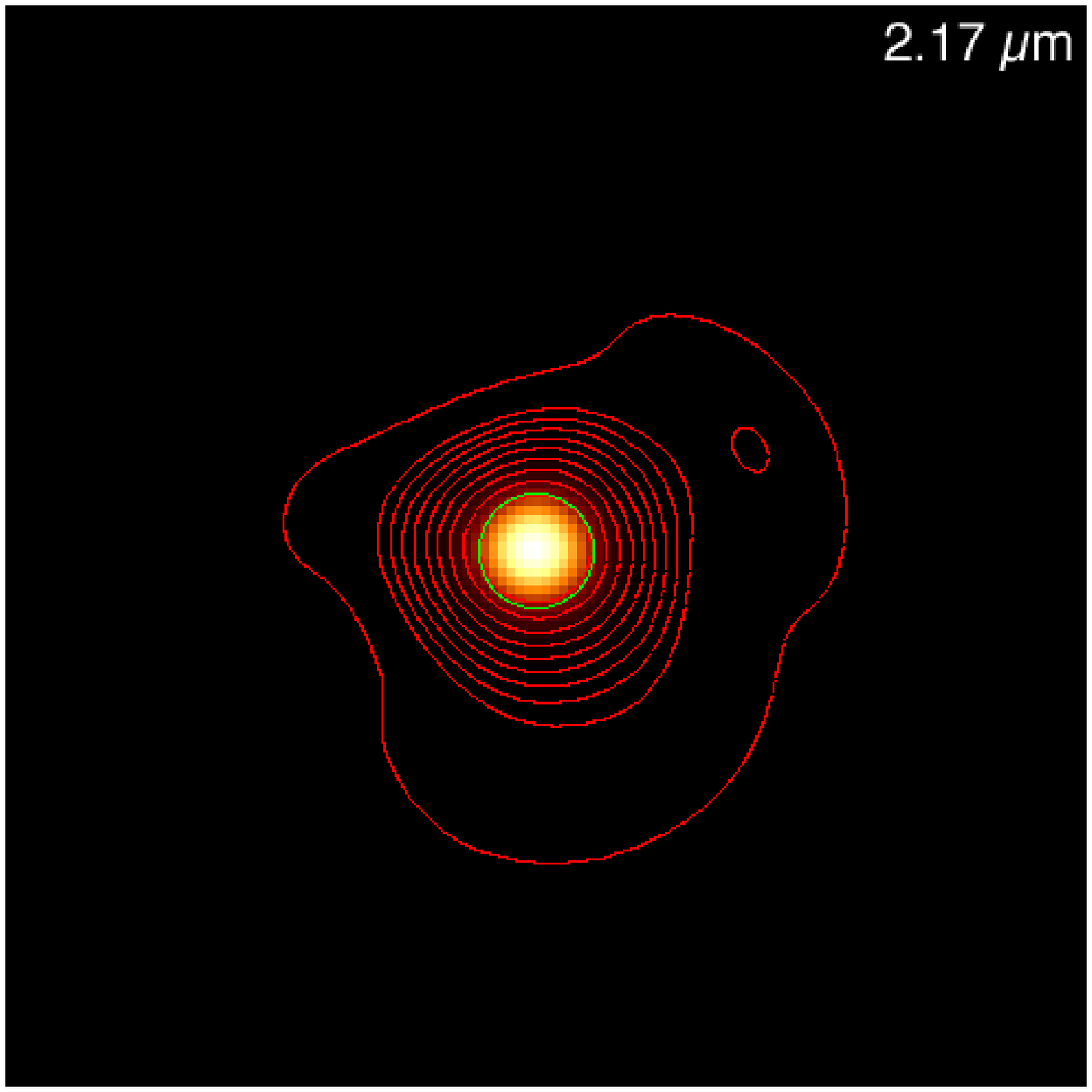} \includegraphics[width=4.3cm]{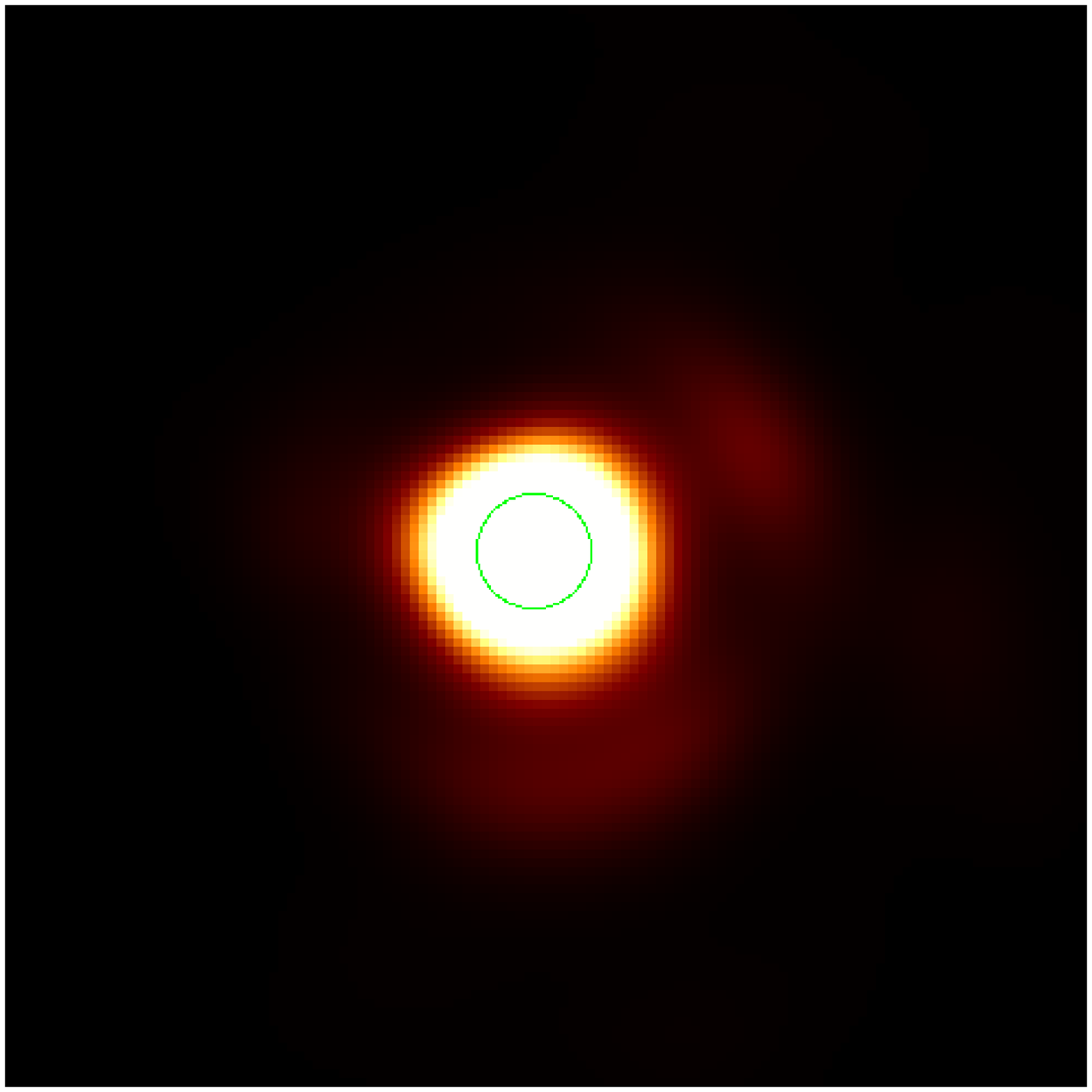} 
\caption{Lucy-Richardson deconvolved images of Betelgeuse (30 iterations). For each wavelength, the left image is the full-dynamic linear color scale image normalized to the maximum and minimum values. The 10 contours plotted in red correspond to flux levels of 1\,000, 2\,000, 4\,000, 8\,000, 16\,000, 32\,000, 64\,000, 128\,000 and 256\,000 W.m$^{-2}$.$\mu$m$^{-1}$.sr$^{-1}$. The right image uses a square root color scale with narrower color cuts (including 95\% of the pixels) to emphasize the faint extensions of the CSE. The green circle corresponds to the 43.7\,mas photospheric angular diameter of the star as measured by Perrin et al.~(\cite{perrin04}) from $K$ band interferometric observations.\label{deconvolved}}
\end{figure*}

We deconvolved the images of Betelgeuse using the PSF images (Sect.~\ref{psf}) as the dirty beams and the Lucy-Richardson (L-R) algorithm.
The images corresponding to the three imaging epochs of Betelgeuse were processed separately with the relevant PSF calibrator observation. The resulting average deconvolved images in the ten NACO narrow-band filters are presented in Fig.~\ref{deconvolved}.  We stopped the L-R deconvolution after 30 iterations, as the resulting images in the $K$ band (2.12 and 2.17\,$\mu$m) show a full width at half maximum of the photosphere of Betelgeuse compatible with the LD size measured by interferometry (Fig.~\ref{deconvolved}, left column), while no significant deconvolution artefact is present. 

In this process, we took into account the difference in rotational position of the pupil of the telescope between the two stars. We rotated the images of Aldebaran obtained in each filter to precisely match the telescope pupil position angle (hereafter abridged as ``PA") of the corresponding images of Betelgeuse. In other words, the secondary mirror support ``spikes" were brought to the same PA in Aldebaran's and Betelgeuse's images. This allowed us to avoid artefacts in the deconvolution of the Betelgeuse images. Although the presence of the spikes has a noticeable effect on the deconvolved images, we checked that the straight deconvolution of the Betelgeuse images by the Aldebaran images (without rotation) gives very similar results.
Photometry on deconvolved images is a tricky task, but McNeil \& Moody~(\cite{mcneil05}) showed that the Lucy-Richardson algorithm preserves the photometric accuracy relatively well compared to other classical algorithms. We therefore normalized photometrically the deconvolved images presented in Fig.~\ref{deconvolved} ($407 \times 407$\,mas field of view) by the total flux of Betelgeuse in each filter (in W.m$^{-2}$.$\mu$m$^{-1}$) as estimated in Sect.~\ref{photometry}, and by the size of the resampled image pixel (3.315\,mas/pix). The resulting physical unit of the images is therefore W.m$^{-2}$.$\mu$m$^{-1}$.sr$^{-1}$ (surface brightness).

%______________ Figure

\subsection{Circumstellar envelope shape}

In the deconvolved images presented in Fig.~\ref{deconvolved}, we observe the photosphere of Betelgeuse as a disk whose size appears mostly constant with wavelength (left columns). We also detect an extension of its envelope in the southwestern quadrant, at a PA of $200 \pm 20$ degrees (counted positively east of North), that we will subsequently refer to as a ``plume". The bright part of this plume extends up to a minimum radius of six times the photosphere of the star ($\approx 90-100$\,mas), and is already visible in the non-deconvolved images of Betelgeuse. At a distance of 4\,R$_*$, its surface brightness in the $1.09\,\mu$m band is $\approx 0.4$\% of the center of the stellar disk, and falls down to $\approx 0.1$\% at 6\,R$_*$ (Fig.~\ref{radius_cuts}).
Two other faint extensions appear in the northeastern (PA~$\approx 75^\circ$) and northwestern (PA~$\approx 310^\circ$) quadrants. Although faint, their presence at all ten wavelengths (processed independently) indicates that they are likely real features.

\subsection{Photospheric and CSE spectral energy distributions}

%______________ Figure
\begin{figure}
\centering
\includegraphics[height=4.9cm]{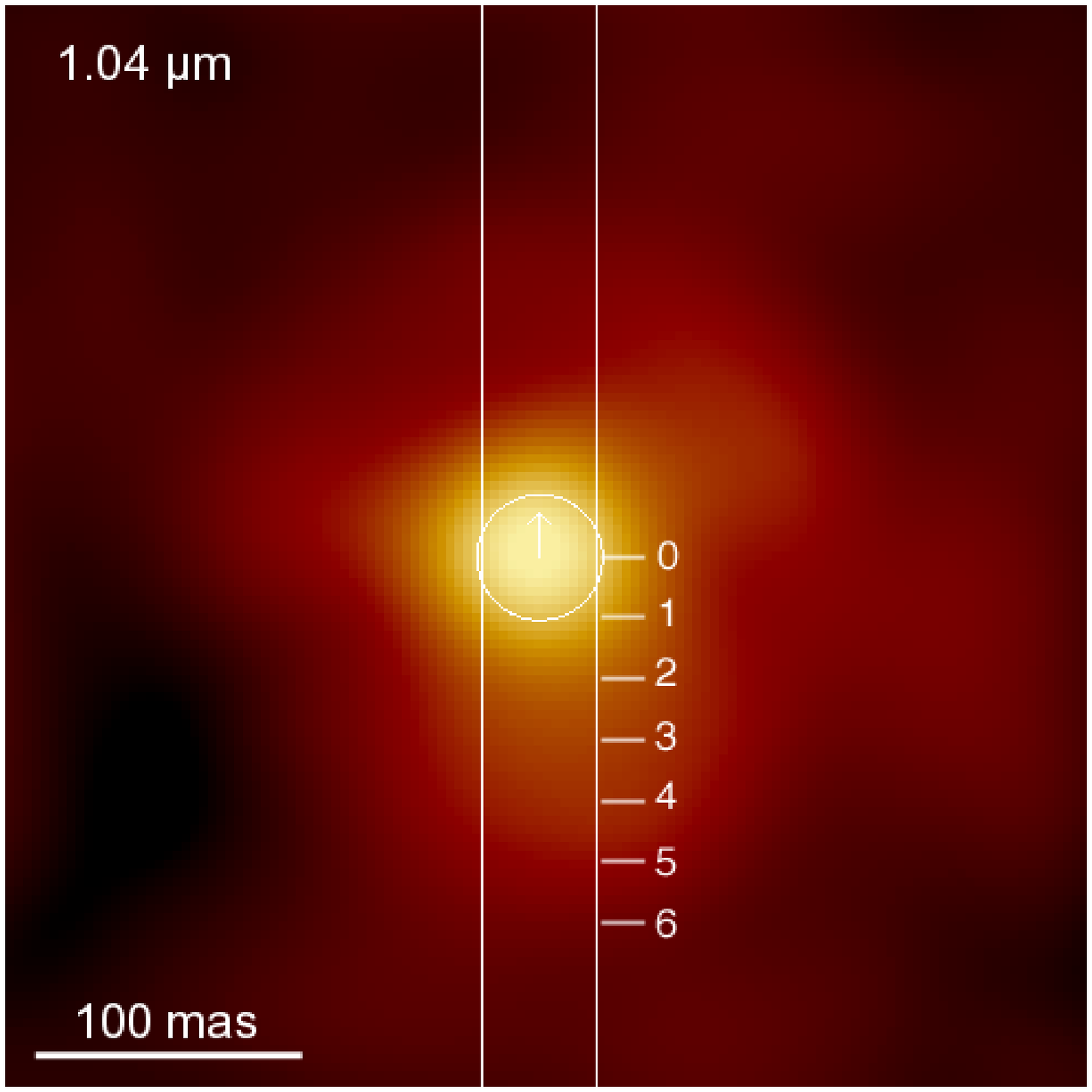}
%\hspace{0.1cm}
\includegraphics[height=4.9cm]{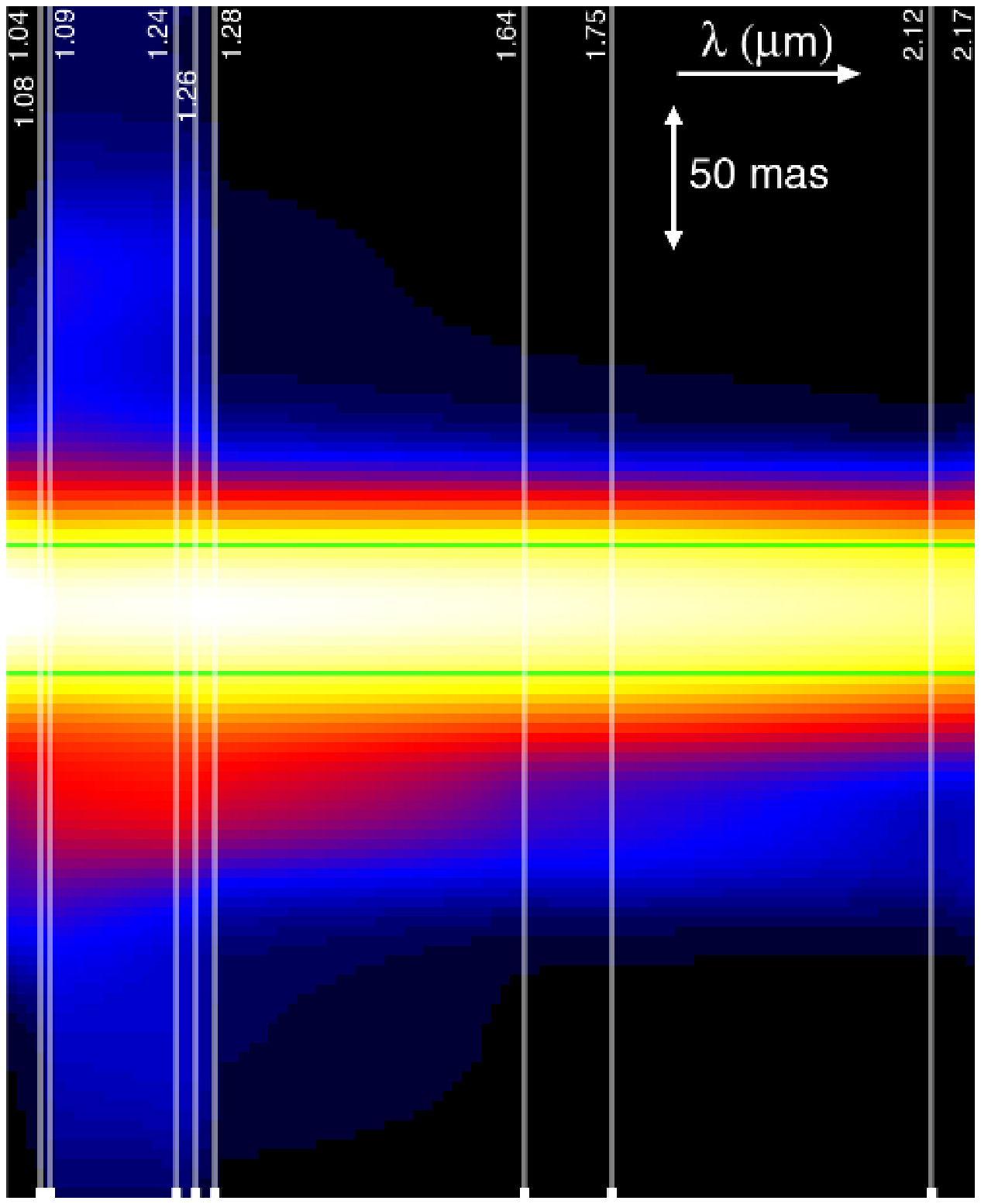} 
\caption{{\it Left panel:} Positions of the 3D pseudo-slit, plotted over the image of Betelgeuse at 1.04\,$\mu$m. The color scale is logarithmic to better show the extensions of the CSE. The central circle represents the photosphere of the star. The graduations from 0 to 6 are distances from the photosphere center in stellar radii for the spectra shown in Fig.~\ref{radius_cuts}. {\it Right panel:} Reconstructed long-slit spectrum of Betelgeuse (linear interpolation from the pseudo-slit cut in the Betelgeuse image cube). The color scale is logarithmic. The vertical lines indicate the central wavelengths of the NACO narrow-band filters (see also Fig.~\ref{nb_filters}), and the 43.7\,mas photospheric angular diameter is shown with green lines.
\label{spectro-slit}}
\end{figure}

%______________ Figure
\begin{figure}
\centering
\includegraphics[width=\hsize]{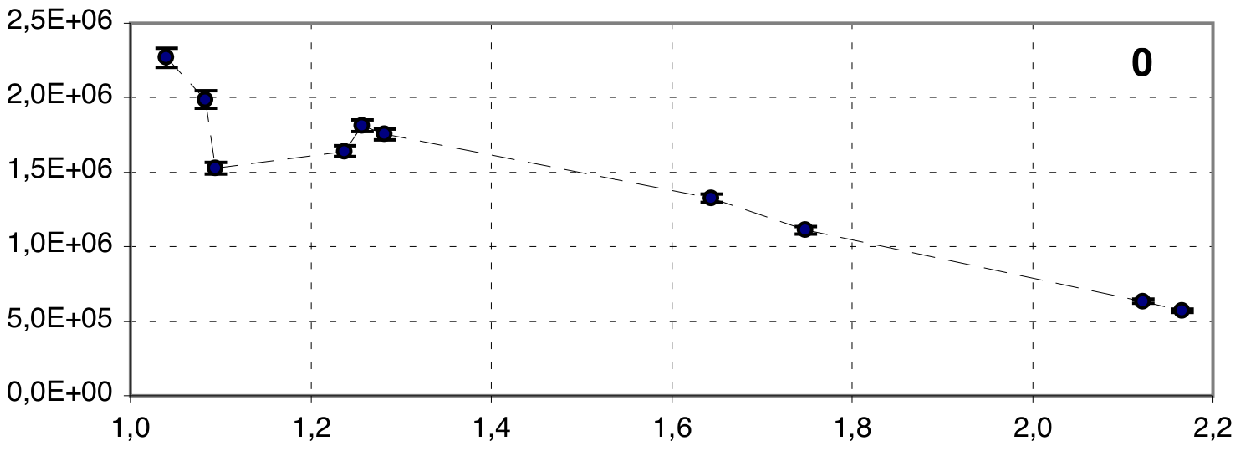}
\includegraphics[width=\hsize]{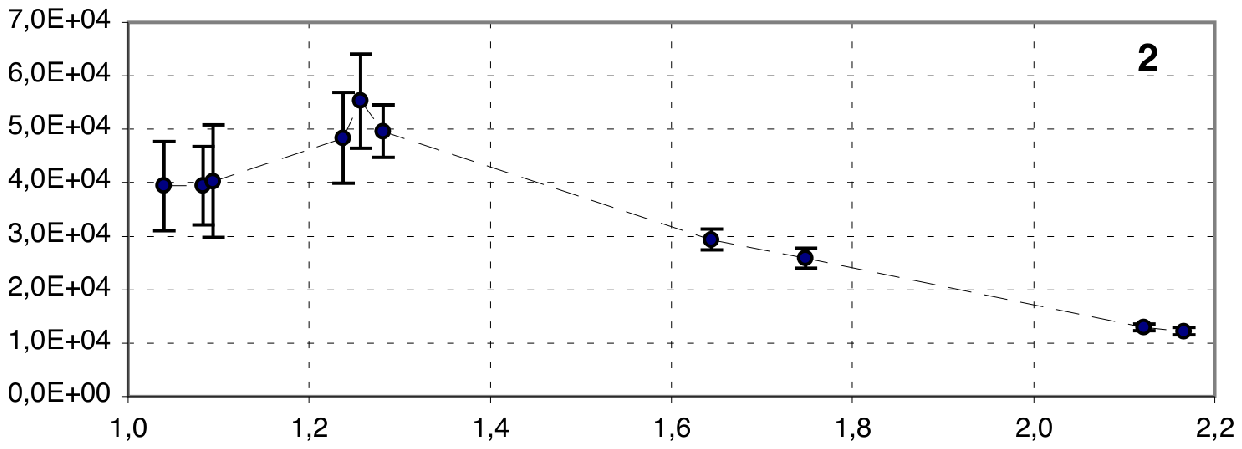} 
\includegraphics[width=\hsize]{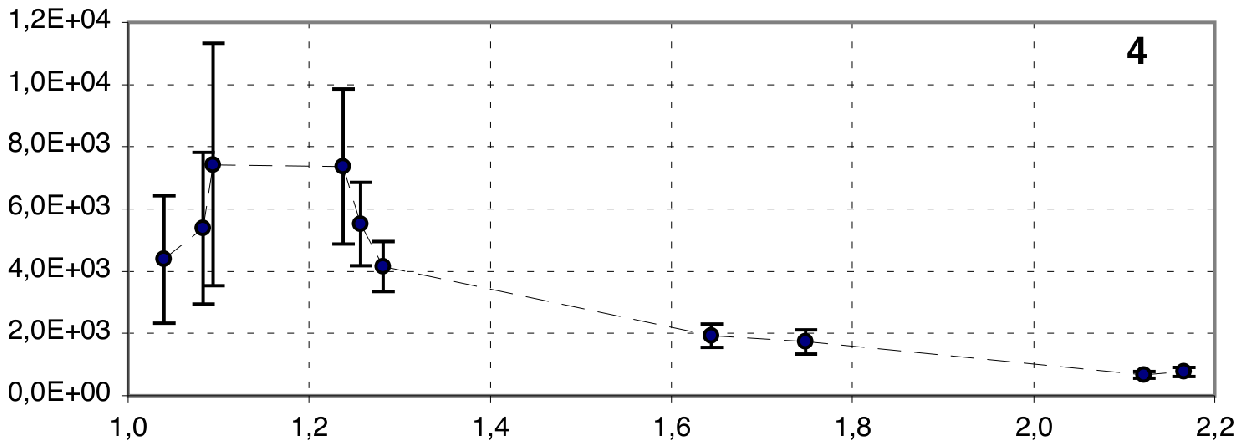} 
\includegraphics[width=\hsize]{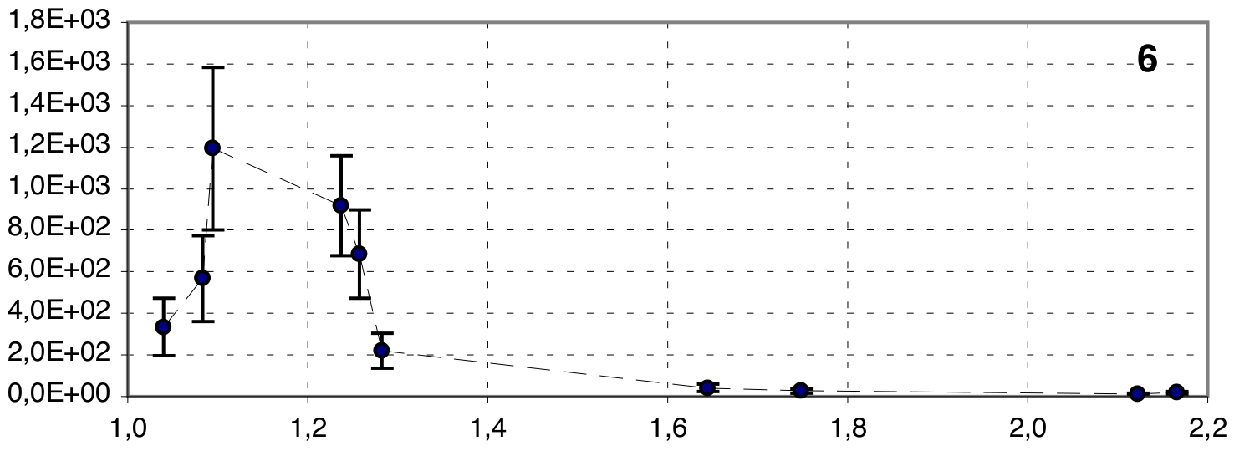} 
\caption{Surface brightness spectra of the close environment of Betelgeuse, averaged over the pseudo-slit width (44\,mas). The vertical scales are in W.m$^{-2}$.$\mu$m$^{-1}$.sr$^{-1}$, and the horizontal scale is the wavelength in $\mu$m. The radius (in $R_*$) from the center of the photosphere is indicated in the upper right corner of each plot.
\label{radius_cuts}}
\end{figure}

Based on the deconvolved and photometrically calibrated images, we can analyze locally the SED of the CSE of Betelgeuse. Fig.~\ref{spectro-slit} shows the pseudo-slit we used for this analysis, positioned in a north-south direction to include the brightest plume. The right part of this Figure shows the average surface brightness spectrum of Betelgeuse and its CSE over the pseudo-slit. The individual spectra observed at different distances from the center of the disk of Betelgeuse along the pseudo-slit are shown in Fig.~\ref{radius_cuts} (they correspond to horizontal cuts in the right part of Fig.~\ref{spectro-slit}). We used the RMS dispersion of the pseudo-slit measurements on the three deconvolved and photometrically calibrated images of Betelgeuse obtained in each filter to estimate the systematic uncertainty of our measurements. This gave us the error bars shown in Fig.~\ref{radius_cuts}.
The two fainter plumes visible in the Betelgeuse images in the northeastern and northwestern quadrants have similar spectra as the southern plume, with a maximum flux in the $J$ band between 1.08 and 1.24\,$\mu$m.

%__________________________________Discussion
\section{Discussion \label{discussion}}

\subsection{Identification of the spectral features}

%_________________________________________ Fig. absorption
%
\begin{figure}
\centering
  \resizebox{\hsize}{!}{\includegraphics{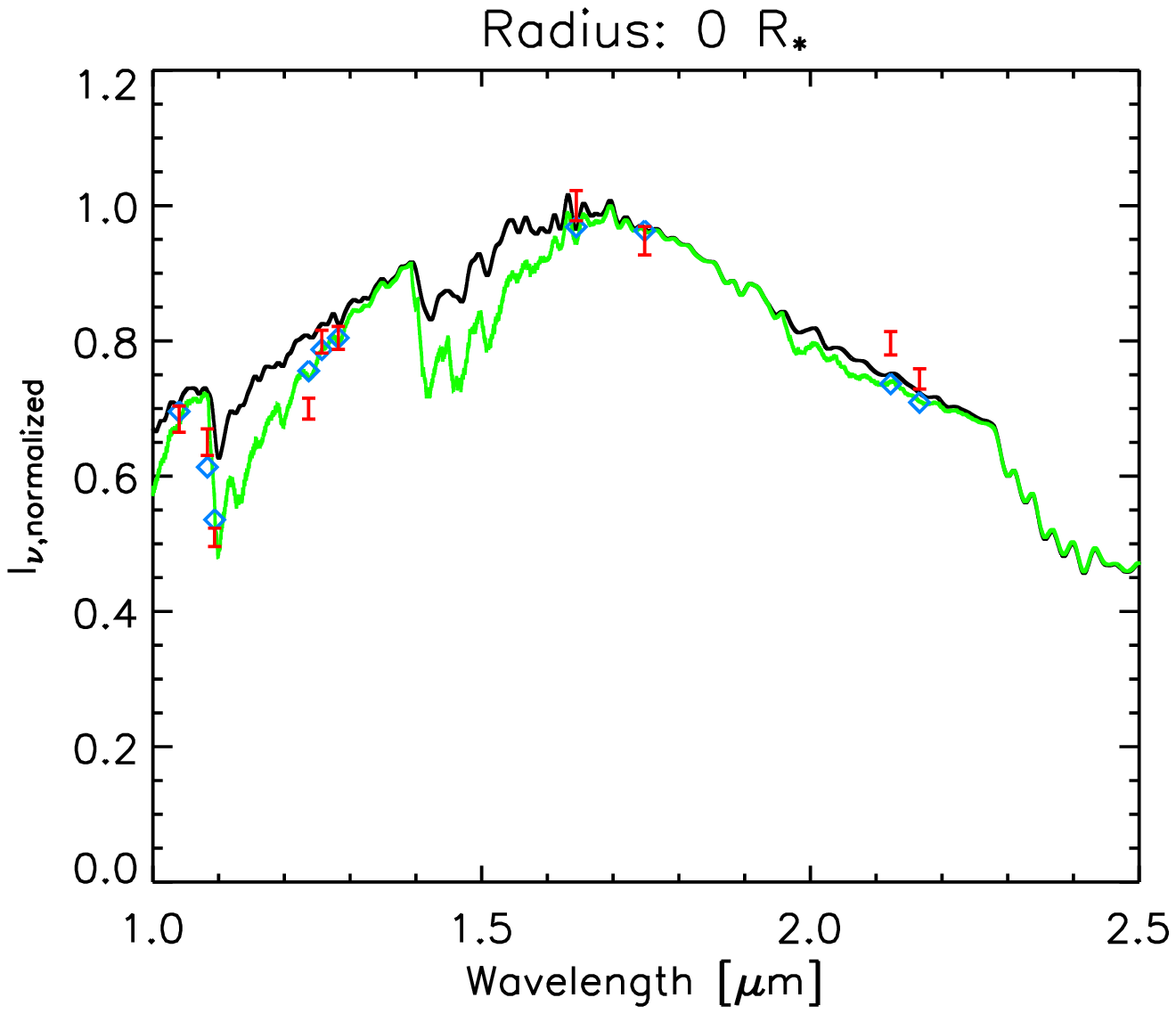}}
  \resizebox{\hsize}{!}{\includegraphics{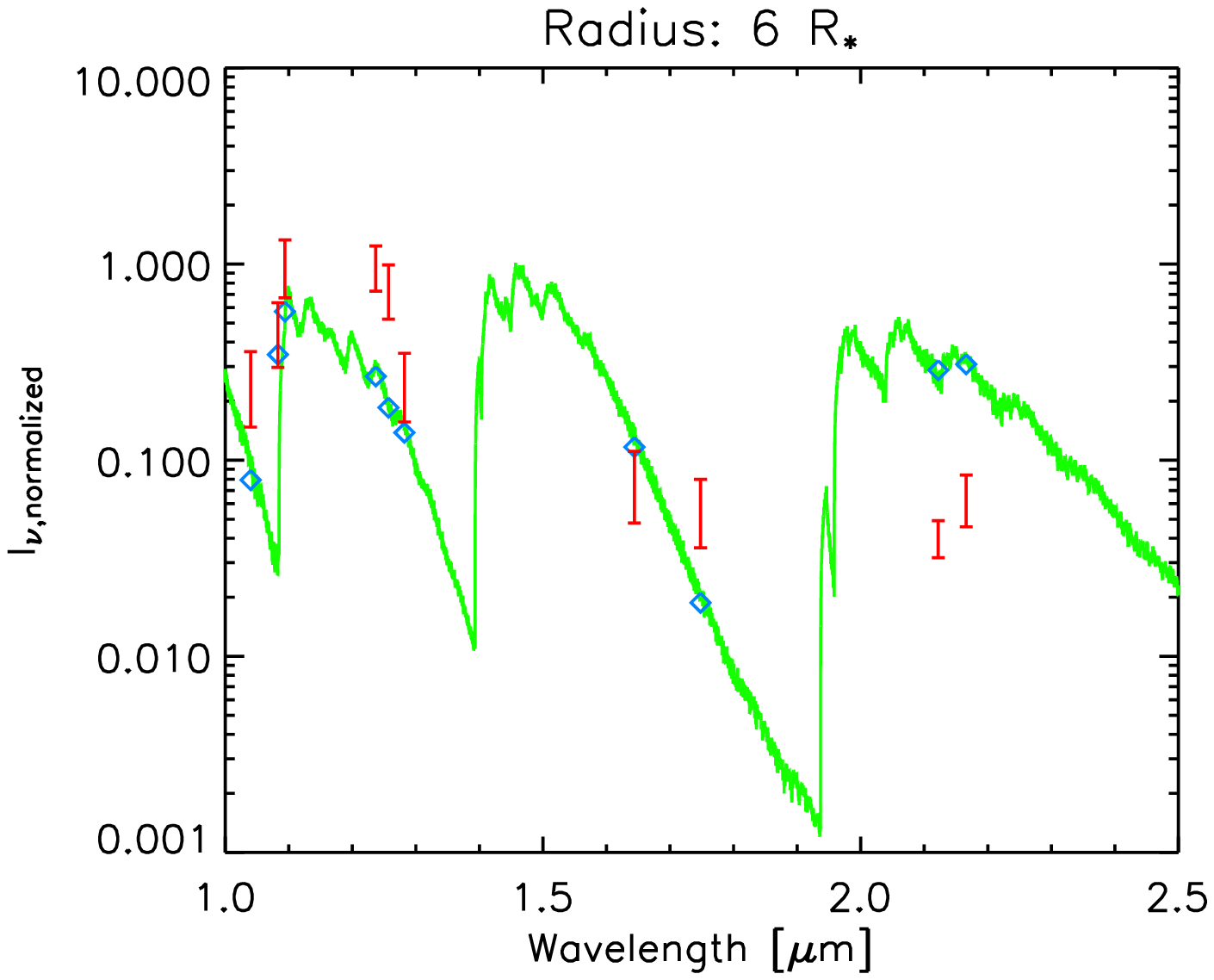}}
  \caption{{\it Upper panel:} The photospheric model spectrum (black), the model + CN slab
    spectrum (green), the predicted inband intensities (blue) and the
    observations (red).
    {\it Lower panel:} The emission spectrum of optically thin CN at 2\,500\,K in
    green, the observations in red at a radius of $6\,R_*$ and the predicted inband
    intensities in blue. Both plots use the same temperature and column density for the CN slab.}
  \label{fig:CN}
\end{figure} 

Both the photospheric spectrum and the spectra of the plume at different
radii show a spectral feature from roughly 1.1\,$\mu$m up to
1.3\,$\mu$m\footnote{As there is a gap in the NACO spectral coverage from
  1.16 to 1.22\,$\mu$m (see Fig.\,\ref{nb_filters}), it is conceivable
that we are in fact dealing with two separate features.}.
Against the photosphere, it is seen in absorption, and
without a strong background source, i.e. offset from the central star
but at the location of the plume, it is observed in emission.

The near-IR spectral range of red supergiant stars such as
$\alpha$\,Ori and $\mu$\,Cep is the topic of a long-standing debate:
several features have alternately been attributed to CN or H$_2$O
(Woolf et al.~\cite{woolf64}, Wing \& Spinrad~\cite{wing70}, Tsuji~\cite{tsuji00}).
The current consensus is that
features due to photospheric CN are present, as predicted by the
self-consistent hydrostatic models, and that water is detected in the
poorly-understood close circumstellar environment
(e.g. Jennings \& Sada~\cite{jennings98}, Verhoelst et al.~\cite{verhoelst06},
Perrin et al.~\cite{perrin07}).  As we are
limited by the transmission of Earth's atmosphere, our spectra do not
cover the water bands which were discussed most (those with bandheads around
1.4 and 1.9\,$\mu$m.)

Regarding the identification of the feature observed here, H$_2$O is
not a candidate since this molecule does not exhibit a clear band at this
particular wavelength range. CN provides an excellent match to the
bandhead, and by a careful choice of temperature (2\,500\,K) and column
density (6$\times10^{18}$ cm$^{-2}$),
it is possible to model this absorption feature with a single slab of CN, without
the introduction of unobserved features at the other wavelengths of CN
opacity (see the upper panel of Fig.\,\ref{fig:CN} for the model fit).

Unfortunately, the same basic modelling approach does not work for the
emission spectra: the prediction of the bluest feature is, in the
optically thin case,
far too narrow (lower panel of Fig.\,\ref{fig:CN}) and any combination of
temperature and column density
that matches this broad feature, predicts a strong
emission feature in the $K$ band starting around 1.9\,$\mu$m, but is not
observed. Neither clumpiness nor a difference between excitation temperature
and source-function temperature or a source of continuous background
emission help to solve this problem. Obscuration of the feature at
1.9\,$\mu$m is in theory possible with cold water vapour
(cfr. the discussion on the identification of the observed absorption
features at those wavelengths), but that would require a large water column
density and a very particular geometry.

An important aspect of our spectro-imaging observations is that the angular resolution
decreases by a factor two in the $K$ band compared to the
$J$ band (where the plume is most visible). In the L-R deconvolution process, the angular
size of the bright photosphere is recovered properly (Fig.~\ref{deconvolved}), but the
detection of the plume is made more difficult by its much lower surface brightness
(typically several hundred times fainter than the photosphere).
We therefore cannot exclude that our marginal detection of the plume in the $K$ band is
caused by a lack of resolving power, that results in a lower sensitivity of the L-R
deconvolved images to extended sources.
As a consequence, as we do not observe the expected emission feature around 1.9\,$\mu$m,
we cannot confirm our idenfication of CN, but cannot propose a viable alternative either.

\subsection{Previous high angular resolution observations}

Thanks to its large angular diameter and correlatively high brightness, Betelgeuse has been an excellent target for high resolution observations using a broad palette of techniques: direct full-pupil imaging, aperture masking, speckle interferometry and long-baseline interferometry. This short summary of the existing observational reports is in no way complete, but is intended to give an overview of the detections of asymmetries and surface features on Betelgeuse. Such asymmetries could point at a link between the photosphere and the envelope of the star as observed in our NACO images.

Early observations in the visible using the rotation shearing interferometer (Roddier \& Roddier~\cite{roddier83}, \cite{roddier85}; Roddier et al.~\cite{roddier86}) revealed that in November 1980 and February 1982, asymmetric emission was present around Betelgeuse, and they proposed that it was caused by scattering in a dusty layer close to the star. The presence of two companions was suggested by Karovska et al.~(\cite{karovska86}), but our NACO observations do not show any point-like source down to a magnitude difference $\gtrsim 5$ in the $JHK$ band over the field of view of our observations ($0.4 \times 0.4\arcsec$). Observations in the visible obtained at the same epoch (February 1981) with the Mayall telescope by Christou et al.~(\cite{christou88}) show the presence of PSF-subtraction residuals in the southwestern quadrant with a typical radius of the order of 100\,mas, a size and position comparable to that of the plume detected with NACO. Hebden et al.~(\cite{hebden87}) observed Betelgeuse in the H$\alpha$ line with the MMT in speckle interferometry at two epochs: December 1983 and November 1985. They identify mostly no asymmetry at both epochs up to a radius of $\approx 95$\,mas, while slightly larger extensions appear in the east-west and north-south directions at larger radii. In February 1989, Buscher et al.~(\cite{buscher90}) observed Betelgeuse using the WHT equipped with a non-redundant aperture mask and three narrow-band filters in the visible around $\lambda=0.7\,\mu$m. These authors detected an asymmetry on the surface of the star, that they interpret as a hot spot located at a PA of $\approx 280^\circ$, with a relative flux contribution of 10-15\%. Using the same instrument and technique, Wilson et al.~(\cite{wilson92}) re-observed Betelgeuse in January 1991, and also detected asymmetries on the disk of the star and a northeast-southwest elongation different from to the 1989 epoch.

Gilliland \& Dupree~(\cite{gilliland96}) resolved the chromosphere of Betelgeuse at ultraviolet wavelengths using the \emph{Hubble Space Telescope} equipped with the \emph{Faint Object Camera} (FOC), and also obtained resolved spectra with the \emph{Goddard High Resolution Spectrograph} (GHRS). These data were obtained in March 1995 and clearly show the resolved chromosphere in UV continuum, extending to more than twice the radius of the visible photosphere. The deconvolved FOC images consistently show an unresolved bright region located in the southwestern quadrant of the disk, where we observed the plume with NACO. Gilliland \& Dupree~(\cite{gilliland96}) estimate that the spot temperature is $\gtrsim 200$\,K hotter than the surrounding chromosphere. The chromosphere observed at $\lambda = 255$\,nm with HST/FOC is on average much hotter ($\approx 5000$\,K) than the overall effective temperature of the star ($\approx 3600$\,K, Perrin et al.~\cite{perrin04}).

In October 1995, Burns et al.~(\cite{burns97}) used the COAST interferometer at $\lambda=0.80\,\mu$m to observe Betelgeuse and found no asymmetry of the stellar disk itself, but their reconstructed image (their Fig.~4) shows an extension along the northeast-southwest direction (PA~$\approx 200^\circ$), aligned with the NACO plume. Further observations obtained in November 1997 with COAST and the WHT telescope at visible and near infrared wavelengths (0.70, 0.91 and $1.29\,\mu$m) by Young et al.~(\cite{young00}) also showed an asymmetry of the light distribution on the disk, that tends to decrease as wavelength increases. Interestingly, their $1.29\,\mu$m observations do not show any asymmetry, while our observations with NACO at $1.28\,\mu$m still show the presence of the southwestern plume. Tuthill et al.~(\cite{tuthill97}) also used the WHT telescope in aperture masking mode around $\lambda=0.70\,\mu$m between January 1992 and December 1993. Depending on the epoch, they identified two to three hot spots with changing positions, and relative flux contributions of up to 20\%, on the surface of Betelgeuse. 

Betelgeuse was observed in early 1998 by Hinz et al.~(\cite{hinz98}) using nulling interferometry in the mid-infrared ($\lambda = 10\,\mu$m) with two mirrors of the Multiple Mirror Telescope separated by a baseline of 5 meters. This interferometric technique allows to cancel out the light from the unresolved star, leaving as a residual the emission from the circumstellar environment. Their resulting nulled image (their Fig.~3) shows a very extended ($\approx 5\arcsec$ in diameter) asymmetric dusty envelope, with extensions in the southwestern, northwestern and northeastern directions, similar to the positions of the three plumes we observed with NACO. Smith et al.~(\cite{smith09}) recently found a mildly clumpy, spherical shell around Betelgeuse in CO emission, from spatially resolved spectroscopy at $\lambda \approx 4.6\,\mu$m. Some interesting structures were detected $\approx 1\arcsec$ west of the star, but their identification to a mass loss event or envelope asymmetry is unclear.

The single-mode two-telescope interferometric observations obtained by Perrin et al.~(\cite{perrin04}) and Chagnon et al.~(\cite{chagnon02}) respectively in the $K$ and $L$ bands provided high accuracy photospheric angular diameters of Betelgeuse ($\theta_{\rm LD\,K} = 43.7 \pm 0.10\,{\rm mas},\ \theta_{\rm UD\,L} = 42.17 \pm 0.05\,{\rm mas}$). However, due to the absence of closure phase measurements, it was not possible to check for the presence of asymmetry on the stellar disk. New observations of Betelgeuse with the three-telescope beam combiner IONIC installed at the IOTA interferometer were obtained in October 2005 and March 2006 by Haubois et al.~(\cite{haubois06}) in the infrared $H$ band. They point at the presence of one very faint hot spot (relative flux 0.5\%) in the northwestern quadrant of Betelgeuse (through model-fitting), or alternatively at two hot spots close to the center of the disk (through image reconstruction).
Spectrally dispersed interferometric observations of Betelgeuse in the $K$ band (obtained in January 2008) are reported by Ohnaka et al.~(\cite{ohnaka09}). These authors find a centrally symmetric emission on the apparent disk of the star in the continuum, but an inhomogeneous velocity field in the 2.303\,$\mu$m CO line, with a North-South preferential direction. This direction appears consistent with the preferential axis of the main plume we detected with NACO.

At thermal infrared wavelengths ($\lambda = 11.15\,\mu$m), Tatebe et al.~(\cite{tatebe07}) identified an asymmetry on the disk of Betelgeuse from three-telescope interferometric observations with the ISI instrument obtained in November and December 2006. They interpreted this asymmetry as the signature of a hot spot located on the southern limb of the star at a PA of $170-180^\circ$. This position is compatible with the PA of the plume we detect with NACO. Perrin et al.~(\cite{perrin07}) observed Betelgeuse with the two-telescope instrument VLTI/MIDI in the $N$ band ($\lambda = 7.5-13.5\,\mu$m), detecting a molecular layer (the MOLsphere) containing H$_2$O, SiO and Al$_2$O$_3$ at a radius of $\approx 1.4\,R_*$ from the star.

\subsection{Rotation}

The rotation of a supergiant star as Betelgeuse is expected to be very slow, but it can have non-negligible effects on the stellar atmosphere structure. Uitenbroek et al.~(\cite{uitenbroek98}) presented a detailed analysis of the HST/GHRS spectra first reported by Gilliland \& Dupree~(\cite{gilliland96}). They showed that chromospheric emission in the UV emission lines of Mg\,{\sc II}  extends to $\approx 270$\,mas from the center of the star, twice further than the UV continuum. In addition, thanks to a scanning of the instrument over the stellar disk, they could determine the projected rotational velocity of ($v \sin i = 5$\,km/s), the PA of its polar axis (PA~$\approx 235^\circ$) and the inclination of the polar axis on the line of sight of ($i = 20^\circ$). From spectroscopy and theoretical considerations, Gray~(\cite{gray00}) obtained a similar rotational velocity. This position of the polar cap is compatible with the hot spot visible in the FOC images and the PA of the southwestern plume observed in our NACO images (PA$= 200 \pm 20^\circ$)

\subsection{Origin of the southwestern plume}

The presence of the bright southwestern plume detected with NACO implies that the spherical symmetry observed on the infrared photosphere of Betelgeuse is not preserved in its close environment. 
Among the plausible mechanisms to explain its presence, two appear as promising: convection and rotation. Following the proposal by Schwarzschild~(\cite{schwarzschild75}) and the simulation results by Freytag \& H{\"o}fner~(\cite{freytag08}), a first hypothesis is that the very large convection cells enhance locally the mass-loss from the star, creating the observed plumes of molecules. However, although many studies have shown that the photosphere of Betelgeuse presents a significant degree of asymmetry, variable with time, the link with the chromosphere and the envelope remains to be established.
A second hypothesis is that mass loss may be enhanced above the polar cap of the star, creating the southwestern plume of Betelgeuse. This possibility is supported by the HST imaging and spectroscopy by Uitenbroek et al.~(\cite{uitenbroek98}), as it reveals the presence of a hot spot that could be located at the pole of the star. An interaction between this hot spot and the extended chromosphere of the star could result in a preferential production of molecules above the pole and the creation of a plume.
 
%__________________________________Conclusion
\section{Conclusion}

From diffraction-limited observations in ten narrow-band filters spread over the $JHK$ spectral domain, we find evidence for the presence of a complex CSE around Betelgeuse. The most conspicuous feature of this envelope is a bright plume extending in the southwestern quadrant up to a minimum distance of six photospheric radii. Its SED in emission appears complementary to a strong absorption feature present in the $J$ band (1.08-1.24\,$\mu$m) on the photospheric spectrum. This points at the presence of a molecular layer, possibly CN, although the formal identification of the spectral signature of this molecule in the $K$ band is not established. Two hypotheses could explain the formation of the plume: convection causing an enhanced mass loss above a large upwards moving convective cell, or rotation, through the presence of a hot spot at the location of the polar cap of the star. Further spectro-imaging observations resolving the photosphere of Betelgeuse (e.g. by interferometry) should allow to discriminate between these possibilities, if a link can be established between photospheric hot spots (characteristic of convection) and the detected plume.

%__________________________________Acknowledgements
\begin{acknowledgements}
We thank Dr. C. Lidman for his help in the preparation of these observations,
and the ESO Paranal staff for their perfect execution in visitor mode at the telescope.
STR acknowledges partial support by NASA grant NNH09AK731.
This work also received the support of PHASE, the high angular resolution
partnership between ONERA, Observatoire de Paris, CNRS and University Denis Diderot Paris 7.
This research took advantage of the SIMBAD and VIZIER databases at the CDS, Strasbourg (France), and NASA's Astrophysics Data System Bibliographic Services.
\end{acknowledgements}

%__________________________________Bibliography
{}

%__________________________________Table of observations
\onllongtab{1}{
% Beginning of online table
\begin{longtable}{rllrcccccc}
\caption{Log of Betelgeuse observations with NACO. \label{naco_log}}\\
\hline \hline
Code & Date \& UT & Object  & Filter & ND$^{\mathrm{a}}$ & FOV & DIT$^{\mathrm{b}}$ & N & $\sigma$$^{\mathrm{c}}$  & AM$^{\mathrm{c}}$\\
 &  &   &  & & (pixels) & (ms) &  & ($\arcsec$)  & \\
\hline
\endfirsthead
\caption{continued.}\\
\hline \hline
\noalign{\smallskip}
\endhead
\hline
\noalign{\smallskip}
001 & 2009-01-03T01:42:15.951  & Aldebaran  & NB1.04  & $\bullet$ &  64 & 7.2 & 10000 & 0.74 & 1.36\\
002 & 2009-01-03T01:44:38.612  & Aldebaran  & NB1.08  & $\bullet$ &  64 & 7.2 & 10000 & 0.74 & 1.35\\
003 & 2009-01-03T01:46:34.673  & Aldebaran  & NB1.09  & $\bullet$ &  64 & 7.2 & 10000 & 0.71 & 1.35\\
004 & 2009-01-03T01:48:35.599  & Aldebaran  & NB1.24  & $\bullet$ &  64 & 7.2 & 10000 & 0.59 & 1.35\\
005 & 2009-01-03T01:50:45.023  & Aldebaran  & NB1.26  & $\bullet$ &  64 & 7.2 & 10000 & 0.57 & 1.35\\
006 & 2009-01-03T01:52:40.849  & Aldebaran  & NB1.28  & $\bullet$ &  64 & 7.2 & 10000 & 0.75 & 1.34\\
007 & 2009-01-03T01:54:33.505  & Aldebaran  & NB1.64  & $\bullet$ &  64 & 7.2 & 10000 & 0.68 & 1.34\\
008 & 2009-01-03T01:56:57.299  & Aldebaran  & NB1.75  & $\bullet$ &  64 & 7.2 & 10000 & 0.63 & 1.34\\
009 & 2009-01-03T01:59:28.687  & Aldebaran  & NB2.12  & $\bullet$ &  64 & 7.2 & 10000 & 0.60 & 1.34\\
010 & 2009-01-03T02:01:26.846  & Aldebaran  & NB2.17  & $\bullet$ &  64 & 7.2 & 10000 & 0.65 & 1.34\\
011 & 2009-01-03T02:11:42.448  & Betelgeuse  & NB1.75  & $\bullet$ &  64 & 7.2 & 30000 & 0.58 & 1.29\\
012 & 2009-01-03T02:15:35.585  & Betelgeuse  & NB1.64  & $\bullet$ &  64 & 7.2 & 30000 & 0.63 & 1.28\\
013 & 2009-01-03T02:19:43.274  & Betelgeuse  & NB1.28  & $\bullet$ &  64 & 7.2 & 30000 & 0.62 & 1.27\\
014 & 2009-01-03T02:23:51.357  & Betelgeuse  & NB1.26  & $\bullet$ &  64 & 7.2 & 20000 & 0.63 & 1.26\\
015 & 2009-01-03T02:26:46.654  & Betelgeuse  & NB1.24  & $\bullet$ &  64 & 7.2 & 20000 & 0.62 & 1.26\\
016 & 2009-01-03T02:29:42.335  & Betelgeuse  & NB1.09  & $\bullet$ &  64 & 7.2 & 20000 & 0.66 & 1.25\\
017 & 2009-01-03T02:32:36.369  & Betelgeuse  & NB1.08  & $\bullet$ &  64 & 7.2 & 20000 & 0.76 & 1.25\\
018 & 2009-01-03T02:35:30.800  & Betelgeuse  & NB1.04  & $\bullet$ &  64 & 7.2 & 20000 & 0.82 & 1.24\\
019 & 2009-01-03T02:38:31.072  & Betelgeuse  & NB2.12  & $\bullet$ &  64 & 7.2 & 20000 & ... & 1.24\\
020 & 2009-01-03T02:41:59.465  & Betelgeuse  & NB2.17  & $\bullet$ &  64 & 7.2 & 20000 & 0.80 & 1.23\\
021 & 2009-01-03T02:44:26.466  & Betelgeuse  & NB2.17  & $\bullet$ &  64 & 7.2 & 20000 & 0.77 & 1.23\\
\hline
\noalign{\smallskip}
022 & 2009-01-04T00:55:29.183  & $\delta$\,Phe  & NB1.75 & $\circ$ &  64 & 7.2 & 15000 & 0.95 & 1.17\\
023 & 2009-01-04T00:58:11.645  & $\delta$\,Phe  & NB1.64 & $\circ$ &  64 & 7.2 & 15000 & 1.02 & 1.17\\
024 & 2009-01-04T01:00:30.726  & $\delta$\,Phe  & NB2.12 & $\circ$ &  64 & 7.2 & 15000 & 1.13 & 1.17\\
025 & 2009-01-04T01:02:46.968  & $\delta$\,Phe  & NB2.17 & $\circ$ &  64 & 7.2 & 15000 & 1.03 & 1.18\\
026 & 2009-01-04T01:13:17.267  & Aldebaran  & NB1.64 & $\bullet$ &  64 & 7.2 & 15000 & ... & 1.40\\
027 & 2009-01-04T01:16:46.775  & Aldebaran  & NB1.26 & $\bullet$ &  64 & 7.2 & 15000 & ... & 1.39\\
028 & 2009-01-04T01:19:04.804  & Aldebaran  & NB1.08 & $\bullet$ &  64 & 7.2 & 15000 & 1.18 & 1.39\\
029 & 2009-01-04T01:21:27.836  & Aldebaran  & NB2.17 & $\bullet$ &  64 & 7.2 & 15000 & 1.20 & 1.38\\
030 & 2009-01-04T01:23:59.410  & Aldebaran  & NB1.64 & $\bullet$ &  64 & 7.2 & 15000 & 1.68 & 1.38\\
031 & 2009-01-04T01:32:15.918  & Betelgeuse  & NB1.75 & $\bullet$ &  64 & 7.2 & 15000 & 1.41 & 1.41\\
032 & 2009-01-04T01:34:21.554  & Betelgeuse  & NB1.64 & $\bullet$ &  64 & 7.2 & 15000 & 1.43 & 1.41\\
033 & 2009-01-04T01:36:42.290  & Betelgeuse  & NB1.28 & $\bullet$ &  64 & 7.2 & 15000 & 1.42 & 1.40\\
034 & 2009-01-04T01:38:59.301  & Betelgeuse  & NB1.26 & $\bullet$ &  64 & 7.2 & 15000 & 1.23 & 1.39\\
035 & 2009-01-04T01:41:16.077  & Betelgeuse  & NB1.24 & $\bullet$ &  64 & 7.2 & 15000 & 1.22 & 1.38\\
036 & 2009-01-04T01:44:34.612  & Betelgeuse  & NB1.09 & $\bullet$ &  64 & 7.2 & 15000 & 1.17 & 1.37\\
037 & 2009-01-04T01:46:53.081  & Betelgeuse  & NB1.08 & $\bullet$ &  64 & 7.2 & 15000 & 1.01 & 1.36\\
038 & 2009-01-04T01:49:09.422  & Betelgeuse  & NB1.04 & $\bullet$ &  64 & 7.2 & 15000 & 0.99 & 1.35\\
039 & 2009-01-04T01:51:33.722  & Betelgeuse  & NB2.12 & $\bullet$ &  64 & 7.2 & 15000 & 0.86 & 1.34\\
040 & 2009-01-04T01:54:28.303  & Betelgeuse  & NB2.17 & $\bullet$ &  64 & 7.2 & 15000 & 0.75 & 1.33\\
041 & 2009-01-04T02:02:04.380  & Aldebaran  & NB1.64 & $\bullet$ &  64 & 7.2 & 15000 & 0.82 & 1.33\\
042 & 2009-01-04T02:04:48.343  & Aldebaran  & NB1.75 & $\bullet$ &  64 & 7.2 & 15000 & 0.73 & 1.33\\
043 & 2009-01-04T02:07:31.592  & Aldebaran  & NB1.28 & $\bullet$ &  64 & 7.2 & 15000 & 0.66 & 1.33\\
044 & 2009-01-04T02:09:49.050  & Aldebaran  & NB1.26 & $\bullet$ &  64 & 7.2 & 15000 & 0.83 & 1.33\\
045 & 2009-01-04T02:12:05.597  & Aldebaran  & NB1.24 & $\bullet$ &  64 & 7.2 & 15000 & 0.73 & 1.33\\
046 & 2009-01-04T02:14:22.059  & Aldebaran  & NB1.09 & $\bullet$ &  64 & 7.2 & 15000 & 0.64 & 1.33\\
047 & 2009-01-04T02:16:38.434  & Aldebaran  & NB1.08 & $\bullet$ &  64 & 7.2 & 15000 & 0.57 & 1.33\\
048 & 2009-01-04T02:18:55.784  & Aldebaran  & NB1.04 & $\bullet$ &  64 & 7.2 & 15000 & 0.74 & 1.33\\
049 & 2009-01-04T02:22:17.955  & Aldebaran  & NB2.12 & $\bullet$ &  64 & 7.2 & 15000 & 0.69 & 1.33\\
050 & 2009-01-04T02:24:33.229  & Aldebaran  & NB2.17 & $\bullet$ &  64 & 7.2 & 15000 & 0.60 & 1.33\\
051 & 2009-01-04T02:32:50.331  & Betelgeuse  & NB1.64 & $\bullet$ &  64 & 7.2 & 15000 & 0.57 & 1.24\\
052 & 2009-01-04T02:34:58.410  & Betelgeuse  & NB1.75 & $\bullet$ &  64 & 7.2 & 15000 & 0.59 & 1.24\\
053 & 2009-01-04T02:37:17.272  & Betelgeuse  & NB1.28 & $\bullet$ &  64 & 7.2 & 15000 & 0.60 & 1.23\\
054 & 2009-01-04T02:39:29.761  & Betelgeuse  & NB1.26 & $\bullet$ &  64 & 7.2 & 15000 & 0.65 & 1.23\\
055 & 2009-01-04T02:41:46.478  & Betelgeuse  & NB1.24 & $\bullet$ &  64 & 7.2 & 15000 & 0.63 & 1.22\\
056 & 2009-01-04T02:44:04.635  & Betelgeuse  & NB1.09 & $\bullet$ &  64 & 7.2 & 15000 & 0.65 & 1.22\\
057 & 2009-01-04T02:46:21.058  & Betelgeuse  & NB1.08 & $\bullet$ &  64 & 7.2 & 15000 & 0.75 & 1.22\\
058 & 2009-01-04T02:48:37.938  & Betelgeuse  & NB1.04 & $\bullet$ &  64 & 7.2 & 15000 & 0.65 & 1.21\\
059 & 2009-01-04T02:51:02.391  & Betelgeuse  & NB2.12 & $\bullet$ &  64 & 7.2 & 15000 & 0.59 & 1.21\\
050 & 2009-01-04T02:53:19.617  & Betelgeuse  & NB2.17 & $\bullet$ &  64 & 7.2 & 15000 & 0.61 & 1.21\\
051 & 2009-01-04T03:11:04.218  & 31\,Ori  & NB1.64 & $\circ$ &  64 & 7.2 & 15000 & 0.69 & 1.09\\
052 & 2009-01-04T03:13:24.544  & 31\,Ori  & NB1.26 & $\circ$ &  64 & 7.2 & 15000 & 0.69 & 1.09\\
053 & 2009-01-04T03:15:41.604  & 31\,Ori  & NB1.08 & $\circ$ &  64 & 7.2 & 15000 & 0.73 & 1.09\\
054 & 2009-01-04T03:18:05.278  & 31\,Ori  & NB2.17 & $\circ$ &  64 & 7.2 & 15000 & 0.79 & 1.09\\
\hline
\end{longtable}
\begin{list}{}{}
\item[$^{\mathrm{a}}$] The ``$\bullet$" symbol indicates that the neutral density filter was used (``$\circ$" indicates it was not used).
\item[$^{\mathrm{b}}$] DIT is the detector integration time.
\item[$^{\mathrm{c}}$] $\sigma$ is the DIMM seeing at Paranal in the visible and AM is the airmass.
\end{list}
} % End of online table

\end{document}